\documentclass[journal]{IEEEtran}  

\IEEEoverridecommandlockouts

\usepackage{enumitem}

\usepackage{mathtools, cuted}
\usepackage{nicefrac}							% Fractions
\usepackage{siunitx}							% Measurement Units
\usepackage{amssymb}
\usepackage{amsmath}
\usepackage{gensymb}
\usepackage{mathtools}							% For new math symbols such as coloneqq 
\usepackage{trfsigns}							% add abbreviated symbols for Laplace and Fourier trafos

\usepackage{graphicx}							% Already loaded in the class 
\usepackage{xcolor}
\usepackage{tikz}
\usepackage{pgfplots}
\usepackage{grffile}
\usepackage[american, nooldvoltagedirection]{circuitikz}				% for circuit diagrams (Bertus) for more info, see guide at (http://texdoc.net/texmf-dist/doc/latex/circuitikz/circuitikzmanual.pdf)
\usepackage[german,USenglish,UKenglish]{babel}
\usepackage[T1]{fontenc}
\usepackage{bm}									% bold and italic for symbols 
\usepackage{dsfont}								% nicer set notations 
\usepackage{microtype}
\usepackage{xspace}
\usepackage{enumerate}							% personalized enumeration 
\usepackage{cancel}								% cancel out parts of an equation
\usepackage[normalem]{ulem}						% \sout{} for strikethrough text, modifier keeps normal \emph{}
\newtheorem{theorem}{\bf Theorem}			% optional: [section]
\newtheorem{assumption}{\bf Assumption}
\newtheorem{problem}{\bf Problem}

\newtheorem{remark}{\bf Remark}
\newtheorem{proposition}{\bf Proposition}
\newtheorem{lemma}{\bf Lemma}
\newtheorem{definition}{\bf Definition}

\usetikzlibrary{external}
\usetikzlibrary{shapes,arrows}
\usetikzlibrary{calc}
\usetikzlibrary{angles}
\usetikzlibrary{positioning}
\usetikzlibrary{fit}
\usetikzlibrary{backgrounds}
\usetikzlibrary{mindmap,trees}

\pgfplotsset{compat=newest}
\usetikzlibrary{plotmarks}
\usetikzlibrary{arrows.meta}
\usepgfplotslibrary{patchplots}

\usepackage{arydshln}					% for dashed horizontal lines 
\usepackage{hyperref}					% Must be loaded last

	\definecolor{purpleDark}{RGB}{118, 4, 205}
	\definecolor{purpleLight}{RGB}{186, 102, 250}
	\definecolor{blueDark}{RGB}{52, 78, 243}
	\definecolor{blueLight}{RGB}{118, 135, 244}
	\definecolor{redDark}{RGB}{197, 34, 0}
	\definecolor{redLight}{RGB}{255, 91, 57}
	\definecolor{yellowDark}{RGB}{255, 183, 0}
	\definecolor{yellowLight}{RGB}{255, 204, 77}
	\definecolor{greenDark}{RGB}{56, 187, 105}
	\definecolor{greenLight}{RGB}{42, 189, 97}
	
	\colorlet{greenFaint}{green!10!white}
	\colorlet{redFaint}{red!10!white}

\definecolor{colorDGU1}{rgb}{0.00000, 0.24000, 0.72000}
\definecolor{colorDGU2}{rgb}{0.85000, 0.32500, 0.09800}
\definecolor{colorDGU3}{rgb}{0.929, 0.649000, 0.125000}

\definecolor{colorEU4}{rgb}{0.00000, 0.24000, 0.72000}
\definecolor{colorEU5}{rgb}{0.00000, 0.61000, 0.52870}
\definecolor{colorEU6}{rgb}{0, 0 0}
\definecolor{colorEU7}{rgb}{0.4660, 0.6740, 0.1880}
\definecolor{colorEU8}{rgb}{0.49400, 0.18400, 0.55600}

\definecolor{colorHold}{rgb}{0.49400, 0.18400, 0.55600}
\definecolor{colorBoost}{rgb}{0.00000 0.61000 0.52870}
\definecolor{colorMix}{rgb}{0.6350 0.0780 0.1840}

	\renewcommand{\vec}[1]{\bm{#1}}				% Vector underline
	\renewcommand{\matrix}[1]{\bm{#1}}		% Matrix underline			
	
	\newcommand{\Reals}{\mathbb{R}}		% Reals
	\newcommand{\posDef}{\succ}			% Operator for positive definite matrices
	\newcommand{\Hamil}{H}

	\newcommand{\x}{\vec{x}}
	
	\newcommand{\xdot}{\dot{\x}}

	\newcommand{\z}{\vec{z}}

	\newcommand{\zero}{\vec{0}}
	
	\newcommand{\G}{\matrix{G}}
	
	\newcommand{\J}{\matrix{J}}
	\newcommand{\K}{\matrix{K}}
	
	\newcommand{\Q}{\matrix{Q}}
	\newcommand{\R}{\matrix{R}}
	
	\newcommand{\CC}{{\mathcal{C}}}
    \newcommand{\DD}{{\mathcal{D}}}
    \newcommand{\EE}{{\mathcal{E}}}
    \newcommand{\GG}{{\mathcal{G}}}
    \newcommand{\HH}{{\mathcal{H}}}
    
    \newcommand{\KK}{{\mathcal{K}}}
    \newcommand{\LL}{{\mathcal{L}}}
    \newcommand{\MM}{{\mathcal{M}}}
    \newcommand{\NN}{{\mathcal{N}}}
    \newcommand{\PP}{{\mathcal{P}}}
    \newcommand{\TT}{{\mathcal{T}}}
    \newcommand{\VV}{{\mathcal{V}}}

\title{\LARGE \bf
A Unifying Passivity-Based Framework for Pressure and Volume Flow Rate Control in District Heating Networks}

\author{Felix Strehle, Juan E.\ Machado, Michele Cucuzzella, Albertus J.\ Malan, Jacquelien M.A.\ Scherpen, and S{\"o}ren Hohmann% <-this % stops a space
\thanks{$^{**}$~F. Strehle and J. Machado contributed equally.}%
\thanks{$^{***}$~The work of J. Machado was supported by the Dutch
Research Council (NWO) under Grant ESI.2019.005 and  the NWO, ERA-Net Smart Energy Systems and European Union’s Horizon 2020 research and innovation programme under Grant 775970.}
\thanks{F.\ Strehle, A.J.\ Malan, S.\ Hohmann are with the Institute of Control Systems, Karlsruhe Institute of Technology (KIT), 76131, Karlsruhe, Germany. 
	{\tt\small felix.strehle(albertus.malan, soeren.hohmann)@kit.edu}}%
\thanks{J.E.\ Machado, M.\ Cucuzzella  J.M.A.\ Scherpen are with the Jan C. Willems Center for Systems and Control, ENTEG, Faculty of Science and
	Engineering, University of Groningen, Nijenborgh 4, 9747 AG Groningen, the
	Netherlands. {\tt\small j.e.machado.martinez(j.m.a.scherpen)@rug.nl}}%
\thanks{M.\ Cucuzzella is also with the Department of Electrical, Computer and Biomedical Engineering, University of Pavia, via Ferrata 5, 27100 Pavia, Italy. {\tt \small michele.cucuzzella@unipv.it}}
}%

\begin{document}

% TO DO LIST:

% \begin{itemize}
%     \item Remove $K_{\mathrm{I},i}$ from the document. 
%     \item Change shifted passivity to EIP.
%     \item Remove some references and check for ? references
%     \item Streamline presentation of parameters after equations
% \end{itemize}
%\newpage

\maketitle
\thispagestyle{empty}
\pagestyle{empty}
\allowdisplaybreaks

%%%%%%%%%%%%%%%%%%%%%%%%%%%%%%%%%%%%%%%%%%%%%%%%%%%%%%%%%%%%%%%%%%%%%%%%%%%%%%%%
\begin{abstract}
% \todo{Please edit; my ideas are at an end. It is too long and maybe not precise enough yet}

%%%% Purpose; reason for writing; problem
A fundamental precondition for the secure and efficient operation of district heating networks (DHNs) is a  stable hydraulic behavior. 
However, the ongoing transition towards a sustainable heat supply, especially the rising integration of distributed heat sources and the increasingly meshed topologies, introduce complex and potentially destabilizing  hydraulic dynamics.
In this work, we propose a unifying, passivity-based framework which guarantees asymptotic stability of any forced hydraulic DHN equilibrium while allowing for meshed, time-varying topologies and different, dynamically interacting distributed heat sources. To establish the desired hydraulic equilibria, we propose decentralized, passivity-based pressure and volume flow rate controllers for the pumps and valves in the actuated DHN subsystems.
% In this work, we propose decentralized, passivity-based pressure and volume flow rate controllers that guarantee asymptotic stability of any forced hydraulic DHN equilibrium.
%
In particular, we leverage the equilibrium-independent passivity (EIP) properties of the DHN subsystems, the skew-symmetric nature of their interconnections, and LaSalle's Invariance principle to assess asymptotic stability in a modular manner.
% In particular, we use the fact that the equilibrium-independent passivity (EIP) properties of DHN subsystems, the skew-symmetric nature of the subsystem interconnection structure, and LaSalle's Invariance principle can be combined to guarantee asymptotic stability in a modular manner.
%
% {\color{gray}This establishes a unifying framework in which DHN subsystems can be connected and disconnected in a plug-and-play fashion without endangering stability.}
%
The obtained results hold for the state-of-the-art as well as future DHN generations featuring, for example, multiple distributed heat sources, asymmetric pipe networks, and multiple temperature layers. 
%
% as they are based on a comprehensive DHN model featuring multiple distributed heat sources, distributed variable-speed pump (DVSP) configurations with pumps at consumers and at pipes, asymmetric, meshed pipe networks, multiple temperature layers as well as temperature boosting.
% Moreover, we incorporate dynamic models for the pumps, consider valves as actuators, and explicitly account for the presence of pressure holding units.
%
We verify our findings by means of simulations. 
%The theoretical findings are verified by means of simulations. 

% %%%% Methodology
% In this work, we show how the EIP properties of DHN subsystems, the skew-symmetric nature of the subsystem interconnection structure, and LaSalle's Invariance principle can be combined to guarantee asymptotic stability of any forced, hydraulic DHN equilibrium in a modular manner.
% %%%% Results
% This establishes a unifying framework in which subsystems, if EIP, can connect and disconnect in a plug-and-play fashion without endangering stability.
% Following this EIP requirement, we then design decentralized, passivity-based pressure and volume flow rate controllers for the DHN actuators, i.e., pumps and valves.
% %%%% Setup
% For our considerations, we provide a comphrehensive DHN model which covers current as well as future DHN generations featuring multiple distributed heat sources, distributed variable-speed pump (DVSP) configurations with pumps at consumers and at pipes, asymmetric, meshed pipe networks, multiple temperature layers as well as temperature boosting.
% Moreover, we incorporate dynamic models for the pumps, consider valves as actuators, and explicitly account for the presence of pressure holding units. 

\end{abstract}

%%%%%%%%%%%%%%%%%%%%%%%%%%%%%%%%%%%%%%%%%%%%%%%%%%%%%%%%%%%%%%%%%%%%%%%%%%%%%%%%
\section{Introduction} \label{sec:intro}

District heating networks (DHNs) are a key element for a holistic energy transition, particularly in densely populated areas \cite{Lund14,Vandermeulen18,Novitsky20,Wang17conversion}.
For their operation, well-defined and stable hydraulic conditions are a fundamental requirement as they form the basis for the actual thermal power flows \cite{Novitsky20,Pan16}. 
% Particularly in the transient regime, i.e., few seconds after the occurrence of a disturbance or alteration of the desired operating point, the hydraulic processes govern the system dynamics \cite{Pan16,Chertkov19}.

%%%%%%%%%%%% SOTA hydraulics
In the operation of traditional 2nd or 3rd generation DHNs, the hydraulics and thus the thermal power flows are well-understood (see e.g., \cite[pp.~52--54]{Nussbaumer20}). 
%
%%%%%%%%%% Problems new DHNs
However, emerging 4th generation DHNs bring about challenges that call for new strategies and methods of operating, controlling and analyzing DHNs \cite{Lund14,Vandermeulen18,Novitsky20, Strehle21dhn}.\footnote{See \cite{Lund14} for a comparison and overview of the different DHN generations.}
%
%%%%%%%%%% Decentralization with distributed actuators as main trend
Most prominently, we can observe a \emph{decentralization} trend with several interacting subsystems and controllable components (pumps, control valves).
Primarily, this is due to the integration of ever more renewable and  distributed heat generation units (DGUs) such as heat pumps, waste heat recycling, combined heat and power, waste/biomass-to-energy, solar, and geothermal heat plants, and  \cite{Lund14,Vandermeulen18}.
%
% DVSPs
Additionally, distributed variable-speed pumps (DVSPs) installed at every DGU and some (up to all) consumer substations, have shown considerable potential to reduce the overall electrical energy required to operate DHN pumps \cite{Wang17conversion,Yan13, Gong19}.

Furthermore, we see a strive towards a more efficient DHN operation with lower water temperatures and smaller pipe diameters, novel topologies with multiple temperature layers, and changing hydraulic conditions \cite{Lund14, Vandermeulen18}. 
%
% Explanation/comparison DHN generations
Supply/return temperatures, for example, are decreasing from around \SIrange{80}{120}{\degreeCelsius}/\SIrange{40}{70}{\degreeCelsius} in 2nd or 3rd generation DHNs to \SIrange{40}{70}{\degreeCelsius}/\SIrange{20}{40}{\degreeCelsius} in 4th generation DHNs \cite{Lund14, Novitsky20}\cite[p.~44]{Nussbaumer20}. This, together with decreasing the pipe diameters, allows to efficiently integrate renewable heat sources and new consumers (e.g., low-energy buildings) and  reduces the heat distribution losses \cite[p.~44]{Nussbaumer20} \cite{Koefinger17, Volkova20, Volkova22}.
Besides that, different temperature layers can be combined to multi-layer topologies in future DHNs to increase the efficiency
(see temperature cascading in e.g., \cite[p.~44]{Nussbaumer20} \cite{Koefinger17, Volkova20, Volkova22}). Indeed, the return pipes of a traditional 2nd or 3rd generation DHN may serve as supply for a (new) low-temperature DHN part. 
%
% Booster pumps
However, to ensure a proper heat supply, such new operation modes often require additional booster pumps, i.e., controllable devices placed at consumers and at strategic points in the pipe network  \cite{Lund14,Persis11}\cite[p.~54]{Nussbaumer20}.

%%%% Summary from control point of view: topology changes, PnP, more system dynamics and interactions on small time scales 
In summary, from a control point of view, the decentralization trend with the increasingly actuated subsystems, the naturally intermittent supply behavior of DGUs \cite{Vandermeulen18,Lorenzi20}, and the new operation strategies lead to more frequent \emph{topology changes} and \emph{plug-in and plug-out operations} of units. Additionally, \emph{complex pressure and volume flow dynamics} and \emph{interactions on small time scales} are introduced \cite{Novitsky20}.
For example, it is possible to observe more frequent volume flow reversals in pipes  \cite{Heijde17dynamic,Mohring21} and interactions between the pump and valve controllers, which may lead to severe hydraulic oscillations \cite{Wang17conversion,Yan13,Sommer19}.

%%%%%%%%%%%% Analogy DHN and MG -> look there 
Upon closer examination, the discussion above reveals parallels to the trends and challenges in modern power systems such as microgrids (see for example \cite{Guerrero13} or the discussion in \cite{Strehle21dhn}).
To provide stabilization in the face of intermittent generation and distributed units, the power system control community is looking for decentralized control solutions.
Thus, the similar trends and challenges motivate to explore \emph{decentralized} control designs for novel pressure and volume flow rate controllers.
Such decentralized controllers provide \emph{scalability} and allow for an easy addition or removal of subsystems in a \emph{plug-and-play} fashion without requiring communication, without adapting the other controllers, and without endangering pressure and volume flow rate stability.
%
%%%% Passivity and PHS is promising for MGs -> also here
Recent works on microgrids (see e.g., \cite{Strehle22, Watson21}) have shown that the compositional properties of passive systems present a promising, unifying framework for designing decentralized controllers and realizing modular energy systems in which topologies change frequently and different controllers interact dynamically via the network.

%%%%%%%%%%%%%% SOTA
%%%% Decentralized passivity-based hydraulic control
In the literature, the field of decentralized, passivity-based hydraulic control of DHNs has recently been explored in \cite{Machado22adaptive}. However, the considered DHN model exhibits a number of restrictions such as symmetric DHN topologies\footnote{In symmetric DHN topologies, supply and return pipes are laid in parallel. This excludes practically relevant cases with meshed supply pipe networks and tree-like return pipe networks or more complex structures arising in multi-layer topologies.},
two temperature layers only, static pump models, no pressure holding units, and valves modelled as non-actuated components. Furthermore, pumps are assumed to be installed at every producer and every consumer, which excludes traditional DHNs in which consumers regulate their volume flow rates only via control valves.
%
%%%% Pressure control
The same restrictions underlie the DHN models used in \cite{Persis11,Jensen12,Persis14,Scholten17}, where, additionally, only single producer DHNs are considered.
%
%%%% Preliminary works that contribute towards PBC design due to PHS modeling 
Further noteworthy works that contribute towards a passivity-based control design and analysis of DHNs by introducing port-Hamiltonian system (PHS) models are \cite{Hauschild20,Strehle22dhn}; see also \cite{Perryman22}, where water distribution networks are modeled as PHSs. However, \cite{Hauschild20} exhibits  the same setup restrictions as \cite{Persis11,Jensen12,Persis14,Scholten17}, while \cite{Strehle22dhn} overcomes most of these restrictions with the exception of modeling  valves  as non-actuated components and considering pumps installed at every producer and consumer.

%%%%% Own contribution
\subsection{Contributions}

In this work, we present a unifying framework based on equilibrium-independent passivity (EIP) which allows for decentralized pressure and volume flow rate control in DHNs with meshed, time-varying topologies and different, dynamically interacting DGUs.

Firstly, we formalize the considered DHN setups as digraphs. 
Our considerations overcome the restrictions in the above-referenced literature and are based on a comprehensive dynamic hydraulic DHN model that encompasses traditional 2nd and 3rd generation DHNs, future 4th generation DHNs as well as intermediate stages.
That is, we allow for an arbitrary number of DGUs and consumers to be connected in a meshed DHN topology with multiple temperature layers (temperature cascading), DVSP configurations, and optional booster pumps at consumers and pipes.
Moreover, we incorporate dynamic models for the pumps, consider valves as actuators, and explicitly account for the presence of pressure holding units.

% {\color{red}
Secondly, we derive explicit input-state-output port-Hamiltonian system (ISO-PHS) models for each of the DHN subsystems.
Although not strictly necessary for the subsequent control design and modular stability analysis, the systematic PHS modeling is nevertheless useful during these steps. In particular, it gives a clear perspective on which input-output ports are accessible for control and over which ports subsystems interact with each other. Furthermore, the passivity properties with respect to these ports and the Hamiltonian as storage function candidate are directly visible.
% }

Thirdly, we propose pressure controllers for pumps, based on an algebraic interconnection and damping assignment (IDA) \cite{Ortega04} extended with integral action on the non-passive output \cite{Donaire09}; volume flow rate controllers for pumps, based on a state-feedback with integral action on the passive output; and volume flow rate controllers for valves, based on a proportional-integral (PI) control of a modified passive output inspired by \cite{Monshizadeh19}.

Fourthly, by leveraging the EIP properties of the subsystems, the skew-symmetric nature of their interconnection structure, and LaSalle's invariance principle, we prove asymptotic stability of any forced hydraulic DHN equilibrium in a modular, bottom-up manner.
This establishes a unifying framework for pressure and volume flow rate control, where multiple DHN subsystems, if EIP, can enter or leave the DHN in a plug-and-play fashion without having any impact on the stability properties of the hydraulic equilibrium.
Simulations based on realistic DHN data validate our theoretical results.

% {\color{gray}
% \subsection{Preliminaries and Notation}
% \subsubsection{Sets, vectors, and functions} 
% We let $\mathbb{R}$ (resp. $\mathbb{R}_{>0}$) denote the set of real (resp. strictly positive real) numbers. 
% Given $ \x \in \mathbb{R}^{n}$, $\text{diag}(\x) \in \mathbb{R}^{n \times n}$ is the associated diagonal matrix with $\x$ on the diagonal. 
% For an ordered, finite index set $A$, we denote by $\text{diag}(x_i)_{i\in A}$ a diagonal matrix with the elements of its main diagonal given by the components of $x_i\in\mathbb{R}^{n_i}$, for each $i\in A$.
% The notation $A \succ 0$ ($A \succeq 0$) represents a positive definite  (positive semidefinite) matrix or function. 
% Throughout, 
% % $\vec{1}_n$ and $\vec{0}_n$ are $n$-dimensional vectors of unit and zero entries, whereas 
% $\zero_n$, $\one_n$ are $n$-dimensional vectors of zero and unit entry, whereas $\Zero_{n\times n}$, $\One_{n\times n}$ are $n\times n$-dimensional zero and identity matrices.
% A setpoint to be established in equilibrium is denoted by $(\cdot)^*$, whereas a generally unknown equilibrium is denoted by $\bar{(\cdot)}$. 
% % The double notation $\bar{(\cdot)}^*$ indicates that an equilibrium comprises both assigned and unknown parts.
% % \sout{Functions and matrices of the desired \emph{closed-loop} system are denoted with the subscript $\sDesired$.} 
% For any $x\in \mathbb{R}^n$ and any symmetric, positive-definite matrix $P\in \mathbb{R}^{n\times n}$, $\Vert x \Vert_P^2=x^\top P x$.}
\section{System Setup} \label{sec:setup}
%%%% New
% - asymmetric, meshed topologies
% - multiple temperature layers
% - multiple DGUs
% - possible booster pumps in pipes and consumers;     DVSP configurations
% - mixing connections for temperature boosting
%
In this section, we outline the general DHN setups that are covered by our approach and formally describe them as weakly connected digraphs.
We allow for DHNs with asymmetric, meshed topologies, multiple temperature layers, multiple DGUs, pressure holding units, booster pumps at consumers and pipes, DVSP configurations as well as mixing connections used to increase the water temperature in topologies with multiple temperature layers.
To the best of our knowledge, such a comprehensive collection of DHN setups covering 2nd, 3rd, 4th generation DHNs as well as intermediate development stages has not been considered before in any control-oriented literature.

\subsection{DHN Setup and Digraph Representation} \label{sec:modeling:dhn_graph}

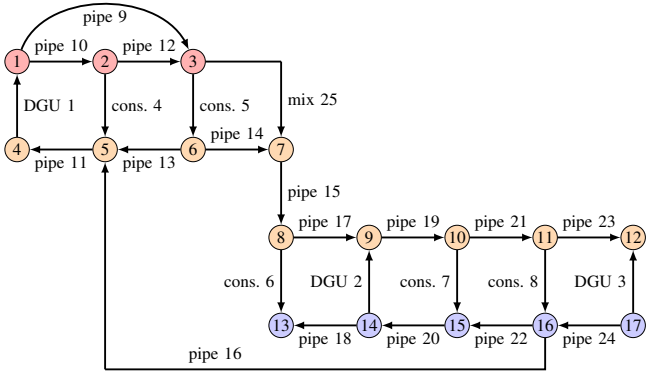
\begin{figure} 
	\centering
	\scalebox{0.65}{
		\begin{tikzpicture}[node_high/.style={circle,draw, minimum size=0.5cm,inner sep=0pt,fill=red!30},
                            node_med/.style={circle,draw, minimum size=0.5cm,inner sep=0pt,fill=orange!30},
                            node_cold/.style={circle,draw, minimum size=0.5cm,inner sep=0pt,fill=blue!20}]

			\def\xDist{1.8}
			\def\yDist{1.8}
			\def\lineWidth{1.1pt}
			
			% Supply nodes
			\node[node_high] (n1) at (0,0) {1};
			\node[node_high] (n2) at (1*\xDist,0) {2};
			\node[node_high] (n3) at (2*\xDist,0) {3};

			% Return layer 1 nodes
			\node[node_med] (n4) at (0, -1*\yDist)  {4};
			\node[node_med] (n5) at (1*\xDist, -1*\yDist)  {5};
			\node[node_med] (n6) at (2*\xDist, -1*\yDist)  {6};
			\node[node_med] (n7) at (3*\xDist, -1*\yDist)  {7};
   
			\node[node_med] (n8) at (3*\xDist, -2*\yDist)  {8};
			\node[node_med] (n9) at (4*\xDist, -2*\yDist)  {9};
            \node[node_med] (n10) at (5*\xDist, -2*\yDist)  {10};
            \node[node_med] (n11) at (6*\xDist, -2*\yDist)  {11};
            \node[node_med] (n12) at (7*\xDist, -2*\yDist)  {12};
			
            % Return layer 2 nodes
			\node[node_cold] (n13) at (3*\xDist, -3*\yDist)  {13};
            \node[node_cold] (n14) at (4*\xDist, -3*\yDist)  {14};
            \node[node_cold] (n15) at (5*\xDist, -3*\yDist)  {15};
            \node[node_cold] (n16) at (6*\xDist, -3*\yDist)  {16};
            \node[node_cold] (n17) at (7*\xDist, -3*\yDist)  {17};
			
			%Layers
			% \node[left=of n1,anchor=center] {\textbf{high}};
			% \node[left=of n4,anchor=center] {\textbf{medium}};
			% \node(naux) at (0, -2*\yDist){};
			% \node[left=of naux ,anchor=center] {\textbf{low}};
			
			%Pipes				
			\foreach \from/\to/\l in {n1/n2/10,n5/n4/11,n2/n3/12,n6/n5/13,n6/n7/14,n7/n8/15,n8/n9/17,n14/n13/18,n9/n10/19,n15/n14/20,n10/n11/21,n16/n15/22,n11/n12/23,n17/n16/24}
			\draw[->, -latex,line width=\lineWidth] (\from) -- (\to) node [midway, auto] () {pipe \l}; 
			% Extra pipe 9 with arc
			\draw[->,-latex,line width=\lineWidth] (n1) to[out=70,in=110] node[midway,auto,below] (){pipe 9} (n3);
            % Extra pipe 16 with right angle
			\draw[line width=\lineWidth] (n16) -- (6*\xDist, -3.5*\yDist) node [midway, auto] (){} -- (1*\xDist, -3.5*\yDist) node [midway, near end,above] () {pipe 16};
			\draw[->, -latex,line width=\lineWidth] (1*\xDist, -3.5*\yDist)-- (n5);
			
			% DGUs
            % Right label
            \foreach \from/\to/\l in {n4/n1/1}
			\draw[->, -latex,line width=\lineWidth] (\from) -- (\to) node [midway, right] () {DGU \l};
            % Left label
			\foreach \from/\to/\l in {n14/n9/2,n17/n12/3}
			\draw[->, -latex,line width=\lineWidth] (\from) -- (\to) node [midway, auto] () {DGU \l}; 
			
			% consumer
            % Left label
			\foreach \from/\to/\l in {n8/n13/6,n10/n15/7,n11/n16/8}
			\draw[->, -latex,line width=\lineWidth] (\from) -- (\to) node [midway, left] () {cons. \l}; 
            % Right label
            \foreach \from/\to/\l in {n2/n5/4,n3/n6/5}
			\draw[->, -latex,line width=\lineWidth] (\from) -- (\to) node [midway, right] () {cons. \l};
			
			% Mixing (right angle)
			\draw[line width=\lineWidth] (n3) -- (3*\xDist, 0);
			\draw[->, -latex,line width=\lineWidth] (3*\xDist, 0) -- (n7)  node [midway,right] () {mix 25}; 
	\end{tikzpicture}}
	\caption{Digraph representation of an exemplary DHN containing three DGUs $\DD=\{1,2,3\}$, five consumers $\LL=\{4,5,6,7,8\}$, 16 pipes $\PP=\{9,\dots,24\}$, and one mixing connection $\MM=\{25\}$ in a three temperature layer topology indicated by the three different colors; the 17 nodes represent one pressure holding unit $\HH=\{4\}$, one ideal junction $\KK=\{7\}$, and 15 elasticity capacitors $\CC=\NN\backslash \{4,7\}$.}
	\label{fig:dhn_graph}
\end{figure}

We describe a DHN as a weakly connected digraph $\GG=(\mathcal{N}, \EE)$ without self-loops as shown in Fig.~\ref{fig:dhn_graph}.  The edges $\EE$ are partitioned into four sets: $\DD=\{1,\dots,D\}, D\geq1,$ represents the DGUs, $\LL= \{D+1,\cdots,D+L\}, L\geq1,$ the consumers (loads), $\PP= \{D+L+1,\cdots,D+L+P\},P\geq2,$ the pipes, and $\MM= \{D+L+P+1,\cdots,D+L+P+M\},M\geq0,$ the mixing connections.
Conventionally, the  nodes $\NN$ correspond to ideal system junctions interconnecting DGUs, consumers, and mixing connections with the pipe network of the DHN. At an ideal junction, all volume flow rates sum up to zero.
However, in this work, we also view pressure holding units and elasticity capacitors arising from equipment in the DGU and consumer circuits as nodes. 
Therefore, we assume that $\NN=\mathcal{H}\cup \mathcal{C} \cup \mathcal{K}$, where $\mathcal{H}$ are pressure holding units, $\mathcal{C}$ are elasticity capacitors, and $\mathcal{K}$ are the remaining ideal junctions.

The orientation of the edges represents the arbitrary reference direction of positive flows. Moreover, for any  $i\in\mathcal{E}$, $\NN_i^-$ and  $\NN_i^+$ denote its source and target node, respectively. Analogously, for a given node $j\in \NN$, $\mathcal{E}_j^-$ and $\mathcal{E}_j^{+}$ denote the sets of edges with $j$ as source node and $j$ as target node, respectively. 
\footnote{
% {\color{red} 
In practice DGUs and consumers are always connected via pipes and never directly connect to the same node (see Fig.~\ref{fig:dhn_graph}).} 
% }

%

% Layers
%
\subsection{Temperature and Hydraulic Layers} \label{sec:modeling:layers}
\begin{figure} 
	\centering
	\scalebox{0.75}{
		\begin{tikzpicture}[main_node/.style={circle,draw, fill=white,minimum size=0.5cm,inner sep=0pt,outer sep=0.15cm}, temp_cold/.style={rectangle, draw, rounded corners, minimum width=3cm,fill=blue!30, anchor=west}]	
			
			\def\xDist{6}
			\def\yDist{1.1}
			\def\lineWidth{1.1pt}
			\def\layerDist{0.15cm}
			
			%high 3rd gen
			\node[rectangle, draw, rounded corners, minimum width=5cm,fill=red!30, anchor=west] (high1) at (0,0.5*\yDist) {high (\SIrange{80}{120}{\degreeCelsius})};
			
			%Medium 3rd gen
			\node[rectangle, draw, rounded corners, minimum width=5cm,fill=orange!30, anchor=west] (medium1) at (0,-\yDist) {medium (\SIrange{40}{70}{\degreeCelsius})};
			
			%DGUs
			\draw[->,-latex,line width=\lineWidth] (1.375,|-medium1.north)--(1.375,|-high1.south); 
			\draw[->,-latex,line width=\lineWidth] (1.625,|-medium1.north)--(1.625,|-high1.south);
			\node[main_node](dgus) at (1.5,-0.25*\yDist){$\DD$};
			%Loads
			\draw[->,-latex,line width=\lineWidth] (3.375,|-high1.south)--(3.375,|-medium1.north); 
			\draw[->,-latex,line width=\lineWidth] (3.625,|-high1.south)--(3.625,|-medium1.north);
			\node[main_node](loads) at (3.5,-0.25*\yDist){$\LL$};

			%Medium 4th gen
			\node[rectangle, draw, rounded corners, minimum width=5cm,fill=orange!30, anchor=west] (medium2) at (\xDist,0.5*\yDist) {medium (\SIrange{40}{70}{\degreeCelsius})};
			
			%Low 4th gen
			\node[rectangle, draw, rounded corners, minimum width=5cm,fill=blue!20, anchor=west] (low2) at (\xDist,-\yDist) {low (\SIrange{20}{40}{\degreeCelsius})};
			
			%DGUs
			\draw[->,-latex,line width=\lineWidth] (1.375+\xDist,|-medium1.north)--(1.375+\xDist,|-high1.south); 
			\draw[->,-latex,line width=\lineWidth] (1.625+\xDist,|-medium1.north)--(1.625+\xDist,|-high1.south);
			\node[main_node](dgus) at (1.5+\xDist,-0.25*\yDist){$\DD$};
			%Loads
			\draw[->,-latex,line width=\lineWidth] (3.375+\xDist,|-high1.south)--(3.375+\xDist,|-medium1.north); 
			\draw[->,-latex,line width=\lineWidth] (3.625+\xDist,|-high1.south)--(3.625+\xDist,|-medium1.north);
			\node[main_node](loads) at (3.5+\xDist,-0.25*\yDist){$\LL$};
			
			%%%%%%%%%%%%%%%
			%Hydraulic Layers Auxiliary coordinates
			\coordinate[above left=\layerDist of high1](alh1);
			\coordinate[above right=\layerDist of high1](arh1);
			\coordinate[below right=\layerDist of high1](brh1);
			\coordinate[below left=\layerDist of high1](blh1);
			
			\coordinate[above left=\layerDist of medium1](alm1);
			\coordinate[above right=\layerDist of medium1](arm1);
			\coordinate[below right=\layerDist of medium1](brm1);
			\coordinate[below left=\layerDist of medium1](blm1);

			\coordinate[above left=\layerDist of medium2](alm2);
			\coordinate[above right=\layerDist of medium2](arm2);
			\coordinate[below right=\layerDist of medium2](brm2);
			\coordinate[below left=\layerDist of medium2](blm2);	
			
			\coordinate[above left=\layerDist of low2](all2);
			\coordinate[above right=\layerDist of low2](arl2);
			\coordinate[below right=\layerDist of low2](brl2);
			\coordinate[below left=\layerDist of low2](bll2);
			
			% Hydraulic layers
			\begin{scope}[on background layer]
				\draw[dotted,line width=\lineWidth,fill=lightgray,rounded corners](arh1)--(brh1)--(blh1)--(alh1)--cycle;
			\end{scope}
			
			\begin{scope}[on background layer]
				\draw[dotted,line width=\lineWidth,fill=lightgray,rounded corners](arm1)--(brm1)--(blm1)--(alm1)--cycle;
			\end{scope}	
		
			\begin{scope}[on background layer]
				\draw[dotted,line width=\lineWidth,fill=lightgray,rounded corners](arm2)--(brm2)--(blm2)--(alm2)--cycle;
			\end{scope}	
		
			\begin{scope}[on background layer]
				\draw[dotted,line width=\lineWidth,fill=lightgray,rounded corners](arl2)--(brl2)--(bll2)--(all2)--cycle;
			\end{scope}
	\end{tikzpicture}}
	\caption{Illustration of a traditional 2nd or 3rd generation DHN with high temperature supply and medium temperature return (left) and a 4th generation DHN with medium temperature supply and low temperature return (right); between the temperature layers there may be an any number of $D\geq1$ DGU and $L\geq1$ consumer edges; the temperature layers coincide with the two hydraulic layers (dashed grey bubbles).}
	\label{fig:dhn_3rd_4th_gen}
\end{figure}
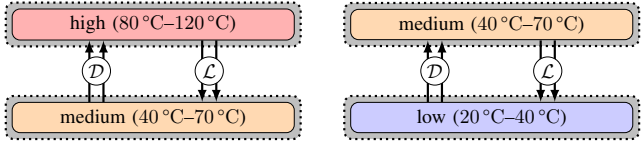
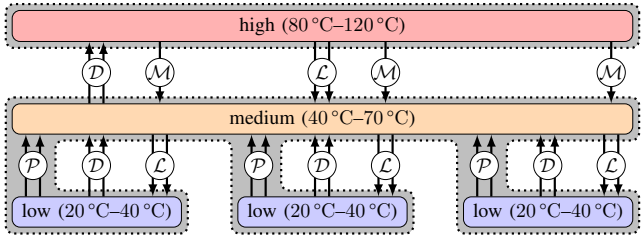
\begin{figure} 
	\centering
	\scalebox{0.75}{
		\begin{tikzpicture}[main_node/.style={circle,draw, fill=white,minimum size=0.5cm,inner sep=0pt,outer sep=0.15cm}, temp_cold/.style={rectangle, draw, rounded corners, minimum width=3cm,fill=blue!20, anchor=west}]	
			
		\def\xDist{4}
		\def\yDist{1.1}
		\def\lineWidth{1.1pt}
		\def\layerDist{0.15cm}
		
		%High
		\node[rectangle, draw, rounded corners, minimum width=11cm,fill=red!30, anchor=west] (high) at (0,\yDist) {high (\SIrange{80}{120}{\degreeCelsius})};
		
		%%%%%%%%%%%		
		%Medium
		\node[rectangle, draw, rounded corners, minimum width=11cm,fill=orange!30, anchor=west] (medium) at (0,-0.5*\yDist) {medium (\SIrange{40}{70}{\degreeCelsius})};
		%DGUs
		\draw[->,-latex,line width=\lineWidth] (1.375,|-medium.north)--(1.375,|-high.south); 
		\draw[->,-latex,line width=\lineWidth] (1.625,|-medium.north)--(1.625,|-high.south);
		\node[main_node](dgus) at (1.5,0.25*\yDist){$\DD$};
		%Loads
		\draw[->,-latex,line width=\lineWidth] (1.375+\xDist,|-high.south)--(1.375+\xDist,|-medium.north); 
		\draw[->,-latex,line width=\lineWidth] (1.625+\xDist,|-high.south)--(1.625+\xDist,|-medium.north);
		\node[main_node](loads) at (1.5+\xDist,0.25*\yDist){$\LL$};
		%Mixing
		\draw[->,-latex,line width=\lineWidth] (2.625,|-high.south)--(2.625,|-medium.north); 
		\node[main_node](dgus2) at (2.625,0.25*\yDist){$\MM$};
		\draw[->,-latex,line width=\lineWidth] (2.625+\xDist,|-high.south)--(2.625+\xDist,|-medium.north); 
		\node[main_node](dgus2) at (2.625+\xDist,0.25*\yDist){$\MM$};
		\draw[->,-latex,line width=\lineWidth] (2.625+2*\xDist,|-high.south)--(2.625+2*\xDist,|-medium.north); 
		\node[main_node](dgus2) at (2.625+2*\xDist,0.25*\yDist){$\MM$};
		
		%%%%%%%%%%%%%%%%%%
		%Cold 1
		\node[temp_cold] (cold1) at (0,-2*\yDist) {low (\SIrange{20}{40}{\degreeCelsius})};
		%Pipes to medium
		\draw[->,-latex,line width=\lineWidth] (0.25,|-cold1.north)--(0.25,|-medium.south); 
		\draw[->,-latex,line width=\lineWidth] (0.5,|-cold1.north)--(0.5,|-medium.south);
		\node[main_node](pipes1) at (0.375,-1.25*\yDist){$\PP$};
		%DGUs
		\draw[->,-latex,line width=\lineWidth] (1.375,|-cold1.north)--(1.375,|-medium.south); 
		\draw[->,-latex,line width=\lineWidth] (1.625,|-cold1.north)--(1.625,|-medium.south);
		\node[main_node](dgus1) at (1.5,-1.25*\yDist){$\DD$};
		%Loads
		\draw[->,-latex,line width=\lineWidth] (2.75,|-medium.south)--(2.75,|-cold1.north); 
		\draw[->,-latex,line width=\lineWidth] (2.5,|-medium.south)--(2.5,|-cold1.north);
		\node[main_node](loads1) at (2.625,-1.25*\yDist){$\LL$};
		
		%%%%%%%%%%%%%%%%
		%Cold 2
		\node[temp_cold] (cold2) at (\xDist,-2*\yDist) {low (\SIrange{20}{40}{\degreeCelsius})};
%		%Pipes2 to medium
		\draw[->,-latex,line width=\lineWidth] (0.25+\xDist,|-cold2.north)--(0.25+\xDist,|-medium.south); 
		\draw[->,-latex,line width=\lineWidth] (0.5+\xDist,|-cold2.north)--(0.5+\xDist,|-medium.south);
		\node[main_node](pipes2) at (0.375+\xDist,-1.25*\yDist){$\PP$};
		%DGUs
		\draw[->,-latex,line width=\lineWidth] (1.375+\xDist,|-cold2.north)--(1.375+\xDist,|-medium.south); 
		\draw[->,-latex,line width=\lineWidth] (1.625+\xDist,|-cold2.north)--(1.625+\xDist,|-medium.south);
		\node[main_node](dgus2) at (1.5+\xDist,-1.25*\yDist){$\DD$};
		%Loads
		\draw[->,-latex,line width=\lineWidth] (2.75+\xDist,|-medium.south)--(2.75+\xDist,|-cold2.north); 
		\draw[->,-latex,line width=\lineWidth] (2.5+\xDist,|-medium.south)--(2.5+\xDist,|-cold2.north);
		\node[main_node](loads2) at (2.625+\xDist,-1.25*\yDist){$\LL$};
		
		%%%%%%%%%%%%%%%
		%Cold 3
		\node[temp_cold] (cold3) at (2*\xDist,-2*\yDist) {low (\SIrange{20}{40}{\degreeCelsius})};
		%Pipes3 to medium
		\draw[->,-latex,line width=\lineWidth] (0.25+2*\xDist,|-cold2.north)--(0.25+2*\xDist,|-medium.south); 
		\draw[->,-latex,line width=\lineWidth] (0.5+2*\xDist,|-cold2.north)--(0.5+2*\xDist,|-medium.south);
		\node[main_node](pipes3) at (0.375+2*\xDist,-1.25*\yDist){$\PP$};
		%DGUs
		\draw[->,-latex,line width=\lineWidth] (1.375+2*\xDist,|-cold2.north)--(1.375+2*\xDist,|-medium.south); 
		\draw[->,-latex,line width=\lineWidth] (1.625+2*\xDist,|-cold2.north)--(1.625+2*\xDist,|-medium.south);
		\node[main_node](dgus3) at (1.5+2*\xDist,-1.25*\yDist){$\DD$};
		%Loads
		\draw[->,-latex,line width=\lineWidth] (2.75+2*\xDist,|-medium.south)--(2.75+2*\xDist,|-cold2.north); 
		\draw[->,-latex,line width=\lineWidth] (2.5+2*\xDist,|-medium.south)--(2.5+2*\xDist,|-cold2.north);
		\node[main_node](loads3) at (2.625+2*\xDist,-1.25*\yDist){$\LL$};
		
		%%%%%%%%%%%%%%%
		%Hydraulic Layers Auxiliary coordinates
		\coordinate[above left=\layerDist of high](alh);
		\coordinate[above right=\layerDist of high](arh);
		\coordinate[below right=\layerDist of high](brh);
		\coordinate[below left=\layerDist of high](blh);
		
		\coordinate[above left=\layerDist of medium](alm);
		\coordinate[above right=\layerDist of medium](arm);
		\coordinate[below right=\layerDist of medium](brm);
		\coordinate[below left=\layerDist of medium](blm);
		
		\coordinate[above right=\layerDist of cold1](arc1);
		\coordinate[below right=\layerDist of cold1](brc1);
		\coordinate[below left=\layerDist of cold1](blc1);
		
		\coordinate[above right=\layerDist of cold2](arc2);
		\coordinate[below right=\layerDist of cold2](brc2);
		\coordinate[below left=\layerDist of cold2](blc2);
		
		\coordinate[above right=\layerDist of cold3](arc3);
		\coordinate[below right=\layerDist of cold3](brc3);
		\coordinate[below left=\layerDist of cold3](blc3);
		
		\coordinate (arpipes1) at (pipes1.east|-brm);		
		\coordinate (brpipes1) at (pipes1.east|-arc1);
		
		\coordinate (arpipes2) at (pipes2.east|-brm);		
		\coordinate (brpipes2) at (pipes2.east|-arc2);
		\coordinate (alpipes2) at (blc2|-brm);
		
		\coordinate (arpipes3) at (pipes3.east|-brm);		
		\coordinate (brpipes3) at (pipes3.east|-arc3);
		\coordinate (alpipes3) at (blc3|-brm);
		
		% Hydraulic layer1
		\begin{scope}[on background layer]
			\draw[dotted,line width=\lineWidth,fill=lightgray,rounded corners](arh)--(brh)--(blh)--(alh)--cycle;
		\end{scope}
		
		%Hydraulic layer2
		\begin{scope}[on background layer]
			\draw[dotted,line width=\lineWidth,fill=lightgray,rounded corners](alm)--(arm)--(brm)--(arpipes3)--(brpipes3)--(arc3)--(brc3)--(blc3)--(alpipes3)--(arpipes2)--(brpipes2)--(arc2)--(brc2)--(blc2)--(alpipes2)--(arpipes1)--(brpipes1)--(arc1)--(brc1)--(blc1)--cycle;
		\end{scope}

	\end{tikzpicture}}
	\caption{Illustration of a DHN with the three temperature layers high (red), medium (orange), low (blue), and the two hydraulic layers (dashed grey bubbles); between the temperature layers there may be any number of $D\geq1$ DGU and $L\geq1$ consumer edges; between the high and medium layer may be any number of $M\geq0$ mixing connections.}
	\label{fig:dhn_3_layers}
\end{figure}
As illustrated in Figs.~\ref{fig:dhn_graph}--\ref{fig:dhn_3_layers}, a DHN may comprise different temperature layers. 
We distinguish between three temperature layers, i.e., \emph{high temperature} (\SIrange{80}{120}{\degreeCelsius}), \emph{medium temperature} (\SIrange{40}{70}{\degreeCelsius}), and \emph{low temperature} (\SIrange{20}{40}{\degreeCelsius}) \cite{Lund14, Volkova20, Volkova22, Novitsky20}\cite[pp.~16--17]{Schmidt17}\cite[p.~44]{Nussbaumer20}. 
%
% Explanatory remark
The high and medium temperature layers form the supply and return of the dominating 2nd and 3rd generation DHNs, while the medium and low temperature layers form the supply and return of the emerging 4th generation DHNs (see Fig.~\ref{fig:dhn_3rd_4th_gen}) \cite{Lund14}\cite[pp.~16--17]{Schmidt17}.

In future DHNs, the medium temperature return of a 2nd or 3rd generation DHN may additionally serve as supply for (new) low-temperature DHN sections, yielding a three temperature layer topology as illustrated in Figs.~\ref{fig:dhn_graph} and \ref{fig:dhn_3_layers}.
Such low-temperature DHN sections allow to efficiently use the heat energy in a DHN (temperature cascading) and integrate renewable heat sources (e.g., waste heat, solar thermal, heat pumps) and new consumers (e.g., low-energy buildings) into existing DHNs \cite[p.~44]{Nussbaumer20} \cite{Koefinger17, Volkova20, Volkova22}. However, due to the ongoing trend of decreasing DHN temperatures, particularly in summer, the medium temperature might not be sufficiently high to cover the heat demand of some low-temperature consumers. Thus, low-temperature DHN sections in a three layer topology typically have at least one mixing connection, i.e., an edge  $i\in\MM$, that allows to boost the temperature by mixing high with medium temperature water (see, e.g., node 7 in Fig.~\ref{fig:dhn_graph}) \cite{Volkova22,Koefinger17,Volkova20}.

Furthermore, in a three temperature layer topology, the low temperature water is typically fed directly into the medium temperature layer (see node 5 in Fig.~\ref{fig:dhn_graph} and the set of pipe edges $\PP$ between low and medium temperature in Fig.~\ref{fig:dhn_3_layers}) \cite{Koefinger17, Volkova20, Volkova22}. This implies that despite there possibly being three temperature layers, there are  exactly two hydraulic layers (see Fig.~\ref{fig:dhn_3_layers}).
The number of hydraulic layers can be defined as follows:
\begin{definition}\label{def:layers}
	A DHN has $n_l\geq2$ hydraulic layers, where $n_l$ is the number of weakly connected subgraphs $\GG_1,\dots,\GG_{n_l}$ obtained by removing all edges $i\in\DD\cup\LL\cup\MM$, i.e., all DGUs, consumers, and mixing connections, from $\GG$.
\end{definition}
%%
% {\color{red}
\begin{remark}
	We want to emphasize that our setup does not require to have three temperature layers. Traditional DHNs with high temperature supply and medium temperature return are covered as well as standalone 4th generation DHNs with medium temperature supply and low temperature return. In any case, there are always exactly two hydraulic layers containing all pipe edges.
\end{remark}
% }
%

%%%%%%%%%%%%%%%%%%%%%%%%%%%%%%%%%%%%%%%%%%%%%%%%%%%%%%%%%
\section{Hydraulic Modeling} \label{sec:modeling}

With the DHN setup formalized as a digraph, we now present the mathematical models describing the hydraulic dynamics of the edges and nodes. In Sections~\ref{sec:modeling:pumps} and \ref{sec:modeling:valve}, we first model the main actuators responsible for pressure and volume flow rate control, i.e.,  \emph{pumps} and \emph{valves}. They serve as building blocks for the hydraulic models of the \emph{DGU}, \emph{consumer}, \emph{pipe}, and \emph{mixing connection} edges in Sections~\ref{sec:modeling:dgu}--\ref{sec:modeling:mixing}, 
as well as the \emph{pressure holding} nodes in Section~\ref{sec:modeling:pressure_holding}. Lastly, the \emph{capacitive}, and \emph{simple junction} nodes are modeled in sections~\ref{sec:modeling:capacitor} and \ref{sec:modeling:simple_junction}.
For the modeling, we make the following assumptions which are valid under normal operating conditions (see also \cite{Persis11,Chertkov19}):
\begin{assumption} \label{assumption:incompressibility_pos_values}
	The compressibility of water is neglected. Any reference and nominal pressure values as well as all model parameters are strictly positive. Pressure losses inside pipes $\lambda(q):\Reals\to\Reals$ and valves $\mu(q,s):\Reals\times\Reals_{\geq0}\to\Reals$ caused by volume flow rates $q\in\Reals$ are continuously differentiable functions that are strictly monotonically increasing and satisfy  $\lambda(0)=0$ and $\mu(0,s)=0$ for all valve stem positions $s\in\Reals_{\geq0}$, respectively.
\end{assumption}

%%%%%%%%%%%%%%%%%%%

\subsection{Hydraulic Actuators}

\subsubsection{\bf Pumps}\label{sec:modeling:pumps}

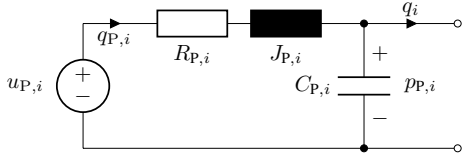
\begin{figure}
	\centering
    \scalebox{0.83}{
	\begin{circuitikz}[european, american voltages]
		%		\ctikzset{voltage=straight}
		\draw
		(-1,-1) to [V,v_>=$u_{\mathrm{P},i}$] (-1,-3)
		(-1,-1) to [short,i_>=$q_{\mathrm{P},i}$] (0,-1)
		to [R, l_=$R_{\text{P},i}$] (1.5,-1)
		to [L, l_=$J_{\text{P},i}$] (3,-1)
		to [short,-*] (3.5,-1)
		to [short,i^>=$q_{i}$,-o] (5,-1)
		(3.5,-1) to [C,l_=$C_{\text{P},i}$,v^=$p_{\text{P},i}$] (3.5,-3)
		(-1,-3) to [short,-*] (3.5,-3)			
		(3.5,-3) to [short,-o] (5,-3);
	\end{circuitikz}}
	\caption{Equivalent circuit of a linear, second-order approximation of pump dynamics (cf.\ \cite{Goppelt18}). 
 }\label{fig:pumps_rlc}
\end{figure}
Pumps are the essential actuated components in DHNs. They are used for  controlling the absolute pressure at specific points (pressure holding) (see \cite[pp.~54--55]{Nussbaumer20}\cite{Wang17conversion}), 
for (differential) pressure and volume flow rate control in DGUs (see \cite{Persis09,Persis11,Persis14,Jensen12,Machado22automatica,Wang17conversion,Yan13, Gong19,Schmidt17,Vandermeulen18}), 
for  boosting the pressure in consumers and pipes (see \cite{Persis09, Persis11, Lund14}),
and for direct volume flow rate control in consumers (see \cite{Wang17conversion,Yan13, Gong19}). 
In the prevalent literature (see, e.g., the references aobve), pumps are  considered as ideal pressure sources, which are analogous to voltage sources in DC networks. However, the dynamics of pumps, particularly the ones of centrifugal pumps that are widely used in DHNs \cite{Persis11,Scholten17}, lie in the range of several hundred milliseconds (see \cite[Figs.~8 and 9]{Goppelt18}). Since this is a time scale comparable to that of the overall DHN hydraulics (see, e.g., \cite{Novitsky20,Chertkov19}), a more accurate control design and system analysis must be performed if 
increasing numbers of pumps are integrated into DHNs.

As a starting point for such an improved design and analysis, we follow \cite{Strehle22dhn,Goppelt18} and model each pump by a linear equivalent RLC circuit as shown in Fig.~\ref{fig:pumps_rlc}. 
% Thus, following \cite{Strehle22dhn}, we model each pump by a linear equivalent RLC circuit as shown in Fig.~\ref{fig:pumps_rlc}. 
The RLC circuit arises by approximating the complex arrangement of power electronics, speed-controlled AC motor, and pump hydraulics by a linear second-order system.
By applying Kirchhoff's voltage law (KVL) and KCL to Fig.~\ref{fig:pumps_rlc}, we obtain the model  
\begin{subequations}\label{eq:modeling:pumps_rlc_equations}
        \begin{equation}
   \resizebox{0.96\hsize}{!}{$\tfrac{\mathrm{d}}{\mathrm{d}t}\underbrace{\begin{bmatrix}
J_{\mathrm{P},i}q_{\mathrm{P},i}\\
C_{\mathrm{P},i}p_{\mathrm{P},i}
    \end{bmatrix}}_{\vec{x}_i} =\underbrace{ \begin{bmatrix}
        -p_{\mathrm{P},i}-R_{\mathrm{P},i}q_{\mathrm{P},i}\\
        q_{\mathrm{P},i}
    \end{bmatrix}}_{\vec{f}_i(\vec{x}_i)} + \underbrace{\begin{bmatrix}
      1\\
      0
\end{bmatrix}}_{\vec{G}_i(\vec{x}_i)}\underbrace{\begin{bmatrix}
        u_{\mathrm{P},i}
    \end{bmatrix}}_{{u}_i}+\underbrace{\begin{bmatrix}
        0\\
        1
    \end{bmatrix}}_{\vec{K}_i}
   \underbrace{[-q_i]}_{d_i},$}
\end{equation}
\begin{align}
% \vec{y}_i  = \underbrace{\vec{x}_i}_{\vec{h}_i(\vec{x}_i)},
\vec{y}_i  = \underbrace{\begin{bmatrix} q_{\mathrm{P},i} & p_{\mathrm{P},i} \end{bmatrix}^\top}_{\vec{h}_i(\vec{x}_i)},~~
    {z}_i  =\underbrace{\begin{bmatrix}
          0 & \tfrac{1}{C_{\mathrm{P},i}}
    \end{bmatrix}}_{\vec{T}_i}\vec{x}_i=p_{\mathrm{P},i},
\end{align}
\end{subequations}
where $\vec{x}_i$ is the state vector,  ${u}_i$  the control input, and $\vec{y}_i$ the measurable output vector. 
% Note that $\vec{y}_i$ depends only on locally available signals. 
The additional input ${d}_i$ and output ${z}_i$  model the interaction or physical interconnection between the pump and other subsystems, e.g., a DGU. Furthermore, $R_{\text{P},i}$, $J_{\text{P},i}$, $C_{\text{P},i}$ are the model parameters, $p_{\mathrm{P},i}$ is the pressure difference produced by the pump between its terminals, $q_i$ is the volume flow rate through the pump, and $q_{\mathrm{P},i}$ is an auxiliary variable without physical interpretation. The control input  $u_{\mathrm{P},i}$ can be interpreted as a pressure source originating  from the rotational speed of the pump produced by an AC motor \cite{Goppelt18}.\footnote{The choice of the variables $\vec{x}_i$, $d_i$ and $z_i$ in \eqref{eq:modeling:pumps_rlc_equations} is based on the port-Hamiltonian \cite[p.~114]{vdS17} representation of the system done in \cite{Strehle22dhn} (see \cite{Perryman22} for a similar approach). We follow an analogous reasoning for the remaining DHN subsystems. 
% {\color{red} 
See Section~\ref{sec:control}  and Appendix~\ref{APP:ISO-PHS fomulation} for more details.}
% } 
% {\color{blue} In the arXiv version of our paper, we explicitly identify the port-Hamiltonian form of each of the DHN subsystems. Such a representation, which we skipped in this article due to space reasons,  gives a clear perspective on which input-output ports are accessible for control and over which ports subsystems interact with each other. Furthermore, the passivity properties with respect to these ports and the Hamiltonian as storage function are directly visible.}}

% {\color{red}
\begin{remark}
In \eqref{eq:modeling:pumps_rlc_equations},  $R_{\text{P},i}$, $J_{\text{P},i}$, and $C_{\text{P},i}$ are black box parameters without physical interpretation. This is  due to the fact that in the RLC circuit representation, the speed control and AC motor dynamics are merged with the hydraulic pump dynamics comprising fluid mass inertia, pressure losses, and hydraulic capacities due to fluid compressibility and fluid volume.  % %
In practice, $R_{\text{P},i}$, $J_{\text{P},i}$, and $C_{\text{P},i}$ can be identified from measurement data obtained by operating the respective pump in typical scenarios (see e.g., \cite{Goppelt18}). Alternatively, they can be fitted in simulations to match characteristic curves provided in data sheets. 
\end{remark}
% }
\medskip
\subsubsection{\bf Control valves} \label{sec:modeling:valve}
Besides pumps, valves are the main actuators in DHNs. Their main task is the regulation of volume flow rates \cite[pp.~143--145,151]{Nussbaumer20}\cite[p.~19,29]{Schmidt17}\cite{Vandermeulen18}\cite{Wang17opti}. In order to establish a desired volume flow rate, valves adjust their pressure drop by varying their stem position between fully closed ($s_{\text{v},i}=0$) and fully open ($s_{\text{v},i}=1$). Thus, they behave  as variable, nonlinear flow resistors. In order to avoid volume surge behavior around their closing point, valves are typically designed such that the stem has a lower limit just above zero in normal operation \cite[pp.~145]{Nussbaumer20}. Consequently, the following assumption can be made:
\begin{assumption} \label{assumption:stem_position}
	Valves are designed appropriately such that in normal operation the stem position never reaches the zero value, i.e., $s_{\text{v},i}\in\left[s_{\text{v},i}^{\text{min}},1\right]$, where  $0< s_{\text{v},i}^\text{min}\leq \epsilon_{\mathrm{v},i}$, for some fixed positive constant $\epsilon_{\mathrm{v},i}$.
\end{assumption}

The nonlinear characteristic pressure drop equation of any valve is given by \cite[Eq.~(18)]{Wang17opti}, \cite[Eq.~(5)]{Gong19}
\begin{subequations} \label{eq:modeling:valve_full}
\begin{equation} \label{eq:modeling:valve}
	\mu_i(s_{\text{v},i},q_i)=\frac{1}{(C_{\text{v},i}f_{\text{v},i}(s_{\text{v},i}))^2}  |q_i|q_i,
\end{equation}
where $s_{\text{v},i}\in\left[s_{\text{v},i}^{\text{min}},1\right], 0< s_{\text{v},i}^{\text{min}}\leq \epsilon_{\mathrm{v},i}$ is the stem position, $q_i\in\Reals$ the volume flow rate through the valve, $C_{\text{v},i}>0$ the flow capacity of the valve, and
\begin{equation} \label{eq:modeling:valve_characteristic}
	f_{\text{v},i}(s_{\text{v},i})=
	\begin{cases}
		R_i^{s_{\text{v},i}-1}, & \text{(equal percentage)}\\
		s_{\text{v},i},		 & \text{(linear)}
	\end{cases}
\end{equation}
the valve characteristic with rangeability $R_i>0$ (see also the static orifice law \cite[Eq.~(12)]{Boysen03} or the definition of the so-called $k_{\text{v}}$-value \cite[p.~144]{Nussbaumer20}).
By substituting
\begin{equation} \label{eq:modeling:valve_substitution}
	u_{\text{v},i}(s_{\text{v},i})\coloneqq\frac{1}{f_{\text{v},i}(s_{\text{v},i})^2}, \quad \hat{\mu}_i(q_i)\coloneqq \frac{1}{C_{\text{v},i}^2} |q_i|q_i
\end{equation}
\end{subequations}
in \eqref{eq:modeling:valve}, the pressure drop can be written as 
\begin{equation} \label{eq:valve_subsituted}
		\mu_i(s_{\text{v},i},q_i)=u_{\text{v},i}(s_{\text{v},i}) \hat{\mu}_i(q_i),
\end{equation}
where $u_{\text{v},i}(s_i):\left[s_{\text{v},i}^{\text{min}},1\right]\to\Reals_{\geq0}$ is a bijective mapping of the actual stem position $s_{\text{v},i}$ to the virtual control input $u_{\text{v},i}$, and $\hat{\mu}_i(q_i)$ a continuously differentiable, strictly monotonically increasing function satisfying $\hat{\mu}_i(0)=0$. Note that \eqref{eq:valve_subsituted} is affine in the virtual control input $u_{\text{v},i}$.

% {\color{red}
\begin{remark}
    It might not be intuitively clear that $f(s_{\text{v},i})=s_{\text{v},i}$ represents a linear valve characteristic $q_i=v_i \cdot s_{\text{v},i}$ with some constant $v_i>0$. However, when considering that valve characteristics are specified assuming a fixed pressure drop $\mu_i(s_{\text{v},i},q_i)=\bar{\mu}_i=const.$ \cite[pp.~143--145,151]{Nussbaumer20}, we can directly rewrite \eqref{eq:modeling:valve} as $q_i=v_i \cdot f_{\mathrm{v},i}(s_{\text{v},i})=v_i \cdot s_{\text{v},i}$ with $v_i = C_{\mathrm{v},i} \sqrt{\bar{\mu}_i}$.
\end{remark}
\begin{remark}\label{remark:valve_sufficient_pressure}
    Since valves can only function as variable, nonlinear flow resistors, valve-based volume flow rate control requires sufficient differential pressure over the hydraulic DGU or consumer circuit the valve is part of. In future DHNs with frequently changing hydraulic conditions, this motivates the addition of booster pumps in some consumer circuits or pipes (see Sections~\ref{sec:modeling:consumer}, \ref{sec:modeling:pipe} and Assumption~\ref{assumption:valve_equilibrium}). 
\end{remark}
\begin{remark}\label{remark:constraint_u_v}
    Assumption~\ref{assumption:stem_position} implies that $u_{\text{v}} \in \left[1, u_{\text{v},i}^\mathrm{max}\right]$. However, for the sake of simplicity and in line with classical feedback control design, we do not consider the possibility of control input saturation explicitly during the control design stage. Instead, we discuss the well-behaved nature of input saturation by means of our numerical simulations in Section~\ref{sec:simulation}.
\end{remark}
% }

% * ignore column margin and plots over complete page
\begin{figure}
	\centering
	\scalebox{0.83}{
	\begin{circuitikz}[european, american voltages]
%		\ctikzset{voltage=straight}
		\draw
			(-2,0) node[label={above:$j$}] {} to [short, o-*] (-1,0)
			to [short] (0.7,0)
			(-1,0) to  [C, l=$C_j$,v=$p_{j}$,-*] (-1,-3) 
			to [short] (0.7,-3) 
			(-1,-3) [short] (-1,-3) node[ground] {} {} node [anchor=north west ] {\phantom{d}$p_0=\SI{1}{\bar}$};
		\draw[color=black] 	
			(0.7,0) to [V<=$p_{\text{P},h}$,-*] (0.7,-3) 
			(0.7,-1) to [short,i^>=~] (0.7,0);	
		\draw[color=red]
			(0.7,0) to [short,*-] (0.8,0)
			(0.8,0) to [V<=$p_{\text{P},i}$,invert] (1.9,0) 
			to [short,i_>=$q_i$] (2.5,0)
			to [short] (2.5,0)
			to [vR, l^=$\hat{\mu}_i(q_i)u_{\text{v},i}(s_{\text{v},i})$] (4,0)
			to [R, l_=$\lambda_{i}(q_i)$] (5.5,0)
			to [L, l_=$J_{i}$,-*] (7,0);
		\draw[color=black]
			(7,0) to [C,l_=$C_k$,v^=$p_k$] (7,-3);
		\draw[color=black]
			(7,-3) [short] (7,-3) node[ground] {} node [anchor=north east] {$p_0=\SI{1}{\bar}$\phantom{d}}
			(7,0) to [short, -o] (8,0) node[label={above:$k$}] {};
	\end{circuitikz}}
	\caption{Equivalent circuit of a  hydraulic DGU model $(i\in \mathcal{D})$ with pressure holding (black voltage source) and circulation circuit (red) \cite[Fig.~2]{Strehle22dhn}; without loss of generality, the capacitance $C_{j}$ may be lumped with the pressure holding (see Section~\ref{sec:modeling:pressure_holding}); $j\in \mathcal{N}_i^-$ and $k\in \mathcal{N}_i^+$.}\label{fig:producer}
 %\todo{state to which sets $a,b,i$ belong. DONE}.
\end{figure}

\subsection{Edge Dynamics} \label{sec:modeling:edges}

\subsubsection{\bf Pipes} \label{sec:modeling:pipe}
\begin{figure}
	\centering
	\begin{circuitikz}[european, american voltages]
%		\ctikzset{voltage=straight}
		\draw    
		(0,2)  node[label={above:$p_j$}]{} to [short,i_=$q_i$, o-] (1,2)
		to [R,l_=$\lambda_{i}(q_i)$] (3,2)
		to [L,l_=$J_{i}$] (5,2)
		to [V,l_=$p_{\text{P},i}$, invert] (6.5,2)
		to [short, -o] (7,2) node[label={above:$p_k$}]{}
		% (0,2) to [open,v^>=$p_{j}$, -o] (0,0.1) to (0,0) node[ground] {} node [anchor=west]{$p_0=\SI{1}{\bar}$}
		% (7,2) to [open,v_>=$p_{k}$, -o] (7,0.1) to (7,0) node[ground] {} node [anchor=east]{$p_0=\SI{1}{\bar}$}
		;
	\end{circuitikz}
	\caption{Equivalent circuit of a hydraulic pipe model $(i\in \mathcal{P})$ with optional booster pump;  $j\in \mathcal{N}_i^-$ and $k\in \mathcal{N}_i^+$.}
	\label{fig:pipe}
\end{figure}
The hydraulic pipe model at an edge $i \in \PP$ is illustrated in the equivalent circuit diagram in Fig.~\ref{fig:pipe}.
Following the literature (see, e.g., \cite{Strehle21dhn,Persis11,Machado22automatica,Persis14,Strehle22dhn}), we model the pipe friction by a nonlinear, volume flow-dependent resistance $\lambda_i(q_i)$ (see Assumption~\ref{assumption:incompressibility_pos_values}) and the volume inertia by the linear inductance $J_i$. In contrast to prior works, we assume that some pipes might have booster pumps in series. Such pumps are represented  by the voltage source in Fig.~\ref{fig:pipe} and modeled  by \eqref{eq:modeling:pumps_rlc_equations}.
% For the hydraulic pipe model at an edge $i \in \PP$, we follow \cite{Strehle21dhn,Persis11,Machado22automatica,Persis14,Strehle22dhn} and represent the pipe as illustrated in Fig.~\ref{fig:pipe}. The pipe friction is modeled by a nonlinear, volume flow-dependent resistance $\lambda_i(q_i)$ and the volume inertia by the linear inductance $J_i$. In contrast to prior works, we assume that some pipes might have booster pumps in series. Such pumps are represented   by the voltage source in Fig.~\ref{fig:pipe} and  modeled   by \eqref{eq:modeling:pumps_rlc_equations}.
%
By applying KVL and KCL to Fig.~\ref{fig:pipe}, we obtain the model for each $i\in \mathcal{P}$ as
\begin{subequations}\label{eq:modeling:pipe_equations}
	% \begin{align}
	% 	J_i\dot{q}_i =& p_j+p_{\mathrm{P},i} - \lambda_{i}(q_i) -p_k  \\
	% 			J_{\text{P},i} \dot{q}_{\mathrm{P},i} &= -p_{\text{P},i} - R_{\text{P},i} q_{\text{P},i}+ u_{\mathrm{P},i},\\
	% 	C_{\text{P},i} \dot{p}_{\text{P},i}&= q_{\mathrm{P},i} -q_{i},
	% \end{align}
\begin{equation}
   \resizebox{0.96\hsize}{!}{$\tfrac{\mathrm{d}}{\mathrm{d}t}\underbrace{\begin{bmatrix}
J_iq_i\\
J_{\mathrm{P},i}q_{\mathrm{P},i}\\
C_{\mathrm{P},i}p_{\mathrm{P},i}
    \end{bmatrix}}_{\vec{x}_i} =\underbrace{ \begin{bmatrix}
        p_{\mathrm{P},i}-\lambda_i(q_i)\\
        -p_{\mathrm{P},i}-R_{\mathrm{P},i}q_{\mathrm{P},i}\\
        q_{\mathrm{P},i}-q_i
    \end{bmatrix}}_{\vec{f}_i(\vec{x}_i)} + \underbrace{\begin{bmatrix}
       0\\
       1\\
      0
\end{bmatrix}}_{\vec{G}_i(\vec{x}_i)}\underbrace{\begin{bmatrix}
        u_{\mathrm{P},i}
\end{bmatrix}}_{{u}_i}+\underbrace{\begin{bmatrix}
        1\\
        0\\
        0
    \end{bmatrix}}_{\vec{K}_i}\underbrace{p_j-p_k}_{{d}_i},$}
\end{equation}
\begin{align}
% \vec{y}_i  = \underbrace{\vec{x}_i}_{\vec{h}_i(\vec{x}_i)},
\vec{y}_i  = \underbrace{\begin{bmatrix}q_i & q_{\mathrm{P},i} & p_{\mathrm{P},i}\end{bmatrix}^\top}_{\vec{h}_i(\vec{x}_i)},~~
    {z}_i  =\underbrace{\begin{bmatrix}
        \tfrac{1}{J_i} & 0 & 0
    \end{bmatrix}}_{\vec{T}_i}\vec{x}_i=q_i,
\end{align}
\end{subequations}
where $\vec{x}_i$ is the state vector,  $u_i$ the control input, $\vec{y}_i$ the measurable output vector, $(d_i,z_i)$ the interaction (coupling) port pair, and $(j,k)\in \mathcal{N}_i^{-}\times\mathcal{N}_i^{+}$ are the source and target nodes of $i$ with pressures  $p_{j}$ and $p_{k}$, respectively.
\begin{remark}\label{remark:pipe_without_booster}
	Any  pipe $i\in\PP$ without a booster pump can be modeled by \eqref{eq:modeling:pipe_equations} by fixing $u_{\text{P},i}=p_{\text{P},i}=0$ and removing the part corresponding to the dynamics of  $q_{\mathrm{P},i}$ and $p_{\mathrm{P},i}$.
\end{remark}

% \bigskip
\medskip

%%%%%%%%%%%%%%%%%%%%
\subsubsection{\bf DGUs}\label{sec:modeling:dgu}
From a hydraulic viewpoint, a DGU may comprise two main parts as illustrated in Fig.~\ref{fig:producer}: a circulation circuit (red) (see \cite{Lennermo19,Lamaison17}) and an optional  pressure holding unit (see \cite[pp.~54--55]{Nussbaumer20}). 
We view the circulation circuit (in red) as the actual edge $i\in \mathcal{D}$,  which comprises a circulation pump and a control valve connected in series with pipes and a heat exchanger. All  pipes are lumped into the nonlinear, volume flow-dependent resistance $\lambda_i(q_{i})$  and the inductance $J_i$, which represent the pipe friction and volume inertia, respectively. The control valve is modeled as a variable, nonlinear resistance  $\hat{\mu}_i(q_{i})u_{\text{v},i}(s_{\text{v},i})$ with control input $u_{\text{v},i}(s_{\text{v},i})$ as in \eqref{eq:modeling:valve_full}.  The circulation pump is modeled by  \eqref{eq:modeling:pumps_rlc_equations}.
By applying KVL and KCL to the red part in Fig.~\ref{fig:producer}, we obtain the model for each $i\in \mathcal{D}$ as
%
% \begin{subequations} 
% 	\begin{align}
% 		J_i\dot{q}_i =& p_j+p_{\mathrm{P},i} - \lambda_{i}(q_i) -  \hat{\mu}_i(q_i)u_{\mathrm{v},i}(s_{\text{v}i}) -p_k  \\
% 				J_{\text{P},i} \dot{q}_{\mathrm{P},i} &= -p_{\text{P},i} - R_{\mathrm{P},i} q_{\text{P},i}+ u_{\mathrm{P},i},\\
% 		%
% 		C_{\text{P},i} \dot{p}_{\text{P},i}&= q_{\mathrm{P},i} -q_{i},
% 	\end{align}
% \end{subequations}
\begin{subequations}\label{eq:modeling:circulation_equations}
\begin{equation}
   \resizebox{0.96\hsize}{!}{$\tfrac{\mathrm{d}}{\mathrm{d}t}\underbrace{\begin{bmatrix}
J_iq_i\\
J_{\mathrm{P},i}q_{\mathrm{P},i}\\
C_{\mathrm{P},i}p_{\mathrm{P},i}
    \end{bmatrix}}_{\vec{x}_i} =\underbrace{ \begin{bmatrix}
        p_{\mathrm{P},i}-\lambda_i(q_i)\\
        -p_{\mathrm{P},i}-R_{\mathrm{P},i}q_{\mathrm{P},i}\\
        q_{\mathrm{P},i}-q_i
    \end{bmatrix}}_{\vec{f}_i(\vec{x}_i)} + \underbrace{\begin{bmatrix}
      -\hat{\mu}_i(q_i)  & 0\\
      0 & 1\\
      0 & 0
\end{bmatrix}}_{\vec{G}_i(\vec{x}_i)}\underbrace{\begin{bmatrix}
        u_{\mathrm{v},i}\\
        u_{\mathrm{P},i}
    \end{bmatrix}}_{\vec{u}_i}+\underbrace{\begin{bmatrix}
        1\\
        0\\
        0
    \end{bmatrix}}_{\vec{K}_i}\underbrace{p_j-p_k}_{{d}_i},$}
\end{equation}
\begin{align}
% \vec{y}_i  = \underbrace{\vec{x}_i}_{\vec{h}_i(\vec{x}_i)},
\vec{y}_i  = \underbrace{\begin{bmatrix}q_i & q_{\mathrm{P},i} & p_{\mathrm{P},i}\end{bmatrix}^\top}_{\vec{h}_i(\vec{x}_i)},~~
    {z}_i  =\underbrace{\begin{bmatrix}
        \tfrac{1}{J_i} & 0 & 0
    \end{bmatrix}}_{\vec{T}_i}\vec{x}_i=q_i,
\end{align}
\end{subequations}
where $\vec{x}_i$ is the state vector, $\vec{u}_i$ the control input vector, $\vec{y}_i$ the measurable output vector, $(d_i,z_i)$ the interaction (coupling) port pair, and $(j,k)\in \mathcal{N}_i^{-}\times\mathcal{N}_i^{+}$ are the source and target nodes of $i$ with pressures $p_{j}$ and $p_{k}$, respectively.

\begin{remark}
    The pressure holding unit of a given DGU is represented by the (black) voltage source $p_{\mathrm{P},h}$ shown in Fig.~\ref{fig:producer}. The  capacitances $C_j$ and $C_k$ model the  hydraulic elasticity of all the components in the DGU circulation circuit, particularly of the heat exchanger (see \cite{Straede95,Boysen03}). For simplicity, we assume that $C_j$  is lumped with $C_{\mathrm{P},h}$ of the pressure holding (see Fig.~\ref{fig:pumps_rlc}). Furthermore, to clearly describe the network interconnection among all the subsystems in the DHN, we view the pressure holding and the elasticity capacitances as nodes of the DHN graph; we elaborate on their models in Section~\ref{sec:node_models}. 
\end{remark}

% \bigskip
\medskip

%%%%%%%%%%%%%%%%%%%%%%%%%%%%%%

\subsubsection{\bf Consumers} \label{sec:modeling:consumer}

Most of nowadays consumers are connected indirectly to a DHN via heat exchangers in series with pipes and a control valve for volume flow rate control \cite[pp.~87,143]{Nussbaumer20}.
In future DHNs, however, additional pumps are expected to be included in some (up to all) consumer circuits: either for pressure boosting to ensure a proper functioning of the control valves under unclear and changing hydraulic conditions \cite{Lund14,Persis11} or in DVSP DHN configurations \cite{Wang17conversion,Yan13,Gong19}. 
%On the one hand, this allows to compensate higher pressure losses in the network due to smaller pipe diameters by boosting the pressure \cite{Persis11,Lund14}. On the other hand, results from \cite{Yan13,Wang17conversion,Gong19} and references therein suggest that such a setup considerably reduces the electrical energy required to operate the pumps in DHNs. 
%
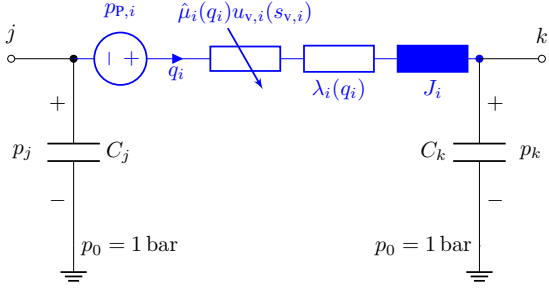
\begin{figure}
	\centering
    \scalebox{0.83}{
	\begin{circuitikz}[european, american voltages]
%		\ctikzset{voltage=straight}
		\draw[color=black]
		(-0.5,0) node[label={above:$j$}] {} to [short, o-*] (0.5,0) 
		to [C=$C_{j}$,v=$p_j$] (0.5,-3)
		to [short] (0.5,-3) node[ground] {} node [anchor=west] {$p_0=\SI{1}{\bar}$};
		\draw[color=blue]
		(0.5,0) to [V<=$p_{\text{P},i}$, invert] (2,0) 
		to [short,i_>=$q_{i}$] (2.3,0)
		to [short] (2.5,0)
		to [vR, l^=$\hat{\mu}_i(q_{i})u_{\text{v},i}(s_{\text{v},i})$] (4,0)
		to [R, l_=$\lambda_{i}(q_{i})$] (5.5,0)
		to [L, l_=$J_{i}$,-*] (7,0);
		\draw[color=black]
		(7,0) to [C,l_=$C_{k}$,v^=$p_k$] (7,-3)
		to [short] (7,-3) node[ground] {} node [anchor=east] {$p_0=\SI{1}{\bar}$}
		(7,0) to [short, -o] (8,0) node[label={above:$k$}] {};
	\end{circuitikz}}
	\caption{Equivalent circuit of a hydraulic consumer model $(i\in \mathcal{L})$ \cite[Fig.~2]{Strehle21dhn}; $j\in \mathcal{N}_i^-$ and $k\in \mathcal{N}_i^+$. } \label{fig:consumer}
\end{figure}
Consequently, the hydraulic consumer circuit at an edge $i\in\LL$ is modeled similarly to the hydraulic DGU circulation circuit (red part in Fig.~\ref{fig:consumer}, \cite[Fig.~2]{Strehle21dhn}). 
The only differences are on the working direction of the pump and the sign convention of the volume flow rate, which is reflected in the edge orientation in the DHN digraph (see Fig.~\ref{fig:dhn_graph} and \cite[pp.~87,143]{Nussbaumer20}).
% A difference is on the sign conventions of the volume flow rates \cite[pp.~87,143]{Nussbaumer20}. 
% This allows for a more intuitive perspective of water flowing from the supply nodes to the consumers and out at the return nodes for positive volume flow rate values.  
%
%Similar to the DGU model, the elasticity capacitances $C_{j}$ and $C_{k}$ are considered to be part of the set of nodes $\NN$. 
Furthermore, pressure holding units are typically not installed at consumers. By applying KVL and KCL to Fig.~\ref{fig:consumer}, we obtain the model for each $i\in \mathcal{L}$ as in \eqref{eq:modeling:circulation_equations}.
\begin{remark}\label{remark:consumers_without_pumps}
	Any consumer $i\in\LL$ without a pump can be modeled by \eqref{eq:modeling:circulation_equations} 
% 	\eqref{eq:modeling:consumer_equations} 
	by fixing $u_{\text{P},i}=p_{\text{P},i}=0$ and removing the part corresponding to the dynamics of  $q_{\mathrm{P},i}$ and $p_{\mathrm{P},i}$. Such consumers regulate their flow rate through their respective control valve.
% 	If a given consumer $i\in\LL$ does not have a pump installed, we model it as in  but set $p_{\text{P2},j}=0$ and remove the $(q_{\mathrm{P},i},p_{\mathrm{P},i})$-dynamics.
\end{remark}

\medskip

\subsubsection{\bf Mixing connection} \label{sec:modeling:mixing}
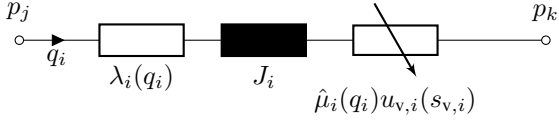
\begin{figure}
	\centering
	\begin{circuitikz}[european, american voltages]
		%		\ctikzset{voltage=straight}
		\draw    
		(0,2)node[label={above:$p_j$}]{} to [short,i_=$q_i$, o-] (1,2)
		to [R,l_=$\lambda_{i}(q_i)$] (2.25,2)
		to [L,l_=$J_i$] (4.25,2)
		to [vR, l_=$\hat{\mu}_i(q_i)u_{\text{v},i}(s_{\text{v},i})$] (5.75, 2)
		to [short, -o] (7,2) node[label={above:$p_k$}]{}
		% (0,2) to [open,v^>=$p_{j}$, -o] (0,0.1) to (0,0) node[ground] {} node [anchor=west]{$p_0=\SI{1}{\bar}$}
		% (7,2) to [open,v_>=$p_{k}$, -o] (7,0.1) to (7,0) node[ground] {} node [anchor=east]{$p_0=\SI{1}{\bar}$}
		;
	\end{circuitikz}
	\caption{Equivalent circuit of a hydraulic mixing connection model $(i\in \mathcal{M})$;  $j\in \mathcal{N}_i^-$ and $k\in \mathcal{N}_i^+$.}
	\label{fig:mixing}
\end{figure}
As outlined in Section~\ref{sec:modeling:layers}, future DHNs may have a topology with three temperature layers. 
In order to guarantee sufficient heat supply of the low-temperature sections, the medium temperature water is typically mixed with high temperature water via a mixing connection before it is fed into the low-temperature section (see, e.g., node 7 in Fig.~\ref{fig:dhn_graph}).
The hydraulic circuit of a mixing connection at an edge $i \in\MM$ is illustrated in Fig.~\ref{fig:mixing}. It comprises a pipe in series with a control valve.  By applying KVL to Fig.~\ref{fig:mixing}, we obtain the  model for each $i\in \mathcal{M}$ as
%
% \begin{equation} \label{eq:modeling:mixing_equations}
% 	J_i \dot{q}_i = p_j- \lambda_i(q_i) -\hat{\mu}_i(q_i)u_{\text{v}i}(s_{\text{v}i}) -p_k,\\
% \end{equation}
\begin{subequations}\label{eq:modeling:mixing_equations}
    \begin{equation}
   \resizebox{0.96\hsize}{!}{$\tfrac{\mathrm{d}}{\mathrm{d}t}\underbrace{\begin{bmatrix}
J_iq_i\\
    \end{bmatrix}}_{x_i} =\underbrace{ \begin{bmatrix}
       -\lambda_i(q_i)
    \end{bmatrix}}_{{f}_i({x}_i)} + \underbrace{\begin{bmatrix}
       -\hat{\mu}_i(q_i)
\end{bmatrix}}_{{G}_i(\vec{x}_i)}\underbrace{\begin{bmatrix}
        u_{\mathrm{v},i}
\end{bmatrix}}_{{u}_i}+\underbrace{\begin{bmatrix}
        1
    \end{bmatrix}}_{{K}_i}\underbrace{p_j-p_k}_{{d}_i},$}
\end{equation}
\begin{align}
% {y}_i  = \underbrace{{x}_i}_{{h}_i({x}_i)},
{y}_i  = \underbrace{q_i}_{h_i(x_i)},~~
    {z}_i  =\underbrace{\begin{bmatrix}
        \tfrac{1}{J_i} 
    \end{bmatrix}}_{{T}_i}{x}_i=q_i,
\end{align}
\end{subequations}
where $x_i$ is the state,  $u_i$ the control input, $y_i$ the measurable output, $(d_i,z_i)$ the interaction (coupling) port pair, and $(j,k)\in \mathcal{N}_i^{-}\times\mathcal{N}_i^{+}$ are the source and target nodes of $i$ with pressures  $p_{j}$ and $p_{k}$, respectively.

%%%%%%%%%%%%%%%%%%%%

\subsection{Node Models}\label{sec:node_models}

As outlined in Sections~\ref{sec:modeling:dhn_graph}, \ref{sec:modeling:dgu} and \ref{sec:modeling:consumer},  
the set of nodes $\NN$ of the DHN graph $\GG=(\NN,\EE)$ is the union of three disjoint sets:
% pressure holding units at DGUs, the elasticity capacitances both at DGUs and consumers, and ideal junctions are modeled as nodes of the DHN graph $\GG=(\NN,\EE)$. Thus, $\NN$ is the union of three disjoint sets:
%
\begin{equation}\label{eq:node_splitting}
\NN= \mathcal{H} \cup  \mathcal{C}\cup \mathcal{K},
\end{equation}
where $\mathcal{H}$ is the set of pressure holding units,  $\mathcal{C}$ is the set of elasticity capacitances, 
% (at the supply connections of DGUs and the supply and return connections of consumers), 
and $\mathcal{K}$ is the set of simple junctions. 

\subsubsection{\bf Pressure holding}\label{sec:modeling:pressure_holding}

Pressure holding units are realized technically in two ways: dynamic pressure holding with a pressure dictation pump, and static pressure holding with a closed vessel \cite[pp.~54--56]{Nussbaumer20}. Furthermore, pressure holding units are almost exclusively installed on the suction side of circulation pumps (pre-pressure control) (see Fig.~\ref{fig:producer}) and are instrumental in preventing cavitation \cite[pp.~54--55]{Nussbaumer20}, \cite{Sommer19}.

Dynamic pressure holding is typically conducted in larger DGUs with powerful circulation pumps. It is realized by a pressure dictation pump located between a pressurized container and the DHN \cite[pp.~54--55]{Nussbaumer20}, \cite[Fig.~1]{Buffa21}.  
As outlined in Section~\ref{sec:modeling:pumps}, we approximate the dynamics of  any pump by the linear second-order system \eqref{eq:modeling:pumps_rlc_equations}. 
% Then, a dynamic pressure holding unit installed at DGU $i\in\DD$ is equivalent to 
Thus, the case in which a dynamic pressure holding unit is installed at a DGU $i\in\DD$ is equivalent to replacing the black voltage source in Fig.~\ref{fig:producer} by the RLC circuit in Fig.~\ref{fig:pumps_rlc}. Note that  in contrast to the circulation pump, which is coupled with the circulation circuit (red part in Fig.~\ref{fig:producer}), the black voltage source already represents the entire pressure holding unit, i.e., we assume that the dictation pump is lumped together with the pressurized container.
% 
% From KVL and KCL, we obtain the model 
Thus, the model for each $j\in \mathcal{H}$ is similar to \eqref{eq:modeling:pumps_rlc_equations} and given by
%
% \begin{equation}\label{eq:model_pressure_holding}
% 	\begin{split}
% 		J_{\text{P},j} \dot{q}_{\mathrm{P},j} &= -p_{\text{P},j} - R_{\text{P},j} q_{\text{P},j} + u_{\mathrm{P},j},\\
% 		%
% 		C_{\text{P},j} \dot{p}_{\text{P},j}&=q_{\text{P},j} + \sum_{i\in \mathcal{I}_j}q_i,
% 	\end{split}
% \end{equation}
\begin{subequations}\label{eq:model_pressure_holding}
    \begin{equation}
   \resizebox{0.96\hsize}{!}{$\tfrac{\mathrm{d}}{\mathrm{d}t}\underbrace{\begin{bmatrix}
J_{\mathrm{P},j}q_{\mathrm{P},j}\\
C_{\mathrm{P},j}p_{\mathrm{P},j}
    \end{bmatrix}}_{\vec{x}_j} =\underbrace{ \begin{bmatrix}
        -p_{\mathrm{P},j}-R_{\mathrm{P},j}q_{\mathrm{P},j}\\
        q_{\mathrm{P},j}
    \end{bmatrix}}_{\vec{f}_j(\vec{x}_j)} + \underbrace{\begin{bmatrix}
      1\\
      0
\end{bmatrix}}_{\vec{G}_j(\vec{x}_j)}\underbrace{\begin{bmatrix}
        u_{\mathrm{P},j}
    \end{bmatrix}}_{{u}_j}+\underbrace{\begin{bmatrix}
        0\\
        1
    \end{bmatrix}}_{\vec{K}_j}\underbrace{\sum_{i\in\mathcal{I}_j}q_i}_{{d}_j},$}
\end{equation}
\begin{align}
% \vec{y}_j  = \underbrace{ \vec{x}_j}_{\vec{h}_j(\vec{x}_j)},
\vec{y}_j  = \underbrace{ \begin{bmatrix} q_{\mathrm{P},j} &p_{\mathrm{P},j}\end{bmatrix}^\top}_{\vec{h}_j(\vec{x}_j)},~~
    {z}_j  =\underbrace{\begin{bmatrix}
          0 & \tfrac{1}{C_{\mathrm{P},j}}
    \end{bmatrix}}_{\vec{T}_j}\vec{x}_j=p_{\mathrm{P},j},
\end{align}
\end{subequations}
where $\vec{x}_j$ is the state vector,  $u_j$ the control input, $\vec{y}_j$ the measurable output vector, $(d_j,z_j)$ the interaction (coupling) port pair, and $\mathcal{I}_j\subseteq \mathcal{E}$ the set of edges that are incident to $j$.

\begin{remark}
A static pressure holding is used in smaller DGUs with compact circulation pumps. It is realized by directly adding a closed, pressurized vessel. 
In an equivalent circuit perspective, this can be understood as a preloaded capacitor. Thus, in case of static pressure holding, we simply replace the black voltage source in Fig.~\ref{fig:producer} with a capacitor $C_{\text{P},h}$ that we consider to be lumped with $C_{j}$.
\end{remark}

% {\color{red}
\begin{remark}
    The presented DHN model allows for pressure holding units to be installed at the suction side of each DGU $i\in\DD$.
    In practice, however, it is typically sufficient to operate only one of them at any given time (see, e.g., \cite[p.~55]{Nussbaumer20}\cite[Fig.~1]{Buffa21} and Fig.~\ref{fig:dhn_graph}). This is due to the fact that in a connected and closed hydraulic network, the pressure values at each node $j\in\VV$ are uniquely determined by fixing one node pressure value and expressing all edge subsystems by means of differential pressure dynamics as done in Section~\ref{sec:modeling:edges} (see, e.g., \cite[Sec.~2.3]{Krug21}). 
\end{remark}
% }

\subsubsection{\bf Capacitive nodes} \label{sec:modeling:capacitor}  Invoking the volume balance, we obtain the model for each $j\in \mathcal{C}$ as
\begin{subequations}\label{eq:model_capacitive_nodes}
\begin{align}
% C_j\dot{p_j} & = \sum_{i\in \EE_j^-\cup\EE_j^+} q_i,\\
\tfrac{\mathrm{d}}{\mathrm{d}t}\underbrace{\begin{bmatrix}
    C_jp_j
\end{bmatrix}}_{{x}_j} & = \underbrace{\begin{bmatrix}
    1
\end{bmatrix}}_{{K}_i}\underbrace{\sum_{i\in \mathcal{I}_j} q_i}_{{d}_j},\\
{z}_j & = \underbrace{\begin{bmatrix}
    \tfrac{1}{C_j}
\end{bmatrix}}_{{T}_j}{x}_j=p_j,
\end{align}
\end{subequations}
where $x_j$ is the state, $(d_j,z_j)$ the interaction (coupling) port pair, and $\mathcal{I}_j$ the set of edges that are incident to $j$. 

\subsubsection{\bf Simple junctions}\label{sec:modeling:simple_junction}
 The model for each $j\in\mathcal{K}$ is analogous to \eqref{eq:model_capacitive_nodes} with $C_j$ fixed to zero and treating $z_j=p_j$ as an algebraic variable. Note that there is no state variable to describe the behavior of simple junctions.
\section{Problem Formulation and Approach} \label{sec:problem}

With the DHN formalized as a digraph and its hydraulic model established, we can now formulate the main pressure and volume flow rate control problems addressed in this work. 
%The operation of DHNs requires different pressures and volume flow rates to be stabilized. 
%In view of the parallels outlined in Section~\ref{sec:intro} regarding the trends and control challenges arising in DHNs and power systems, we 
As outlined in Section~\ref{sec:intro}, we focus on decentralized control schemes with plug-and-play capabilities. Furthermore, we keep to the terminology of electrical power systems and suppose that  DGUs may operate either in \emph{grid-forming} or \emph{grid-feeding} mode $i\in\DD=\DD_{\text{form}}\cup\DD_{\text{feed}}$.

%%%%%%%%%%%%%% DGUS
\emph{DGUs in grid-forming mode} ($i\in\DD_{\text{form}}$) actively form the hydraulic conditions required to operate DHNs by regulating the differential pressure generated by their circulation pumps to desired setpoints $p_{\text{P},i}^*$ \cite[p.~47--48]{Nussbaumer20} \cite{KSB06}. In this case, the control valves in their circulation circuits are fully open, i.e., $\bar{u}_{\text{v},i}=\left.u_{\mathrm{v},i}\right\vert_{s_{\text{v},i}=1}$.\footnote{In the following, $\bar{(\cdot)}$ denotes any variable in steady-state, whereas $(\cdot)^*$ denotes a desired setpoint that is to be established in steady-state.} 
\emph{DGUs in grid-feeding mode} ($i\in\DD_{\text{valve}}\subseteq\DD_{\text{feed}}$) regulate  the volume flow rate through their circulation circuits to desired setpoints $q_{i}^*$ by means of their control valves. Under approximately constant water temperature, this is equivalent to controlling the thermal energy they feed into the DHN (see \cite{Lennermo14,Lennermo19}\cite[Sec.~2.3]{Krug21}). Note that for a proper functioning of the control valve, the circulation pump still introduces some differential pressure $p_{\text{P},i}^*$, which is then throttled by the control valve such that the desired $q_{i}^*$ is reached.
%

%%%%%%%%%%%% Consumer
\emph{Consumers} ($i\in \LL$) regulate the thermal energy they consume by controlling their volume flow rates to desired setpoints $q_{i}^*$ \cite[pp.~143--145,151]{Nussbaumer20}\cite[p.~29]{Schmidt17}. Traditionally, this is done by control valves only. The set of consumers for which   $u_{\text{v},i}(s_{\text{v},i})$ is the main control input is thus denoted by $\LL_{\text{valve}}\subseteq\LL$.
However, as discussed in  Section~\ref{sec:modeling:consumer}, booster pumps  might be added to some consumer circuits. We identify these consumers by the set  $\LL_{\text{boost}}\subseteq\LL$. In each consumer $i\in \mathcal{L}_\mathrm{boost}$, the pump pressure is controlled to some desired setpoint $p_{\text{P},i}^*$, which is then throttled by the control valve such that the desired $q_{i}^*$ is reached.
%

%%%%%%%%%%%%% Pipes
For \emph{pipes} $i\in\PP$, we identify by $\mathcal{P}_\mathrm{boost}\subseteq \mathcal{P}$ the subset of pipes that  have a \emph{booster pump} connected in series. These pumps are in charge of counteracting the differential pressure loss over the corresponding  pipe by introducing a  differential pressure $p_{\mathrm{P},i}^*$.
%%%% Mixing connections
In \emph{mixing connections} $i\in\MM$, a desired volume flow rate $q_i^*$ is stabilized such that a desired mixing ratio of high and medium temperature water is achieved (see Section~\ref{sec:modeling:mixing}).

Last but not least, we note that  regardless of the operation mode of the DGUs, each \emph{pressure holding unit} $j\in \mathcal{H}$ associated to a given DGU regulates the pressure $p_j$  at the suction side of the circulation pump $j\in \mathcal{N}_i^{-}$ (see Fig.~\ref{fig:producer}) to a suitable setpoint $p_j^*=p_{\mathrm{P},h}^*$. This pressure also serves as the static pressure in a DHN (see \cite[p.~55]{Nussbaumer20}). 

\begin{remark}
	In some DVSP setups, it is suggested to directly conduct the volume flow rate control (grid-feeding, consumer)  by pumps without including any control valves in the circuits of DGUs $i\in\DD_{\text{VSP}}\subseteq \DD_{\textbf{feed}}$ and consumers $i\in\LL_{\text{VSP}}\subseteq \LL$ \cite{Wang17conversion,Yan13}.
	\footnote{To avoid cluttering notation, we assume that valves in DVSP cases are still there, however fully open, i.e., $\bar{u}_{\text{v},i}=\left.u_{\mathrm{v},i}\right\vert_{s_{\text{v},i}=1}$.}
	In \cite{Gong19}, a hybrid DVSP setup is suggested in which all DGUs and some consumers have only pumps, while some consumers have only control valves. 
	Consequently, it is apparent that depending on the topology and producer-consumer configuration, different hydraulic designs of DGU and consumer circuits might be beneficial. In this work, we thus allow for all possible combinations (see Problem~\ref{problem} below) and show that under the roof of the passivity-based framework presented in the sequel, asymptotic stability of the desired pressures and volume flow rates can always be guaranteed.
\end{remark} 
\smallskip

In summary, we note that the control tasks boil down to \emph{pressure and volume flow rate control of pumps} and \emph{volume flow rate control of valves}.
% Consequently, the main control problems read as follows:
\smallskip

\begin{problem} \label{problem}
Consider a DHN as described in Section~\ref{sec:modeling}. Design  decentralized controllers for pumps and valves that asymptotically stabilize any forced, hydraulic DHN equilibrium with the following characteristics:
\begin{itemize}
\item[{\bf (a)}] For each $i\in\DD_\mathrm{form}$, $s_{\mathrm{v},i}=1$ is fixed and $u_{\mathrm{P},i}$ is such that $\bar{p}_{\mathrm{P},i}=p_{\mathrm{P},i}^*$.
\smallskip
\item[{\bf (b)}] For each $i\in\DD_\mathrm{valve}$, $u_{\mathrm{P},i}$ and $u_{\mathrm{v},i}$ are such that $\bar{p}_{\mathrm{P},i}=p_{\mathrm{P},i}^*$ and   $\bar{q}_i=q_{i}^*$.
\smallskip
\item[{\bf (c)}] For each $i\in \mathcal{D}_\mathrm{VSP}$, $s_{\mathrm{v},i}=1$ is fixed and $u_{\mathrm{P},i}$  is such that  $\bar{q}_i=q_{i}^*$.
\smallskip
\item[{\bf (d)}] For each $i\in \mathcal{L}_\mathrm{boost}$,  $u_{\mathrm{P},i}$ and $u_{\mathrm{v},i}$ are such that  $\bar{p}_{\mathrm{P},i}=p_{\mathrm{P},i}^*$ and   $\bar{q}_i=q_{i}^*$.
\smallskip
\item[{\bf {(e)}}] For each $i\in\mathcal{L}_\mathrm{valve}$, $u_{\mathrm{P},i}=0$ is fixed and  $u_{\mathrm{v},i}$ is such that  $\bar{q}_i=q_{i}^*$.
\smallskip
\item[{\bf (f)}] For each $i\in \mathcal{L}_\mathrm{VSP}$, $s_{\mathrm{v},i}=1$ is fixed and $u_{\mathrm{P},i}$ is such that  $\bar{q}_i=q_{i}^*$.
\smallskip
\item[{\bf (g)}] For each $i\in \mathcal{P}_\mathrm{boost}$, $u_{\mathrm{P},i}$ is such that  $\bar{p}_{\mathrm{P},i}=p_{\mathrm{P},i}^*$. 
\smallskip
\item[{\bf (h)}] For each $i\in \mathcal{M}$, $u_{\mathrm{v},i}$ is such that $\bar{q}_i=q_{i}^*$.
\smallskip
\item[{\bf (i)}] For each $j\in \mathcal{H}$, $u_{\mathrm{P},h}$ is such that $\bar{p}_{\mathrm{P},j}=p_{\mathrm{P},j}^*$. 
\end{itemize}
\end{problem}
%
%%%%%%%%%% Remark about where desired values come from
\begin{remark}\label{remark:higher_level_control}
	Note that similar to the hierarchical control of power systems, the setpoints of the volume flow rates and pressures are assumed to be known and specified by a higher level control ensuring sensible DHN operation, i.e.,  ensuring that a feasible  hydraulic equilibrium  exists. 
\end{remark}

\subsection{Approach} \label{sec:problem:approach}
Inspired by \cite[Theorem 3.1]{Arcak16}, \cite{Strehle22,Strehle21dhn}, we address Problem~\ref{problem} in a modular, bottom-up manner by means of a passivity-based approach.
The leading idea is that {the hydraulic equilibrium} is \emph{stable} if two conditions are satisfied: 
\begin{itemize}
	\item[C.1] each, possibly closed-loop subsystem at the edges and nodes of the DHN digraph is EIP with positive definite storage function;
	\item[C.2] the interconnnection structure of the DHN subsystems is power-preserving.
\end{itemize}
\emph{Asymptotic stability} of {the hydraulic equilibrium} can be investigated either by invoking LaSalle's Invariance Principle or checking for strict EIP of all subsystems.
\section{Passivity-Based Control Design} \label{sec:control}

In Section~\ref{sec:modeling}, we have established that the dynamics of any edge or node $i\in\EE\cup\NN$ of a DHN, with the exception of simple junctions, can be written in the  
general form
\begin{subequations}
\begin{align*}
\dot{\x}_i & = \vec{f}_i(\vec{x}_i)+\vec{G}_i(\x_i)\vec{u}_i+\vec{K}_i {d}_i,\\
\vec{y}_i & = \vec{h}_i(\vec{x}_i),~
{z}_i  = \vec{T}_i \vec{x}_i.
\end{align*}
\end{subequations}
%
% {\color{red} 
Based on \cite{Strehle22dhn}, it is possible to represent each of these subsystems as an input-state-output port-Hamiltonian system (ISO-PHS), which in general can be written as follows \cite[p.~114]{vdS17}):
\begin{subequations}
		%\label{eq:modeling:phs}
		\begin{align*} %\label{eq:modeling:phs_dynamics}
			\xdot_i &=  \left(\J_i-\R_i\right) \nabla H_i(\vec{x}_i)-\vec{\mathcal{R}}_i(\x_i)+ \G_i(\x_i) \vec{u}_i+\K_i \vec{d}_i,\\
			%
			%\label{eq:modeling:phs_y}
			\vec{\zeta}_i &= \G_i^\top(\x_i) \nabla H_i(\vec{x}_i),\\
			%
			%\label{eq:modeling:phs_z}
			\vec{z}_i &= \K_i^\top\nabla H_i(\vec{x}_i),\\
			%
			%\label{eq:modeling:phs_hamil}
			\hspace{-0.1cm}\Hamil_i(\x_i) &= \frac{1}{2} \x_i^\top \Q_i \x_i,
		\end{align*}
\end{subequations}
where $\x_i$ is the state, $\nabla H_i(\vec{x}_i)$ is  the co-state, $\Hamil_i(\x_i)$ is the Hamiltonian, $(\vec{u}_i,\vec{\zeta}_i)$ is the control port pair, and $(\vec{d}_i,\vec{z}_i)$ is the uncontrolled interaction (coupling) port pair; $\J_i=-\J_i^\top$ is the interconnection matrix, $\R_i=\R_i^\top\geq0$ the damping matrix, $\G_i(\x_i)$ the input matrix, $\K_i$ the interaction matrix, and $\Q_i=\Q_i^\top\posDef0$ the energy matrix; $\vec{\mathcal{R}}_i(\x_i)\geq0$ is the nonlinear, vector-valued damping function. To improve readability all details are included in Appendix~\ref{APP:ISO-PHS fomulation}.

The ISO-PHS representation gives a clear perspective on which input-output ports are accessible---or are convenient for control---and over which ports subsystems interact with each other:  note that the output $\vec{\zeta}_i$ is not necessarily the measurable  output. Furthermore, the passivity properties with respect to these ports and the Hamiltonian as storage function are directly visible \cite{vdS17}. Indeed, it is direct to see that along system solutions the Hamiltonian $H_i$ satisfies the following inequality:
\begin{equation*}
    \dot{H}_i(\vec{x}_i) \leq \vec{\zeta}_i^\top \vec{u}_i+\vec{z}_i^\top \vec{d}_i.
\end{equation*}

% }

% {\color{red} 
Then, considering the approach described in Section~\ref{sec:problem:approach}, the main goal in this section is to design decentralized controllers of the form
\begin{align*}
        \dot{\vec{\xi}}_i & = \vec{\pi}_i(\vec{y}_i,\vec{\xi}_i),~
        \vec{u}_i  = \vec{c}_i(\vec{y}_i,\vec{\xi}_i),
\end{align*}
for the pumps and valves such that each closed-loop subsystem $i\in \DD\cup\LL\cup\PP_{\text{boost}}\cup\MM\cup\HH$ fulfills its respective steady-state characteristic from Problem~\ref{problem} (a)--(i) and is EIP with respect to its interaction port pair $({d}_i,{z}_i)$ and some positive definite storage function.
That is, it is to be shown that for any feasible equilibrium pair $(\bar{\x}_i,\bar{d}_i)$ with $\bar{\x}_i$ fulfilling the respective requirement (a)--(i) in Problem~\ref{problem}, there exists a scalar function $\x_i\mapsto \hat{H}_i(\x_i)$  that is positive definite with respect to $\bar{\x}_i$ and satisfies
\begin{equation*}
    \dot{\hat{H}}_i(\vec{x}_i) \leq (d_i-\bar{d}_i) (z_i-\bar{z}_i).
\end{equation*}
Subsequently, in Section~\ref{sec:stability}, it will be shown that the interconnection among the DHN subsystems is such that
\begin{equation*}
    \sum_{i\in \mathcal{E}\cup \mathcal{N}}(d_i-\bar{d}_i) (z_i-\bar{z}_i)=0,
\end{equation*}
making $\sum_{i\in \mathcal{E}\cup \mathcal{N}}\hat{H}_i(\vec{x}_i)$ a {\em bona fide} Lyapunov function. 
% {\color{red}
Even though the design that we propose in this section for $\vec{u}_i$ follows passivity-based arguments, we  do not restrict ourselves to use only the passive output $\vec{\zeta}_i$. Moreover, the definition of the above-mentioned storage function $\hat{H}_i$ will be inspired by the Hamiltonian $H_i$ of the system, but will not be equal to it: note that $H_i$ is a Lyapunov function of any ISO-PHS only if ${\vec{u}}_i$ and ${\vec{d}}_i$ are zero at steady-state, which is not the situation we face in this paper.
% }  

% {\color{red}
\subsection{Operating Modes of the DHN Subsystems} \label{sec:control:operating_modes}
 
In a first preliminary step, we investigate what kind of degrees of freedom are available in choosing the operating modes of the DHN subsystems. In particular, we are interested in determining (i) if DGUs are subject to any restrictions on their choice of operating modes, i.e., grid-forming or grid-feeding, and  (ii) whether all consumers and mixing connections can in principle, i.e., disregarding any technical constraints, independently control their volume flow rates to desired steady-state values.
%%%

\begin{proposition}
	Consider a DHN as modeled in Section~\ref{sec:modeling} with $D\geq1$ DGUs, $L\geq1$ consumers, $P\geq2$ pipes, $M\geq0$ mixing connections, and two hydraulic layers $\GG_1, \GG_2$. Then, the following holds:
	(i) there must be at least one grid-forming DGU $i\in\DD_{\text{form}}$ connecting the two hydraulic layers $\GG_1, \GG_2$;
	(ii) each steady-state volume flow rate $\bar{q}_i$, $i\in \DD_\mathrm{feed}\cup \LL\cup \mathcal{M}\cup \PP_\mathrm{loop}$, is an independent variable, where each $\PP_\mathrm{loop}\subseteq\PP$ forms an independent loop \emph{within} $\GG_1$, or $\GG_2$.
\end{proposition}
%%%%%%%%%%%%%%%%%

\IEEEproof 
Following \cite[pp.~477--482]{Desoer69} (see also \cite{Persis11, Machado22automatica}), we make use of the fundamental loop analysis of circuit theory. First we note that since the digraph $\GG$ representing a DHN is weakly connected, it admits a generally non-unique spanning tree $\TT$ (see, e.g.,  \cite{Persis11,Wang17opti}). The spanning tree $\TT$ is a weakly connected subgraph of $\GG$ that contains all nodes of $\GG$ and no loops \cite[p.~477]{Desoer69}. Any edge of $\GG$ not in $\TT$ is referred to as \emph{chord} and creates a (fundamental) loop when added to $\TT$. For $\GG=(\NN,\EE)$, there are $|\EE|-|\NN|+1$ chords and loops, respectively. By applying KCL, it can be shown that each steady-state flow through an edge in $\TT$ is the superposition of one or more of the steady-state loop flows \cite[p.~482]{Desoer69}. By setting the loop flows equal to the chord flows, we thus find that the $|\EE|-|\NN|+1$ chord flows form a complete set of independent variables. 

To prove (i), we note that per Definition~\ref{def:layers}, only DGU, consumer, or mixing edges may connect the two hydraulic layers $\GG_1$ and $\GG_2$ (see also Figs.~\ref{fig:dhn_3rd_4th_gen} and \ref{fig:dhn_3_layers}). Additionally, the hydraulic layers may have a meshed structure. Thus, in order to create a spanning tree $\TT$ of $\GG$, we have to connect $\GG_1$ and $\GG_2$ via exactly one edge in the union $\DD\cup\LL\cup\MM$ and might have to remove some pipe edges in $\PP_\text{loop}\subseteq \PP$ that form loops within $\GG_1$ and $\GG_2$, respectively. From the fundamental loop analysis we know that the steady-state flow through the edge connecting $\GG_1$ and $\GG_2$ is not independent. Thus, the edge must be a pressure-controlled subsystem in $\DD\cup\LL\cup\MM$, which can only be fulfilled by a grid-forming DGU $i\in\DD_{\text{form}}$.

In turn, we can choose the set of chords such that it contains all DGU edges except one, all consumer edges, possibly some pipe edges that form loops within a meshed, hydraulic layer $\GG_1$ or $\GG_2$, respectively, and all mixing edges. Consequently, the steady-state flows through these edges are independent variables, which proves (ii). \hfill\IEEEQEDclosed

%%%%%%%%%%%% Remark about independent pipe flows
\begin{remark}
	The independent steady-state pipe flows $\bar{q}_i, i\in\PP_{\text{loop}}$, provide additional degrees of freedom that might be used, e.g.,  to minimize pumping costs (see \cite[Sec.~2.2]{Wang17opti}). %This showcases  why meshed pipe networks might be desirable in some DHNs.
\end{remark}
% }

\subsection{Pressure and Volume Flow Rate Control of Pumps} \label{sec:control_pumps}

Instrumental to solving {Problem~\ref{problem}} is the ability to regulate the pressure $p_{\mathrm{P},i}$ or the  volume flow rate $q_i$ of a given pump in the subsystems $i\in\DD\cup\LL_{\text{boost}}\cup\LL_{\text{VSP}}\cup\PP_{\text{boost}}\cup\HH$ towards desired constant setpoints. For the control design, we initially consider the isolated  pump model~\eqref{eq:modeling:pumps_rlc_equations}  and address pressure and volume flow rate regulation. Then,  the closed-loop pump dynamics are interconnected to the respective subsystems  $\DD\cup\LL_{\text{boost}}\cup\LL_{\text{VSP}}\cup\PP_{\text{boost}}\cup\HH$.

%%%%%%%%%%%%%%%%%%%%%%%
\subsubsection{\bf Pressure Control of Pumps} For the pump pressure $p_{\text{P},i}$, we propose a controller for $u_{\mathrm{P},i}$ that is based on algebraic IDA \cite{Ortega04} passivity-based control extended with integral action on the non-passive output of the pump model \cite{Donaire09}. 
\begin{proposition}\label{prop:pressure_control_pump_general}
Consider the pump model~\eqref{eq:modeling:pumps_rlc_equations}. Assign the  control input ${u}_{\mathrm{P},i}$ as 
\begin{subequations}\label{eq:controller_pump_specific}
\begin{align}
    Q_{\mathrm{I},i}\dot{r}_i & =(p_{\mathrm{P},i}-p_{\mathrm{P},i}^*),\\
    \chi_i & = q_{\mathrm{P},i}+r_i, \label{eq:controller_pump_specific_integral}\\
    %
    % \nu_i & = -\tfrac{1}{J_{\mathrm{P},i}}R_{\mathrm{P},i}\chi_i+(\tfrac{1}{J_{\mathrm{P},i}}K_{\mathrm{I},i}Q_{\mathrm{I},i}-1)p_{\mathrm{P},i}\\
    % & ~~~ + R_{\mathrm{P},i}K_{\mathrm{I},i}Q_{\mathrm{I},i}r_i -(J_{\mathrm{P},i}K_{\mathrm{I},i}Q_{\mathrm{I},i}p_{\mathrm{P},i}^*) \\
    \nu_i & = -p_{\mathrm{P},i}-R_{\mathrm{P},i} q_{\mathrm{p},i} + \tfrac{J_{\mathrm{P},i}}{Q_{\mathrm{I},i}} (p_{\mathrm{P},i}- p_{\mathrm{P},i}^*),  \\
    \label{eq:controller_pump_specific_up}
    u_\mathrm{P,i} & = -\nu_i- R^\mathrm{p}_i\chi_i-p_{\mathrm{P},i}.
    \end{align}
\end{subequations}
with pressure setpoint $p_{\mathrm{P},i}^*>0$ and control parameters $R^\mathrm{p}_i$, $Q_{\mathrm{I},i}>0$. Then, the closed-loop system is given by 
\begin{subequations}\label{eq:closed_loop_dyn_press_control_general}
\begin{align}
    \frac{\mathrm{d}}{\mathrm{d} t}\underbrace{\begin{bmatrix}
            J_{\mathrm{P},i}{\chi}_i \\ C_{\mathrm{P},i}p_{\mathrm{P},i} \\
            Q_{\mathrm{I},i}{r}_i
        \end{bmatrix}}_{\vec{x}_i^\mathrm{p}} & = \underbrace{\begin{bmatrix} 
                    - R^\mathrm{p}_i\chi_i -p_{\mathrm{P},i}\\
				    \chi_i -{r}_i \\
				    p_{\mathrm{P},i}-p_{\mathrm{P},i}^* \end{bmatrix}}_{\vec{f}_i^\mathrm{p}(\vec{x}_i^\mathrm{p})}+\underbrace{\begin{bmatrix}0 \\ 1 \\ 0 \end{bmatrix}}_{\vec{K}_i^\mathrm{p}}{d}_i,\\
    {z}_i &= p_{\mathrm{P},i}.
\end{align}
\end{subequations}
Moreover,  \eqref{eq:closed_loop_dyn_press_control_general} is EIP with supply rate $({z}_i-\bar{{z}}_i) ({d}_i-{\bar{{d}}}_i)$  and positive definite storage function
\begin{equation}\label{eq:H_p_P}
	H^\mathrm{p}_i =\tfrac{1}{2}\Vert \x^\mathrm{p}_i-\vec{\bar{x}}^\mathrm{p}_i\Vert^2_{\text{diag}^{-1}(J_{\mathrm{P},i},C_{\mathrm{P},i},Q_{\mathrm{I},i})},
\end{equation}
for any (feasible) equilibrium pair $(\bar{{d}}_i,\bar{{z}}_i)$ and associated equilibrium state
\begin{equation}\label{eq:equilibrium_pump_close_loop_pressure}
    \bar{\vec{x}}^\mathrm{p}_i= \begin{bmatrix}
   J_{\mathrm{P},i}\bar{\chi}_i & C_{\mathrm{P},i}p_{\mathrm{P},i}^* &  Q_{\mathrm{I},i}\bar{r}_i
    \end{bmatrix}^\top.
\end{equation}
\end{proposition}
\IEEEproof
	See Appendix~\ref{app:proof:pressure_control_pump_general}. \hfill\IEEEQEDclosed

\begin{remark}\label{rem:generic_pump_pressure}
The controller  \eqref{eq:controller_pump_specific} is decentralized as it only requires knowledge of local variables and parameters, such as $q_{\mathrm{P},i}$, $p_{\mathrm{P},i}$ and $J_{\mathrm{P},i}$.
Observe in particular that if we choose $R^{\mathrm{P}}_i=R_{\mathrm{P},i}$, \eqref{eq:controller_pump_specific} becomes independent of $q_{\mathrm{P},i}$. This is desired as $q_{\mathrm{P},i}$ does not represent a physical quantity.
Moreover, note that \eqref{eq:controller_pump_specific_up} can be written as
\begin{align} \label{eq:pump_pressure_controller}
		u_{\text{P},i} & = (-R^\mathrm{p}_i+R_{\text{P},i}) q_{\mathrm{P},i} + R^\mathrm{p}_i Q_{\mathrm{I},i}^{-1} \int_{0}^{t} (p_{\text{P},i}^*-p_{\text{P},i}) \text{d}t \nonumber \\
		&\phantom{=}+ J_{\text{P},i} Q_{\mathrm{I},i}^{-1} (p_{\text{P},i}^*-p_{\text{P},i}),
\end{align}
which  clearly illustrates its composition as the combination of a state feedback term and a PI term. 
\end{remark}

\medskip
\subsubsection{\bf Volume Flow Rate Control via Pumps} \label{sec:control:volume_pump}

Next, we address the task of achieving volume flow rate control of pipes or heat exchangers via pumps. For this, we focus on the model of a pump in series with a pipe element, which is equivalent to \eqref{eq:modeling:pipe_equations} but with ${d}_i$ treated as an arbitrary external input. 

\begin{proposition}\label{prop:flow_control_pump_general}
	 Consider the model of any pump in series with a pipe element (see \eqref{eq:modeling:pipe_equations}). Assign $u_{\mathrm{P},i}$ as 
	\begin{subequations}\label{eq:PI_control_flow_pump_generic}
		\begin{align}
Q_{\mathrm{I},i}\dot{r}_i & = \left(q_i-q_i^* \right) \label{eq:PI_control_flow_pump_integral}\\
u_{\mathrm{P},i} & = -K_{\mathrm{P},i}p_{\mathrm{P},i}-r_\mathrm{i},
		\end{align}
\end{subequations}
with volume flow rate setpoint $q_i^*$ and control parameters $Q_{\mathrm{I},i}$, $K_{\mathrm{P},i}$ satisfying
 \begin{equation}\label{eq:conds_stab_pump_flow}
     Q_{\mathrm{I},i}>0,~~~ 0<Q_{\mathrm{I},i}(K_{\mathrm{P},i}+1)-C_{\mathrm{P},i}=:\kappa_i^\mathrm{f}.
 \end{equation}
Then, the closed-loop system is given by 
\begin{subequations}\label{eq:closed_loop_sys_pump_flow_control_a}
\begin{equation}
    \resizebox{0.98\hsize}{!}{$\frac{\mathrm{d}}{\mathrm{d}t} \underbrace{\begin{bmatrix} 
    J_i q_i\\
    J_{\mathrm{P},i}q_{\mathrm{P},i} \\ C_{\mathrm{P},i}p_{\mathrm{P},i} \\ Q_{\mathrm{I},i}r_i \end{bmatrix}}_{\vec{x}_i^\mathrm{f}}  = \underbrace{\begin{bmatrix}
    p_{\mathrm{P},i}-\lambda_i(q_i)\\
    -p_{\mathrm{P},i}-R_{\mathrm{P},i}q_{\mathrm{P},i}-K_{\mathrm{P},i}p_{\mathrm{P},i}-r_i\\
    q_{\mathrm{P},i}-q_i\\
  q_i-q_i^*
    \end{bmatrix}}_{\vec{f}_i^\mathrm{f}(\vec{x}_i^\mathrm{f})}+\underbrace{\begin{bmatrix}1\\0\\0\\0\end{bmatrix}}_{\vec{K}_i^\mathrm{f}}{d}_i,$}
   \end{equation}
   \begin{equation}
       {z}_i=q_i,
   \end{equation}
\end{subequations}
Moreover, \eqref{eq:closed_loop_sys_pump_flow_control_a} is EIP with supply rate $({z}_i-\bar{{z}}_i) ({d}_i-{\bar{d}}_i)$  and positive definite storage function
\begin{subequations}\label{eq:H_f}
     \begin{align}
	H^\mathrm{f}_i & =\tfrac{1}{2}\Vert \x^\mathrm{f}_i-\vec{\bar{x}}^\mathrm{f}_i\Vert^2_{\vec{Q}_i^\mathrm{f}},\\
 \vec{Q}_i^\mathrm{f} & = \begin{bmatrix}
     \tfrac{1}{J_i} & 0 & 0 & 0\\
     0 & \tfrac{Q_{\mathrm{I},i}}{J_{\mathrm{P},i}\kappa_i^\mathrm{f}} & 0 & 0\\
     0 & 0 & \tfrac{1}{C_{\mathrm{P},i}}+\tfrac{1}{\kappa_i^\mathrm{f}} & \tfrac{1}{\kappa_i^\mathrm{f}}\\
     0 & 0 & \tfrac{1}{\kappa_i^\mathrm{f}} & \tfrac{1}{\kappa_i^\mathrm{f}}
 \end{bmatrix},
 \end{align}
\end{subequations}
for any (feasible) equilibrium pair $(\bar{{d}}_i,\bar{{z}}_i)$ and associated equilibrium state 
% $\bar{\vec{x}}_i^\mathrm{f}$,  which satisfies
%
\begin{equation}\label{eq:equilibrium_pump_close_loop_flow}
    \bar{\x}^\mathrm{f}_i= \begin{bmatrix}
   J_i q_i^* & J_{\mathrm{P},i}q_i^* &  C_{\mathrm{P},i}\bar{p}_{\mathrm{P},i} & Q_{\mathrm{I},i}\bar{r}_i
    \end{bmatrix}^\top.
\end{equation}
\end{proposition}
\IEEEproof
See Appendix~\ref{app:proof:flow_control_pump_general}. \hfill\IEEEQEDclosed

\begin{remark}
The controller \eqref{eq:PI_control_flow_pump_generic}---which is inspired by \cite[Th.~2]{Nahata20}---will be used in the sequel to regulate the flow through some DGUs and some consumers. Note that the dynamics \eqref{eq:modeling:circulation_equations} of DGUs and consumers are equivalent to the dynamics \eqref{eq:modeling:pipe_equations} of a pipe element in series with a pump, excluding the control valve. 
\end{remark}

%%%%%%%%%%%%%%%%%%%%%%%%
\subsection{Volume Flow Rate Control of Valves} \label{sec:control_of_valves}

Problem~\ref{problem} also considers the regulation of  volume flow rates $q_i$ through DGUs, consumers, and mixing connections via control valves. Note that the model of a control valve in series with a pipe element is equivalent to \eqref{eq:modeling:mixing_equations}, but with ${d}_i$ treated as an arbitrary external input.
\begin{proposition}\label{prop:control_law_uv}
Consider the model of any control valve in series with a pipe element (see \eqref{eq:modeling:mixing_equations}). Let
\begin{equation}\label{eq:control_output_valve_generic}
{\hat{y}}_{\mathrm{v},i} = -\hat{\mu}_i(q_i) \left(q_i-q_i^* \right).
\end{equation}
and assign the control input $u_{\mathrm{v},i}$ as 
\begin{subequations}\label{eq:PI_controller_valve_generic}
\begin{align}
Q_{\mathrm{I},i}\dot{r}_i & = -\hat{y}_{\mathrm{v},i}, \label{eq:controller_valve_integral}\\
{u}_{\mathrm{v},i} & = -K_{\mathrm{P},i}\hat{y}_{\mathrm{v},i} +{r}_{i},
\end{align}
\end{subequations}
with volume flow rate setpoint $q_i^*$ and control parameters $Q_{\mathrm{I},i}$, $K_{\mathrm{P},i}>0$. Then, the closed-loop system is given by 
\begin{subequations}\label{eq:cl_generic_valve_control_a}
\begin{equation}
    \resizebox{0.98\hsize}{!}{$\frac{\mathrm{d}}{\mathrm{d}t}\underbrace{ \begin{bmatrix}J_{i}{q}_i \\ Q_{\mathrm{I},i}{r}_i  \end{bmatrix}}_{\vec{x}_i^\mathrm{v}}=
    \underbrace{\begin{bmatrix}  
        - \lambda_i(q_i) -\hat{\mu}_i(q_i)\left( -K_{\mathrm{P},i}\hat{y}_i+r_i\right)\\
	   \hat{y}_{\mathrm{v},i} 
    \end{bmatrix}}_{\vec{f}_i^\mathrm{v}(\vec{x}_i^\mathrm{f})}
    +\underbrace{\begin{bmatrix}
        1 \\ 
        0  
    \end{bmatrix}}_{\vec{K}_i^\mathrm{f}}{d}_i,$}
\end{equation}
\begin{equation}
    {z}_i=q_i,
\end{equation}
\end{subequations}
Moreover, 
% if the function \todo{this sounds like a condition, but it is fulfilled for our models}
% %
% \begin{equation}\label{eq:gamma_i}
%     \gamma_i(q_i)=\lambda_i(q_i)+\bar{r}_i\hat{\mu}_i(q_i)
% \end{equation}
% %
% is monotone, then 
\eqref{eq:cl_generic_valve_control_a} is EIP with supply rate $({z}_i-\bar{{z}}_i) ({d}_i-{\bar{d}}_i)$  and positive definite storage function
\begin{equation}\label{eq:H_v}
	H^\mathrm{v}_i =\tfrac{1}{2}\Vert \x^\mathrm{v}_i-\vec{\bar{x}}^\mathrm{v}_i\Vert^2_{\text{diag}^{-1}(J_{i},Q_{\mathrm{I},i})},
\end{equation}
for any (feasible) equilibrium pair $(\bar{{d}}_i,\bar{{z}}_i)$ and associated equilibrium state 
% $\bar{\x}^\mathrm{v}_i$ satisfying 
%
%
\begin{equation}\label{eq:equilibrium_valve_close_loop_flow}
   \bar{\x}^\mathrm{v}_i= \begin{bmatrix}
   J_i q_i^* & Q_{\mathrm{I},i}\bar{r}_i
    \end{bmatrix}^\top~\text{or}~~\bar{\x}^\mathrm{v}_i=\begin{bmatrix}
   0 & Q_{\mathrm{I},i}\bar{r}_i
    \end{bmatrix}^\top.
\end{equation}
% \begin{equation}
% \bar{q}_i=q_i^*,~\text{or},~\bar{q}_i=0.
% \end{equation}
\end{proposition}
\IEEEproof See Appendix~\ref{app:proof:flow_control_valve_general}

\begin{remark}
The design of the PI controller \eqref{eq:PI_controller_valve_generic} is based on the observation that  the dynamics of a control valve in series with a pipe are EIP with respect to the (control) input-output pair $(u_{\mathrm{v},i},\hat{y}_i)$.
The definition of the output $\hat{y}_i$  originates from the fact that the input matrix $G_i(\vec{x}_i)$ as in~\eqref{eq:modeling:mixing_equations} is state-dependent. This fact complicates the use of a standard PI controller around the shifted passive output  (see also \cite[p.~137]{vdS17}\cite{Monshizadeh19}\cite[p.~26]{Arcak16}). Following \cite[Eq.~(8)]{Monshizadeh19}, we instead propose $\hat{y}_i$ as a new passive output, which is obtained from a suitable, {\em shifted} representation of the dynamics.
\end{remark}

% \begin{remark} 
    In steady-state, $\hat{y}_i=0$ generally allows for either $\bar{q}_i=q_i^*$ or $\bar{q}_i=0$ (see \eqref{eq:control_output_valve_generic} and \eqref{eq:cl_generic_valve_control_a}). However,   $\bar{q}_i=0$ implies $\lambda_i(0)=0$, $\hat{\mu}_i(0)=0$ (see \eqref{eq:modeling:valve_substitution} and Assumption~\ref{assumption:incompressibility_pos_values}) and thus $\bar{d}_i=0$, where $d_i$ is the pressure difference over the serial connection of valve and pipe element (cf.\ \eqref{eq:modeling:mixing_equations}). 
    This makes sense from a practical perspective, as a control valve with zero differential pressure available is not functional.
    In practice, such a situation is thus ruled out by a proper assignment of the pressure setpoints via a higher level control (see Remark~\ref{remark:higher_level_control}), which motivates the following assumption.
% \end{remark}

\begin{assumption} \label{assumption:valve_equilibrium}
    Any control valve in series with a pipe element (see \eqref{eq:modeling:mixing_equations}) is assumed to have a positive differential pressure $d_i>0$ for all time. This implies for DGUs and consumer circuits  $i\in\DD\cup\LL$ that $p_j+p_{\mathrm{P},i}-p_k>0$ (see Figs.~\ref{fig:producer} and \ref{fig:consumer}) and for mixing connections $i\in\MM$ that $p_j-p_k>0$ (see Fig.~\ref{fig:mixing}).
\end{assumption}

%%%%%%%%%%%%%%%%%%%%%%%%%%%
\subsection{Properties of the Closed-Loop Systems}\label{sec:properties_CL_systems}

In this section, we deploy the controlled pumps and valves from Propositions~\ref{prop:pressure_control_pump_general}--\ref{prop:control_law_uv} in the corresponding actuated edges and nodes $i\in\DD\cup\LL\cup\PP_{\text{boost}}\cup\MM\cup\in\HH$. In particular, we show that these closed-loop systems are also EIP, a fact which is central for the subsequent stability analysis of the overall, interconnected DHN model.

%%%% Lemma about EIP for actuated edges and nodes

\begin{lemma}\label{lemma:actuated_property_fufilled}
    Assign the pump controllers \eqref{eq:controller_pump_specific}, \eqref{eq:PI_control_flow_pump_generic}, and the valve controller \eqref{eq:PI_controller_valve_generic} to the respective edge and node models $i\in\DD\cup\LL\cup\PP_{\text{boost}}\cup\MM\cup\HH$  according to the control tasks in Problem~\ref{problem}. Then, the resulting closed-loop subsystems can be written as 
    \begin{subequations}\label{eq:models_cl}
    \begin{align}
    \dot{\vec{\hat{x}}}_i & = \vec{\hat f}_i(\vec{\hat x}_i)+\hat{\K}_id_i,\\
    {z}_i & = \vec{\hat{T}}_i\vec{\hat{x}}_i,
    \end{align}
    \end{subequations}
    with appropriate vectors and matrices. Moreover, for any $i\in\DD\cup\LL\cup\PP_{\text{boost}}\cup\MM\cup\HH$, \eqref{eq:models_cl} is EIP with supply rate $(z_i-\bar{z}_i)(d_i-\bar{d}_i)$ and positive definite storage function 
    \begin{equation}\label{eq:storage_function_close_loop_subsystems}
        \hat{H}_i=\tfrac{1}{2}\Vert \vec{\hat{x}}_i-\bar{\vec{\hat{x}}}_i\Vert^2_{\vec{\hat{Q}}_i},
    \end{equation}
    where $\vec{\hat{Q}}_i$ is a suitable positive definite diagonal matrix, and $\bar{\vec{\hat{x}}}_i$ is any (feasible) equilibrium value of $\vec{\hat{x}}_i$ with associated $\bar{d}_i$. In addition, under Assumption~\ref{assumption:valve_equilibrium}, the equilibrium value $\bar{\vec{\hat{x}}}_i$ of any $i\in\DD\cup\LL\cup\PP_{\text{boost}}\cup\MM\cup\HH$ is such that $\bar{q}_i=q_i^*$ and $\bar{p}_{\mathrm{P},i}=p_{\mathrm{P},i}^*$ in accordance with Problem~\ref{problem}.
\end{lemma}
\IEEEproof
See Appendix~\ref{app:proof_actuated_property_fulfilled} \hfill\IEEEQEDclosed

\begin{remark}
Note that in Lemma~\ref{lemma:actuated_property_fufilled} we have intentionally kept the original names for the interaction input and output variables $d_i$ and $z_i$. This has been done to emphasize that our control designs and the resulting closed-loop systems have not altered the description of the physical interconnection among the DHN subsystems. 
\end{remark}
\section{Modular Stability Analysis}\label{sec:stability}

In this section, we prove that the overall closed-loop dynamics of the DHN admit an asymptotically stable hydraulic equilibrium.

\subsection{EIP of the Unactuated Edges and Nodes}

Firstly, we show that the unactuated pipe edges $i\in\PP\setminus \PP_{\text{boost}}$ and the capacitive nodes $i\in\CC\cup\KK$ are EIP as well, although with non-assignable steady-states. 
%
%%%% Lemma about EIP of unactuated pipe edges and capacitive and simple junction nodes
\begin{lemma} \label{lemma:EIP_unactuated}
    The model of any $i\in \PP \setminus\PP_{\text{boost}}\cup\CC$, i.e., any pipe edge without booster pump and any capacitive node, can be written as \eqref{eq:models_cl} with an appropriate choice of the defining vectors and matrices.  Furthermore, these models are EIP with some positive definite storage function in the form of \eqref{eq:storage_function_close_loop_subsystems}.
    % In addition, any simple junction $i\in \mathcal{K}$ is a (static) EIP system.
\end{lemma}
\IEEEproof
See Appendix~\ref{app:proof_eip_unactuated}.
\hfill\IEEEQEDclosed

\subsection{Interconnection Structure}

Next, we illustrate the power-preserving nature of the interconnection structure that describes the interaction between edge and node subsystems of the DHN.
In order to facilitate subsequent elaborations and provide a compact representation, we assemble the overall system dynamics in vector form.
Recall that  the set of edges  $\mathcal{E}$ is the ordered union of $\mathcal{D}$, $\mathcal{L}$, $\mathcal{P}$ and $\mathcal{M}$. Let $\Delta$ be the ordered union of the sets of nodes $\HH$ and $\mathcal{C}$. For $\Omega= \mathcal{E},\Delta$, let
$\vec{\hat{x}}_\Omega=\text{stack}(\vec{\hat{x}}_i)_{i\in \Omega}$,
$\vec{\hat{f}}_\Omega(\vec{\hat{x}}_\Omega)=\text{stack}(\vec{\hat{f}}_i(\vec{\hat{x}}_i))_{i\in \Omega}$, $\vec{\hat{K}}_\Omega=\text{diag}(\vec{\hat{K}}_i)_{i\in \Omega}$,  $\vec{d}_\Omega=\text{stack}({d}_i)_{i\in \Omega}$,
$\vec{z}_\Omega=\text{stack}({z}_i)_{i\in \Omega}$, $\vec{\hat{T}}_\Omega=\text{diag}(\vec{\hat{C}}_i)_{i\in \Omega}$ and $\hat{H}_\Omega(\vec{\hat{x}}_\Omega)=\sum_{i\in \Omega}\hat{H}_i(\vec{\hat{x}}_i)$, where for each $i\in \Omega$, $\vec{\hat{x}}_i$, $\vec{\hat{f}}_i$, $\vec{\hat{K}}_i$, ${d}_i$, ${z}_i$ and $\hat{H}_i$ are as in Lemmas~\ref{lemma:actuated_property_fufilled} and \ref{lemma:EIP_unactuated}. Then, the overall, closed-loop DHN can be written as
\begin{subequations}\label{eq:DH_vector_form_CL}
    \begin{align}
    % %%% Gamma dynamics
     \dot{\vec{\hat{x}}}_{{\mathcal{E}}} & =\vec{\hat{f}}_{{\mathcal{E}}}(\vec{\hat{x}}_{{\mathcal{E}}})+\vec{\hat{K}}_{{\mathcal{E}}}\vec{d}_{{\mathcal{E}}},~\vec{z}_{{\mathcal{E}}}=\vec{\hat{T}}_{{\mathcal{E}}} \vec{\hat{x}}_{{\mathcal{E}}}, \label{eq:x_Gamma_dynamics_CL_vec}\\
     %%%% Delta dynamics
     \dot{\vec{\hat{x}}}_{{\Delta}} & =\vec{\hat{f}}_{{\Delta}}(\vec{\hat{x}}_{{\Delta}})+\vec{\hat{K}}_{{\Delta}}\vec{d}_{{\Delta}},~\vec{z}_{{\Delta}}=\vec{\hat{T}}_{{\Delta}} \vec{\hat{x}}_{{\Delta}}, \label{eq:x_Delta_dynamics_CL_vec}\\
    % %%% x_D
    % \dot{\vec{\hat{x}}}_\mathcal{D} & =\vec{\hat{f}}_\mathcal{D}(\vec{\hat{x}}_\mathcal{D})+\vec{\hat{K}}_\mathcal{D}\vec{d}_\mathcal{D},~\vec{z}_\mathcal{D}=\vec{\hat{C}}_\mathcal{D} \vec{\hat{x}}_\mathcal{D}, \label{eq:x_D_dynamics_CL_vec}\\
    % %%% x_L
    % \dot{\vec{\hat{x}}}_\mathcal{L} & =\vec{\hat{f}}_\mathcal{L}(\vec{\hat{x}}_\mathcal{L})+\vec{\hat{K}}_\mathcal{L}\vec{d}_\mathcal{L},~\vec{z}_\mathcal{L}=\vec{\hat{C}}_\mathcal{L}  \hat{\vec{x}}_\mathcal{L},\\
    % %%%% x_P
    % \dot{\vec{\hat{x}}}_\mathcal{P} & =\vec{\hat{f}}_\mathcal{P}(\vec{\hat{x}}_\mathcal{P})+\vec{\hat{K}}_\mathcal{P}\vec{d}_\mathcal{P},~\vec{z}_\mathcal{P}=\vec{\hat{C}}_\mathcal{P}  \hat{\vec{x}}_\mathcal{P},\\
    % %%%% x_M
    % \dot{\vec{\hat{x}}}_\mathcal{M} & = \vec{\hat{f}}_\mathcal{M}(\vec{\hat{x}}_\mathcal{M})+\vec{\hat{K}}_\mathcal{M}\vec{d}_\mathcal{M},~\vec{z}_\mathcal{M}=\vec{\hat{C}}_\mathcal{M} \hat{\vec{x}}_\mathcal{M},\\
    % %%%% x_H
    % \dot{\vec{\hat{x}}}_{\HH} & =\vec{\hat{f}}_\HH(\vec{\hat{x}}_\HH) +\vec{\hat{K}}_\HH \vec{d}_\HH,~\vec{z}_\HH=\vec{\hat{C}}_\HH \hat{\vec{x}}_\HH,\\
    %%%% x_C
    % \dot{\vec{\hat{x}}}_\mathcal{C} & = \vec{\hat{K}}_\mathcal{C}\vec{d}_\mathcal{C},~\vec{z}_\mathcal{C}=\vec{\hat{C}}_\mathcal{C}  \hat{\vec{x}}_\mathcal{C},\\
    0 & = \vec{\hat{K}}_\mathcal{K}\vec{d}_\mathcal{K}. \label{eq:algebraic_constraint}
    \end{align}
\end{subequations}
Let us denote by  $\mathbb{B}$ the incidence matrix of the DHN digraph $\GG$ (see Section~\ref{sec:modeling:dhn_graph}). With the considered ordering of the edges and nodes of $\GG$, the incidence matrix can be written as
\begin{equation} \label{eq:indicence_matrix}
\mathbb{B}=\begin{bmatrix}
\mathbb{B}_{\Delta\mathcal{E}} \\
\mathbb{B}_{\mathcal{K}\mathcal{E}} 
\end{bmatrix}.
\end{equation}
% \begin{equation} \label{eq:indicence_matrix}
% \mathbb{B}=\begin{bmatrix}
% \mathbb{B}_{\HH\mathcal{D}} & \mathbb{B}_{\HH\mathcal{L}} & \mathbb{B}_{\HH\mathcal{P}} & \mathbb{B}_{\HH\mathcal{M}}\\
% \mathbb{B}_{\mathcal{C}\mathcal{D}} & \mathbb{B}_{\mathcal{C}\mathcal{L}} & \mathbb{B}_{\mathcal{C}\mathcal{P}} & \mathbb{B}_{\mathcal{C}\mathcal{M}}\\
% \mathbb{B}_{\mathcal{K}\mathcal{D}} & \mathbb{B}_{\mathcal{K}\mathcal{L}} & \mathbb{B}_{\mathcal{K}\mathcal{P}} & \mathbb{B}_{\mathcal{K}\mathcal{M}}
% \end{bmatrix}.
% \end{equation}
%
With \eqref{eq:indicence_matrix}, the interaction inputs of each of the subsystems in \eqref{eq:DH_vector_form_CL} can be represented as 
\begin{subequations}\label{eq:interaction_inputs_vector_form}
\begin{align}
%%% EDGES
\vec{d}_\mathcal{E} & =-\mathbb{B}_{\Delta\mathcal{E}}^\top \vec{z}_\Delta-\mathbb{B}_{\mathcal{K}\mathcal{E}}^\top \vec{z}_\mathcal{K},\\
% \vec{d}_\mathcal{L} & =-\mathbb{B}_{\HH\mathcal{L}}^\top \vec{z}_\HH-\mathbb{B}_{\mathcal{C}\mathcal{L}}^\top \vec{z}_\mathcal{C}-\mathbb{B}_{\mathcal{K}\mathcal{L}}^\top \vec{z}_\mathcal{K},\\
% \vec{d}_\mathcal{P} & =-\mathbb{B}_{\HH\mathcal{P}}^\top \vec{z}_\HH-\mathbb{B}_{\mathcal{C}\mathcal{P}}^\top \vec{z}_\mathcal{C}-\mathbb{B}_{\mathcal{K}\mathcal{P}}^\top \vec{z}_\mathcal{K},\\
% \vec{d}_\mathcal{M} & =-\mathbb{B}_{\HH\mathcal{M}}^\top \vec{z}_\HH-\mathbb{B}_{\mathcal{C}\mathcal{M}}^\top \vec{z}_\mathcal{C}-\mathbb{B}_{\mathcal{K}\mathcal{M}}^\top \vec{z}_\mathcal{K},\\
%%% NODES
\vec{d}_\Delta & = \mathbb{B}_{\Delta\mathcal{E}}\vec{z}_\mathcal{E},\\
% \vec{d}_\mathcal{C} & = \mathbb{B}_{\mathcal{C}\mathcal{D}}\vec{z}_\mathcal{D} +  \mathbb{B}_{\mathcal{C}\mathcal{L}}\vec{z}_\mathcal{L}+ \mathbb{B}_{\mathcal{C}\mathcal{P}}\vec{z}_\mathcal{P}+  \mathbb{B}_{\mathcal{C}\mathcal{M}}\vec{z}_\mathcal{M},\\
\vec{d}_\mathcal{K} & = \mathbb{B}_{\mathcal{K}\mathcal{E}}\vec{z}_\mathcal{E}.
\end{align}
\end{subequations}
Note, e.g., that for any $i\in \mathcal{E}$, ${d}_i=p_j-p_j$, where $(j,k)\in \mathcal{N}_i^{-}\times \mathcal{N}_i^{+}$; see \eqref{eq:modeling:circulation_equations}. That is, ${d}_i$ is expressed as the product of an appropriate column of $\mathbb{B}$ and the vector $\begin{bmatrix}\vec{z}_{\Delta}^\top & \vec{z}_{\mathcal{K}}^\top \end{bmatrix}^\top$. 

Consider the following lemma, which will be useful later to analyze the stability of \eqref{eq:DH_vector_form_CL}.
\begin{lemma} \label{lemma:interconnection}
    The interconnection structure \eqref{eq:interaction_inputs_vector_form} among the subsystems in \eqref{eq:DH_vector_form_CL} is power-preserving.
\end{lemma}
\IEEEproof
As \eqref{eq:interaction_inputs_vector_form} is a skew-symmetric interconnection structure, it holds for all time that
\begin{equation}\label{eq:power_pres_property}
\vec{z}_\mathcal{E}^\top \vec{d}_\mathcal{E}+\vec{z}_\Delta^\top \vec{d}_\Delta+\vec{z}_\mathcal{K}^\top \vec{d}_\mathcal{K}=0.
\end{equation}
% \begin{equation}\label{eq:power_pres_property}
% \sum_{i\in \{\mathcal{D},\mathcal{L},\mathcal{P},\mathcal{M},\HH,\mathcal{C},\mathcal{K} \}}{z}_i^\top {d}_i =0.
% \end{equation}
Hence, the interconnection among the subsystems in \eqref{eq:DH_vector_form_CL} is power-preserving.\footnote{The equivalence \eqref{eq:power_pres_property} also holds if the respective (interaction) input-output pairs are shifted with respect to any feasible equilibrium values.}
\hfill\IEEEQEDclosed

The system conformed by \eqref{eq:DH_vector_form_CL} and \eqref{eq:interaction_inputs_vector_form} is an index-2 differential algebraic system in Hessenberg form (see, e.g., \cite{Harney13numerical}). Based on \cite{Harney13numerical},   the time derivative  of the algebraic constraint~\eqref{eq:algebraic_constraint} can be computed once  to explicitly obtain  $\vec{z}_\mathcal{K}$ in terms of  $\vec{\hat{x}}_\mathcal{E}$ and $\vec{\hat{x}}_\Delta$ as follows:
\begin{subequations}\label{eq:solution_algebraic_eq}
\begin{align}
\vec{z}_\mathcal{K} & =\Phi(\vec{\hat{x}}_\mathcal{E},\vec{\hat{x}}_\Delta),\\
\Phi & := \left(\mathbb{B}_{\mathcal{K}\mathcal{E}}\text{diag}(\tfrac{1}{J_i})_{i\in \mathcal{E}}\mathbb{B}_{\mathcal{K}\mathcal{E}}^\top\right)^{-1} \left(\vec{\hat{K}}_\mathcal{E}\mathbb{B}_{\Delta\mathcal{E}}^\top\vec{\hat{C}}_\Delta\hat{x}_\Delta-\vec{\hat{f}}_\mathcal{E}(\vec{\hat{x}}_\mathcal{E}) \right),
\end{align}
\end{subequations}
where we have used the fact that $\vec{\hat{K}}_\mathcal{K}$ is an identity matrix and $\vec{\hat{C}}_\mathcal{E}\vec{\hat{K}}_\mathcal{E}=\text{diag}(\tfrac{1}{J_i})_{i\in \mathcal{E}}$.
Then, \eqref{eq:DH_vector_form_CL}, \eqref{eq:interaction_inputs_vector_form} is equivalent to the ODE
\begin{subequations}\label{eq:equivalent_ODE}
    \begin{align}
   \dot{\vec{\hat{x}}}_{{\mathcal{E}}} & =\vec{\hat{f}}_{{\mathcal{E}}}(\vec{\hat{x}}_{{\mathcal{E}}})-\vec{\hat{K}}_{{\mathcal{E}}}\mathbb{B}_{\Delta\mathcal{E}}^\top\hat{\vec{T}}_\Delta\hat{\vec{x}}_\Delta-\vec{\hat{K}}_{{\mathcal{E}}}\mathbb{B}_{\mathcal{K}\mathcal{E}}^\top\Phi(\vec{\hat{x}}_\mathcal{E},\vec{\hat{x}}_\Delta),\\
    \dot{\vec{\hat{x}}}_{{\Delta}} & =\vec{\hat{f}}_{{\Delta}}(\vec{\hat{x}}_{{\Delta}})+\vec{\hat{K}}_{{\Delta}}\mathbb{B}_{\Delta\mathcal{E}}\vec{\hat{C}}_\mathcal{E}\vec{\hat{x}}_\mathcal{E},
    \end{align}
\end{subequations}
defined on the invariant manifold
\begin{equation}\label{eq:invariant_set}
    \mathbb{M}=\{(\vec{\hat{x}}_\mathcal{E},\vec{\hat{x}}_\Delta):~0=\vec{\hat{K}}_\mathcal{K}\mathbb{B}_{\mathcal{K}\mathcal{E}}\vec{\hat{C}}_\mathcal{E}\hat{\vec{x}}_\mathcal{E} \}
\end{equation}
The invertibility---and also positive-definiteness---of $\mathbb{B}_{\mathcal{K}\mathcal{E}}\text{diag}(\tfrac{1}{J_i})_{i\in \mathcal{E}}\mathbb{B}_{\mathcal{K}\mathcal{E}}^\top$ in \eqref{eq:solution_algebraic_eq} follows from the fact that $\mathbb{B}_{\mathcal{K}\mathcal{E}}\mathbb{B}_{\mathcal{K}\mathcal{E}}^\top$ is  a principal submatrix of the DHN's graph's Laplacian $\mathbb{L}=\mathbb{B}\mathbb{B}^\top$. This submatrix is invariant under removal of any row of $\mathbb{B}$   not associated to $\mathcal{K}$. Removal of any such row can be understood as a {\em grounding} of $\mathcal{G}$ (see \cite{Miekkala93graph} for a definition). Laplacians of grounded connected graphs, or grounded Laplacians, are positive definite \cite{Miekkala93graph}. Note directly that $\mathbb{B}_{\mathcal{K}\mathcal{E}}\Theta \mathbb{B}_{\mathcal{K}\mathcal{E}}^\top$ is positive definite for any symmetric, positive definite matrix $\Theta$.

\subsection{Asymptotic Stability of the Hydraulic DHN Equilibrium} \label{sec:stability:theorem}

As a last step, we combine the EIP properties of the DHN subsystems analyzed in Lemmas~\ref{lemma:actuated_property_fufilled} and \ref{lemma:EIP_unactuated}, the power-preserving property of their interconnection structure shown in Lemma~\ref{lemma:interconnection}, and LaSalle's Invariance principle to prove asymptotic stability of any feasible, forced hydraulic DHN equilibrium in a modular, bottom-up manner. 

%%%%%%%%%%%%% Theorem DHN Stability
\begin{theorem}\label{thm:stability_analysis}
	Consider a DHN as described in Section~\ref{sec:modeling} with pumps and valves controlled as proposed in Lemma~\ref{lemma:actuated_property_fufilled}. Then, under Assumption~\ref{assumption:valve_equilibrium}, such a DHN admits an asymptotically stable hydraulic equilibrium $(\bar{\vec{\hat{x}}}_\mathcal{E},\bar{\vec{\hat{x}}}_\Delta)$ in which 
    the pump pressures as well as the pump and valve volume flow rates take on the desired steady states specified in Problem~\ref{problem}. 
    That is, 
    for any $i\in\DD\cup\LL\cup\PP_{\text{boost}}\cup\MM\cup\HH$, the respective $\bar{\vec{\hat{x}}}_i$ in $(\bar{\vec{\hat{x}}}_\mathcal{E},\bar{\vec{\hat{x}}}_\Delta)$ are such that $\bar{q}_i=\bar{q}_{\mathrm{P},i}=q_i^*$ and $\bar{p}_{\mathrm{P},i}=p_{\mathrm{P},i}^*$ in accordance with Problem~\ref{problem}.
\end{theorem}
\IEEEproof
See Appendix~\ref{app:proof_stability}.
\section{Simulation} \label{sec:simulation}

% 1) Goal of Simulation: What is to be demonstrated?
In this section, we demonstrate the stabilizing properties, plug-and-play capabilities, and robustness of the proposed pressure and volume flow rate controllers via simulations in {\sc Matlab}/{\sc Simulink} using {\sc Simscape} components.
In Section~\ref{sec:simulation:baseline}, we present a scenario with plug-and-play operations and varying reference values. 
In Section~\ref{sec:simulation:uncertainty}, the first scenario is repeated, albeit with parameter uncertainties and a saturation to the valve input. 
%

% 2) Description of the simulation setup that holds for all scenarios: (Test system, parameters, controller gains)
The simulations are conducted by means of the DHN depicted in Fig.~\ref{fig:dhn_graph} which shows all structural features discussed in Sections~\ref{sec:setup} and \ref{sec:modeling}.
Furthermore, we cover all control problems outlined in Problem~\ref{problem} by assigning appropriate DGU, consumer, and pressure holding configurations, i.e., $\DD_\mathrm{form}=\{1\}$, $\DD_\mathrm{valve}=\{2\}$, $\DD_\mathrm{VSP}=\{3\}$, $\LL_\mathrm{boost}=\{7\}$, $\LL_\mathrm{valve}=\{4,5,6\}$, $\LL_\mathrm{VSP}=\{8\}$, $\PP_\mathrm{boost}=\{15\}$, $\MM =\{25\}$, $\HH =\{4\}$.

The model and controller parameters used in the simulations are reported in 
% {\color{blue} arXiv Tables I and II.}
% {\color{red} 
Tables~\ref{tab:simulation:sim_parameter} and \ref{tab:simulation:pipe_lengths}.

% Pump parameters
The pump parameters follow from \cite[Eq.~(42)]{Goppelt18} by considering that $\frac{R_{\text{P},i}}{J_{\text{P},i}}=7.2878$, $\frac{1}{J_{\text{P},i} C_{\text{P},i}} = 341.4283$, and setting $R_{\text{P},i} = \SI{1e6}{\pascal\second\per\cubic\meter}$ for pressure-controlled pumps, $i\in\DD_\mathrm{form}\cup\DD_\mathrm{valve}\cup\LL_\mathrm{boost}\cup\PP_\mathrm{boost}\cup\HH$ and $R_{\text{P},i} = \SI{1e10}{\pascal\second\per\cubic\meter}$ for volume flow rate-controlled pumps, $i\in\DD_\mathrm{VSP}\cup\LL_\mathrm{VSP}$.\footnote{Note that these hydraulic resistances values are in the order of magnitude of DN 80 and DN 20 pipes, respectively, for a length of \SI{1}{\meter} and an external pressure of \SI{1e5}{\pascal}.}
% Valve parameters; flow capacity
The flow capacity of the control valves is obtained by considering a
maximum volume flow rate of $k_{\text{vs}}=\SI{90}{\cubic\meter\per\hour}=\SI{0.025}{\cubic\meter\per\second}$ at full valve opening $s_{\text{v},i}=1 \rightarrow f_{\text{v},i}(s_{\text{v},i}=1)=1$ and $\mu_i(s_{\text{v},i}=1,k_{\text{vs}}=\SI{0.025}{\cubic\meter\per\second})= \SI{1e5}{\pascal}$ valve pressure (see also \cite[p.~144]{Nussbaumer20}). The value in Table~\ref{tab:simulation:sim_parameter} then follows from
%
% Calculation kvs into Cv
\begin{equation}
    C_{\text{v},i} = \frac{k_{\text{vs}}}{\sqrt{\SI{1e5}{\pascal}}}.
\end{equation}
%
% DGU & consumer parameters, pipes, mixing connection
Any pipe resistance $\lambda_i(q_i)$ and fluid inertia $J_i$, $i\in\EE$, are modelled by using the hydraulic pipe resistance and hydraulic fluid inertia {\sc Simscape} components with standard values and no elevation. The diameters, roughness, and lengths given in Tables~\ref{tab:simulation:sim_parameter} and \ref{tab:simulation:pipe_lengths} are in line with typical values (see \cite{Liu16,Machado22automatica}) and correspond to DN 32 and DN 80 pipes, respectively.
% Capacitive nodes
The elasticities lumped into the capacitive nodes $j\in\CC$ are chosen according to exemplary values for 1-family installations given in \cite{Boysen03,Straede95}.
% Control parameters
Note that for the pressure-controlled pump, we follow Remark~\ref{rem:generic_pump_pressure} and do not assign any damping, i.e., $R_i^\mathrm{P}=R_{\mathrm{P},i}$.
% }

\begin{table}[!t]
	\centering
	\renewcommand{\arraystretch}{1.25}
	\caption{Simulation Parameter Values}
	\label{tab:simulation:sim_parameter}
	\begin{tabular}{lr@{\;}l}
		\noalign{\hrule height 1.0pt}
		Pressure-controlled  & $R_{\text{P},i} =$ & $\SI{1e6}{\pascal\second\per\cubic\meter}$\\
        pumps \eqref{eq:modeling:pumps_rlc_equations} & $J_{\text{P},i} =$ & $\SI{1.37e5}{\pascal\square\second\per\cubic\meter}$  \\
        & $C_{\text{P},i} =$ & $\SI{2.13e-8}{\cubic\meter\per\pascal}$\\
		\hline
        Flow-controlled  & $R_{\text{P},i} =$ & $\SI{1e10}{\pascal\second\per\cubic\meter}$\\
        pumps \eqref{eq:modeling:pumps_rlc_equations} & $J_{\text{P},i} =$ & $\SI{1.37e9}{\pascal\square\second\per\cubic\meter}$  \\
        & $C_{\text{P},i} =$ & $\SI{2.13e-12}{\cubic\meter\per\pascal}$\\
		\hline
		Control valves \eqref{eq:modeling:valve_full} & $C_{\text{v},i} =$ & $\SI{7.9e-5}{\cubic\meter\per\second\pascal\tothe{-0.5}}$ \\
		\hline
		DGUs \& consumers \eqref{eq:modeling:circulation_equations} & length $=$& $\SI{25}{\meter}$\\
        (DN 32) & diameter = & $\SI{0.0359}{\meter}$ \\
        & roughness = & $\SI{4.5e-5}{\meter}$\\
		\hline
		Pipes \eqref{eq:modeling:pipe_equations} & diameter $=$ & $\SI{0.0825}{\meter}$\\
        (DN 80) & roughness $=$ & $\SI{4.5e-5}{\meter}$ \\
        \hline
        Mixing connection \eqref{eq:modeling:mixing_equations} & length $=$& $\SI{25}{\meter}$\\
        (DN 80)& diameter = & $\SI{0.0825}{\meter}$ \\
        & roughness $=$ & $\SI{4.5e-5}{\meter}$ \\
        \hline
        Capactive nodes \eqref{eq:model_capacitive_nodes} & $C_j =$ & $\SI{5e-10}{\cubic\meter\per\pascal}$\\
        \hline
        Pressure controller  & $Q_{\mathrm{I},i}^{-1} =$ & $\SI{3.64e-7}{}$\\
        pump \eqref{eq:pump_pressure_controller}& $R_i^\mathrm{P} =$ & $R_{\mathrm{P},i}$ \\
        \hline
		Flow controller  & $K_{\mathrm{P},i} =$ & $\SI{2e6}{}$ \\
        pump \eqref{eq:PI_control_flow_pump_generic}& $Q_{\mathrm{I},i}^{-1} =$ & $\SI{2e10}{}$ \\ 
        \hline
		Flow controller  & $K_{\mathrm{P},i} =$ & $\SI{1e4}{}$\\
        valve \eqref{eq:PI_controller_valve_generic} & $Q_{\mathrm{I},i}^{-1} =$ & $\SI{1e4}{}$\\
		\noalign{\hrule height 1.0pt}
	\end{tabular}
\end{table}
\begin{table}[!t]
	\centering
	\renewcommand{\arraystretch}{1.25}
	\caption{Pipe Lengths}
	\label{tab:simulation:pipe_lengths}
	\begin{tabular}{cc@{\qquad}cc@{\qquad}cc@{\qquad}cc}
		\noalign{\hrule height 1.0pt}
		Pipe & Length & Pipe & Length & Pipe & Length  \\
		\hline
		9  & $\SI{350}{\meter}$ & 15 & $\SI{50}{\meter}$  & 20 & $\SI{100}{\meter}$ \\
		10 & $\SI{300}{\meter}$ & 16 & $\SI{400}{\meter}$ & 21 & $\SI{100}{\meter}$ \\
		11 & $\SI{300}{\meter}$ & 17 & $\SI{100}{\meter}$ & 22 & $\SI{100}{\meter}$ \\
		12 & $\SI{200}{\meter}$ & 18 & $\SI{100}{\meter}$ & 23 & $\SI{100}{\meter}$ \\
            13 & $\SI{200}{\meter}$ & 19 & $\SI{100}{\meter}$ & 24 & $\SI{100}{\meter}$ \\
            14 & $\SI{50}{\meter}$ &&&&\\
		\noalign{\hrule height 1.0pt}
	\end{tabular}
\end{table}

\begin{table}[!t]
	\centering
	\renewcommand{\arraystretch}{1.25}
	\caption{Pressure and Volume Flow Rate Setpoints}
	\label{tab:simulation:references}
	\begin{tabular}{cccc}
		\noalign{\hrule height 1.0pt}
        Edges && $p_{\mathrm{P},i}^*$ in \SI{e5}{\pascal}   & $q_i^*$ in \SI{e-3}{\cubic\meter\per\second}  \\
        \hline
		1 & \textcolor{colorDGU1}{\rule[.5ex]{2.5em}{1pt}} &  15& --- \\
        2 & \textcolor{colorDGU2}{\rule[.5ex]{2.5em}{1pt}} &   10& 3.5 (4.5)\\
        3 & \textcolor{colorDGU3}{\rule[.5ex]{2.5em}{1pt}} &   ---               & 3\\ 
        \hline
        4 & \textcolor{colorEU4}{\rule[.5ex]{1em}{1pt}\,\rule[.5ex]{0.1em}{1pt}\,\rule[.5ex]{1em}{1pt}} &  ---               & 2\\ 
        5 & \textcolor{colorEU5}{\rule[.5ex]{1em}{1pt}\,\rule[.5ex]{0.1em}{1pt}\,\rule[.5ex]{1em}{1pt}} & ---               & 2\\ 
        6 & \textcolor{colorEU6}{\rule[.5ex]{1em}{1pt}\,\rule[.5ex]{0.1em}{1pt}\,\rule[.5ex]{1em}{1pt}} & ---               & 2 (4)\\ 
        7 & \textcolor{colorEU7}{\rule[.5ex]{1em}{1pt}\,\rule[.5ex]{0.1em}{1pt}\,\rule[.5ex]{1em}{1pt}} & 6                 & 2.5 (5)\\ 
        8 & \textcolor{colorEU8}{\rule[.5ex]{1em}{1pt}\,\rule[.5ex]{0.1em}{1pt}\,\rule[.5ex]{1em}{1pt}} & ---               & 3 (6)\\ 
        \hline
        15& \textcolor{colorBoost}{\rule[.5ex]{2.5em}{1pt}} &  5& ---\\ 
        \hline
        25& \textcolor{colorMix}{\rule[.5ex]{2.5em}{1pt}} &   ---               & 1 (3)\\ 
        \noalign{\hrule height 1.0pt}
        Nodes &&&\\
        \hline
        $\HH=4$ & \textcolor{colorHold}{\rule[.5ex]{2.5em}{1pt}} &2 & ---\\ 
		\noalign{\hrule height 1.0pt}
	\end{tabular}
\end{table}

\subsection{Scenario A: Plug-and-Play and Setpoint Changes} \label{sec:simulation:baseline}
\begin{figure}[t]
	\centering
    \includegraphics{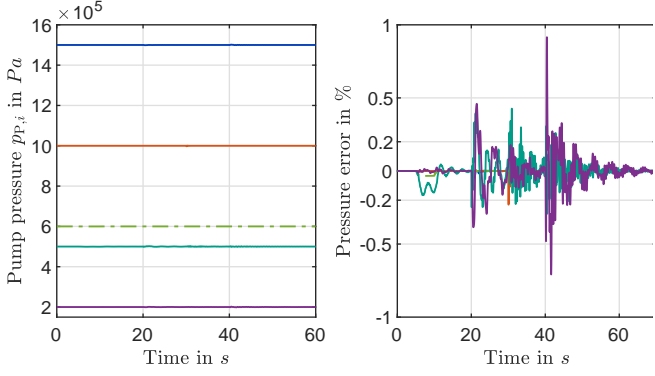}
	\caption{Scenario A: simulated pump pressures in DGUs $\{1,2\}$, consumer $\{7\}$, booster pump $\{15\}$, and dynamic pressure holding unit at node $\{4\}$ with corresponding deviations from the references; the line colors are as per Table~\ref{tab:simulation:references}}
	\label{fig:simulation_results_pressures}
\end{figure}

\begin{figure}[t]
	\centering
    \includegraphics{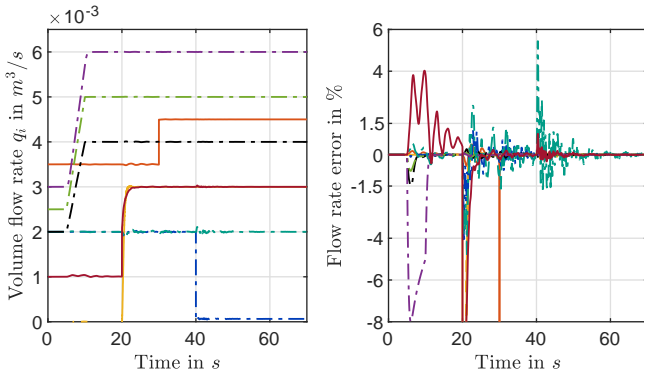}
	\caption{Scenario A: simulated volume flow rates through DGUs $\{2,3\}$, consumers $\{4,5,6,7,8\}$, and mixing valve $\{25\}$ with corresponding deviations from the references; the line colors are as per Table~\ref{tab:simulation:references}}
	\label{fig:simulation_results_flows}
\end{figure}

% 2) Description of the simulation setup: (Test system, parameters, input signals, events)
In this scenario, the simulation starts off with DGU 3 disconnected. 
% Note that the nodes $\CC=\{12,17\}$ are assumed to remain during the DGU disconnect. 
The pressure and volume flow rate setpoints for the controlled pumps and valves are assigned as in Table~\ref{tab:simulation:references}. At the indicated times, the following events occur:
\begin{itemize}
	\item $t = \SI{5}{\second}$: Consumers $\{6,7,8\}$ increase their their volume flow rates by \SI{100}{\percent} until $t = \SI{10}{\second}$.
	\item $t = \SI{20}{\second}$: To help cover the increased demand, DGU 3 connects and the mixing connection 23 increases its volume flow rate to $\SI{3e-3}{\cubic\meter\per\second}$.
	\item $t = \SI{30}{\second}$: DGU 2 increases its input volume flow rate to $\SI{4.5e-3}{\cubic\meter\per\second}$.
	\item $t = \SI{40}{\second}$: Consumer 4 disconnects.
\end{itemize}
%
% 3) Results (only factual) and 4) Interpretation are mixed
The pressure and volume flow rate trajectories shown in Figs.~\ref{fig:simulation_results_pressures} and \ref{fig:simulation_results_flows} confirm the theoretical stability statements. Despite plug-and-play operations and changing operating conditions, the pressures of pressure-controlled pumps and the volume flow rates of flow-controlled pumps and valves are asymptotically stabilized at their desired setpoints. 
For the pressures, the maximum deviations resulting from the events at $t = \SI{5}{\second}, \SI{20}{\second}, \SI{30}{\second}, \SI{40}{\second}$ remain within a \SI{1}{\percent} band with respect to the setpoints and subside below \SI{0.2}{\percent} within approximately \SI{5}{\second}.
For the volume flow rates, larger deviations can be observed. In particular at $t = \SI{20}{\second}$ and $t = \SI{30}{\second}$ during the connection of DGU 3 and the setpoint changes the error plots in Fig.~\ref{fig:simulation_results_flows} shows large outliers. However, from a practical perspective, this is natural as abrupt setpoint changes cannot be realized instantly by the respective volume flow rate controllers.
More importantly, except for the load ramps at $t\in\left[\SI{5}{\second},\SI{10}{\second}\right]$, the volume flow rates settle to within a \SI{1.5}{\percent} band with respect to the setpoints after at most \SI{5}{\second}.
During the load ramps, the errors are slightly higher, but remain below \SI{8}{\percent}. This shows that
% Additionally, the ramps between $t = \SIrange{5}{10}{\second}$ show 
the volume flow rate controllers for both pumps and valves, although not specifically designed for it, are sufficiently fast to adquately track setpoints that vary on a time scale of seconds.
%
% \begin{figure}
% 	\centering
%     \includegraphics{00_Tex/Img/simulation_results_deltas_uncertainty_sat_valve.eps}
% 	\caption{Scenario B: pressure and volume flow rate deviations from the references; the line colors are as per Table~\ref{tab:simulation:references}}
% 	\label{fig:simulation_results_delta_uncertainty_sat_valve}
% \end{figure}

\subsection{Scenario B: Parameter Uncertainty and Valve Saturation} \label{sec:simulation:uncertainty}
% 2) Description of the simulation setup: (Test system, parameters, input signals, events)
Scenario B is similar to Scenario A except for two modification: firstly, a \SI{10}{\percent} uncertainty is added to the pump parameters $R_{\text{P},i},J_{\text{P},i}, C_{\text{P},i}$ and the valve parameters $C_{\text{v},i}$; secondly, the virtual valve control input of all valves is saturated to $u_{\text{v}} \in \left[1, u_{\text{v},i}^\mathrm{max}\right]$ (see Assumption~\ref{assumption:stem_position}).\footnote{In line with classical feedback control design, we did not consider the possibility of control input saturation explicitly during the control design stage in Section~\ref{sec:control}. Instead, we analyze its impact by means of the numerical simulation in this section.}

\begin{figure}[t]
	\centering
 \includegraphics{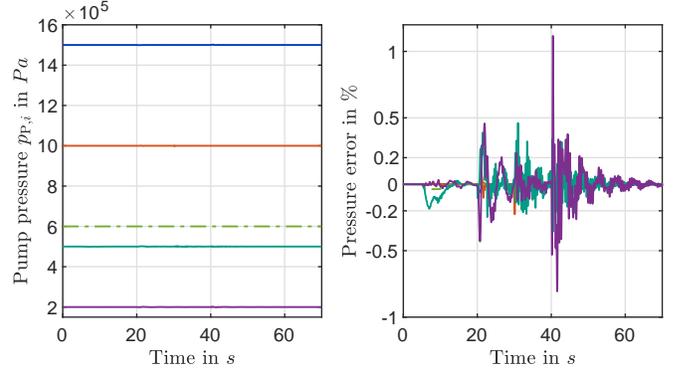}
	\caption{Scenario B: simulated pump pressures in DGUs $\{1,2\}$, consumer $\{7\}$, booster pump $\{15\}$, and dynamic pressure holding unit at node $\{4\}$ with corresponding deviations from the references; the line colors are as per Table~\ref{tab:simulation:references}}
	\label{fig:simulation_results_pressures_unvertainty_sat_valve}
\end{figure}

\begin{figure}[t]
	\centering
    \includegraphics{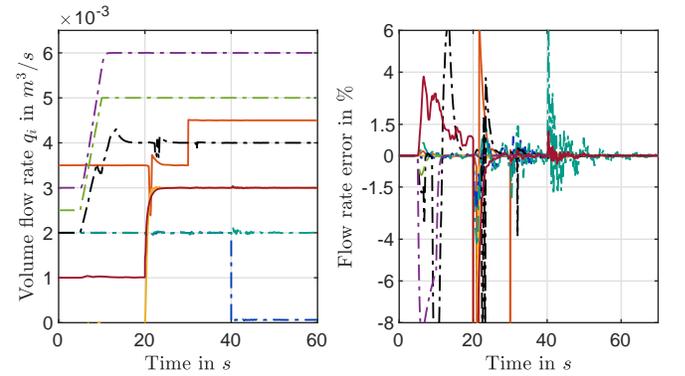}
	\caption{Scenario B: simulated volume flow rates through DGUs $\{2,3\}$, consumers $\{4,5,6,7,8\}$, and mixing valve $\{25\}$ with corresponding deviations from the references; the line colors are as per Table~\ref{tab:simulation:references}}
	\label{fig:simulation_results_flows_unvertainty_sat_valve}
\end{figure}

% 3) Results (only factual)
The resulting pressure and volume flow rate trajectories shown in Figs.~\ref{fig:simulation_results_pressures_unvertainty_sat_valve} and \ref{fig:simulation_results_flows_unvertainty_sat_valve} are similar to those shown in Figs.~\ref{fig:simulation_results_pressures}. The main difference introduced by the valve saturation is an impaired convergence performance of the volume flow rate control via valves, particularly at DGU 2 and Consumer 6 (see the orange and black lines in Fig.~\ref{fig:simulation_results_flows_unvertainty_sat_valve}). In practice, an appropriate redesign of the valves or by increasing the available pressure, e.g., via the booster pump in Pipe 15 or a separate booster pump in the respective consumer, control performance can be improved. 
%
% The maximum absolute pressure difference is approximately \SI{2e3}{\pascal} and the maximum absolute volume flow rate difference is \SI{6.7e-4}{\cubic\meter\per\second} at the reference change of the mixing valve and otherwise \SI{1e-4}{\cubic\meter\per\second}
%

Overall, the results of the two scenarios illustrate that the passivity-based pressure and volume flow rate controllers indeed asymptotically stabilize the hydraulic variables while allowing for plug-and-play operations of the different DHN subsystems. Furthermore, the integral parts of the proposed controllers provides robustness against parameter uncertainties and changing hydraulic conditions naturally occuring during the operation of DHNs.

\section{Conclusion} \label{sec:conclusion}

In this work, we have proposed a unifying control framework that guarantees pressure and volume flow rate stability based on the EIP properties of the DHN subsystem models.
We provided a comprehensive hydraulic model covering state-of-the-art as well as future DHN generations and formalized the hydraulic control problems arising in such systems. 
% {\color{red} 
We demonstrated how each of the subsystem models can be represented in port-Hamiltonian form to facilitate the subsequent passivity-based control design and stability analysis.
% }
Subsequently, we showed how the hydraulic control tasks can be simplified to designing decentralized, passivity-based controllers for the pumps and valves in a DHN and proposed three controllers: two for pressure and volume flow rate control via pumps and one for volume flow rate control via valves.
Based on the EIP properties of the (actuated and unactuated) subsystem models, the skew-symmetry of their interconnection structure, and LaSalle's Invariance principle, we then proved asymptotic stability of any forced, hydraulic DHN equilibrium in a modular manner.
In conclusion, we want to highlight that the presented EIP-based control framework is not restricted to the models and controllers proposed in this work. In fact, the modular approach of the EIP-based stability analysis allows to incorporate more detailed models (e.g., for the pumps and valves) and presents general guidelines for developing other decentralized pressure and volume flow rate controllers.

%%%%%%%%%%% Appendices
\appendices

% {\color{red}
\section{ISO-PHS representation of the DHN subsystems}\label{APP:ISO-PHS fomulation}

\begin{proposition}\label{prop:subsystems_as_pHs}
Consider the following claims:

\begin{itemize}
%[label={\bf (\Roman*)}, wide, labelwidth=!,labelindent=0pt]
    \item[{\bf (i)}] 	The hydraulic model of any pump, or  of any pipe in series with a control valve, can be written as ISO-PHS.

	\smallskip
	
	\item[{\bf (ii)}] The hydraulic model of any DGU, consumer, mixing connection, as well as of any pressure holding or capacitive node (see \eqref{eq:modeling:circulation_equations}, \eqref{eq:modeling:mixing_equations},  \eqref{eq:model_pressure_holding}, and \eqref{eq:model_capacitive_nodes}, respectively), can be written as ISO-PHS. 
\end{itemize}
\end{proposition}
\IEEEproof
%\todo{Juan: Streamline proof. DONE.}
To see that {\bf (i)} holds, consider first the pump dynamics in \eqref{eq:modeling:pumps_rlc_equations}. By direct substitution, it can be verified that this model can be written as an ISO-PHS by defining 
\begin{equation}\label{eq:modeling:pump_phs}
\begin{aligned}
		% States
		\x_{i} &= \begin{bmatrix}J_{\text{P},i} q_{\mathrm{P},i} & C_{\text{P},i} p_{\text{P},i} \end{bmatrix}^\top, \\
		% Control
		{u}_{i} &=u_{\mathrm{P},i},~ \zeta_i= q_{\mathrm{P},i},~
		% Interaction 
		{d}_i=-q_i,
		{z}_i=p_{\mathrm{P},i},\\	
		% J, R, R, G, K 
		\J_{i} &= \begin{bmatrix}
			0 & -1 \\
			1 & 0
		\end{bmatrix}, \,
		\R_{i} = \text{diag}(R_{\mathrm{P},i},0),~
		\vec{\mathcal{R}}_i(x_i) = \zero_2,\\
		\G_i & = \begin{bmatrix}
			1  \\
			0 
		\end{bmatrix},~
		\K_i  =\begin{bmatrix}0\\ 
1 \end{bmatrix},~
		\Q_i = \text{diag}(\tfrac{1}{J_{\text{P},i}}, \tfrac{1}{C_{\text{P},i}}).
\end{aligned}
\end{equation}

Moving on, the model of any pipe in series with a valve is equivalent to \eqref{eq:modeling:mixing_equations}.  Then, the ISO-PHS representation is attained  by defining
\begin{equation}	\label{eq:modeling:valve_phs}
\begin{aligned}
		% States and co-states mixing
		x_i &= J_iq_i,~
		% Control port mixing
		{u}_i=u_{\text{v},i}(s_{\text{v},i}),~ \zeta_i=\frac{1}{C_{\text{v},i}^2}q_{i}^3,\\
		% Interaction port mixing
		{d}_i&=p_j-p_k,~ z_i =q_i,~
		% J, R, R, G, K mixing
		J_i = 0, \,
		R_i = 0, \\
		{\mathcal{R}}_i(x_i) & = \lambda_i(q_i),~G_i(x_i) =-\hat{\mu}_i(q_i),\\
		K_i  & =1,~ Q_i=\frac{1}{J_i},
\end{aligned}
\end{equation}
with $(j,k)\in \mathcal{N}_i^{-}\times \mathcal{N}_i^+$.

\bigskip

\noindent Now, we show that {\bf (ii)} holds. Any  DGU  or  consumer $i\in \mathcal{D}\cup \mathcal{C}$ is modeled by \eqref{eq:modeling:circulation_equations}. Then, ISO-PHS representation is attained by defining
\begin{align}\label{eq:modeling:dgu_phs}
		% States and co-states DGU
		\x_i &= \begin{bmatrix} J_i q_{i} &J_{\mathrm{P}, i} q_{\mathrm{P},i} &C_{\mathrm{P},i}p_{\mathrm{P},i}  \end{bmatrix}^\top,\nonumber \\ 
		% Control port DGU
		\vec{u}_i&=\begin{bmatrix}u_{\text{v},i} &u_{\mathrm{P},i} \end{bmatrix}^\top,~\vec{\zeta}_i=\begin{bmatrix}-\hat{\mu}_i(q_i)q_i & q_{\mathrm{P},i} \end{bmatrix}^\top, \nonumber \\
		% Interaction port DGU
		{d}_i &= p_j-p_k,~
		\z_i =q_i, \nonumber \\
		% J, R, R, G, K DGU
		\J_{i} &= \begin{bmatrix}
			0 & 0 & 1 \\
			0 & 0 & -1\\
-1 & 1 & 0
		\end{bmatrix}, \,
		\R_i=\text{diag}(0,R_{\mathrm{P},i},0), \nonumber \\
		\vec{\mathcal{R}}_{i}(\x_i) & = \begin{bmatrix}
			\lambda_{i}(q_i) \\
			0\\
			0 
		\end{bmatrix},~
		\G_i(\x_i) = \begin{bmatrix}
			-\hat{\mu}_i(q_{\text{P2},i}) & 0\\
			0 & 1\\
			0 & 0
		\end{bmatrix}, \nonumber  \\
		\K_i & = \begin{bmatrix}
			1  \\
			0 \\
			 0
		\end{bmatrix},~ \Q_i = \text{diag}\left(\tfrac{1}{J_i}, \tfrac{1}{J_{\mathrm{P},i}},\tfrac{1}{C_{\mathrm{P},i}}\right),
\end{align}
where $(j,k)\in \mathcal{N}_i^-\times \mathcal{N}_{i}^+$.

The model \eqref{eq:modeling:pipe_equations} of each  pipe $i\in \mathcal{P}$  can be written as a ISO-PHS by defining
\begin{align}\label{eq:modeling:pipe_phs}
		% States and co-states DGU
		\x_i &= \begin{bmatrix} J_i q_{i} &J_{\mathrm{P}, i} q_{\mathrm{P},i} &C_{\mathrm{P},i}p_{\mathrm{P},i}  \end{bmatrix}^\top,\nonumber \\
		% Control port DGU
		{u}_i & =  u_{\mathrm{P},i},~\zeta_i=q_{\mathrm{P},i}, \nonumber \\
		% Interaction port DGU
		{d}_i &= p_j-p_k,~~
		z_i =q_i, \nonumber \\
		% J, R, R, G, K DGU
		\J_{i} &= \begin{bmatrix}
			0 & 0 & 1 \\
			0 & 0 & -1\\
-1 & 1 & 0
		\end{bmatrix},~
		\R_i=\text{diag}(0,R_{\mathrm{P},i},0), \nonumber \\
		\vec{\mathcal{R}}_{i}(\x_i) & = \begin{bmatrix}
			\lambda_{i}(q_i) \\
			0\\
			0 
		\end{bmatrix},~
		\G_i(\vec{x}_i)  = \begin{bmatrix} 0\\1\\ 0 \end{bmatrix}, \nonumber  \\
		\K_i & = \begin{bmatrix}
			1  \\
			0 \\
			 0
		\end{bmatrix},~ \Q_i = \text{diag}\left(\tfrac{1}{J_i}, \tfrac{1}{J_{\mathrm{P},i}},\tfrac{1}{C_{\mathrm{P},i}}\right),
\end{align}
where $(j,k)\in \mathcal{N}_i^-\times \mathcal{N}_{i}^+$.

Each mixing valve $i\in \mathcal{M}$ modeled by \eqref{eq:modeling:mixing_equations} can be represented as an ISO-PHS by defining 
\begin{equation}	\label{eq:modeling:mixing_phs}
\begin{aligned}
		% States and co-states mixing
		{x}_i &= J_iq_i,~
		% Control port mixing
		{u}_i=u_{\text{v},i}(s_{\text{v},i}),~ \zeta_i=\frac{1}{C_{\text{v},i}^2}q_{i}^3,\\
		% Interaction port mixing
		{d}_i&=p_j-p_k,~ z_i =q_i,~
		% J, R, R, G, K mixing
		J_i = 0, \,
		R_i = 0, \\
		{\mathcal{R}}_i(x_i) & = \lambda_i(q_i),G_i(x_i) =-\hat{\mu}_i(q_i),\\
		K_i  & =1, Q_i=\frac{1}{J_i},
\end{aligned}
\end{equation}
where $(j,k)\in \mathcal{N}_i^-\times \mathcal{N}_{i}^+$.

%%% Pump dynamics (or pressure holding node)
Any pressure holding unit $j\in \mathcal{H}$ modeled by \eqref{eq:model_pressure_holding} can be represented as an ISO-PHS by defining 
\begin{equation}\label{eq:d_j:pressure_holding}
\begin{aligned}
		% States
		\x_{j} &= \begin{bmatrix}J_{\text{P},j} q_{\mathrm{P},j} & C_{\text{P},j} p_{\text{P},j} \end{bmatrix}^\top, \\
		% Control
		{u}_{j}&=u_{\mathrm{P},j},~ \zeta_i= q_{\mathrm{P},j},~
		% Interaction 
        {d}_j=\sum_{i\in \mathcal{I}_j}q_i,~
		{z}_j=p_{\mathrm{P},j},\\	
		% J, R, R, G, K 
		\J_{j} &= \begin{bmatrix}
			0 & -1 \\
			1 & 0
		\end{bmatrix}, \,
		\R_{j} = \text{diag}(R_{\mathrm{P},j},0),~
		\vec{\mathcal{R}}_j(\vec{x}_j) = \zero_2,\\
		\G_j & = \begin{bmatrix}
			1  \\
			0 
		\end{bmatrix},~
		\K_j  =\begin{bmatrix}0\\ 
1 \end{bmatrix},~
		\Q_j = \text{diag}(\tfrac{1}{J_{\text{P},j}}, \tfrac{1}{C_{\text{P},j}}),
\end{aligned}
\end{equation}
where $\mathcal{I}_j\subseteq \mathcal{E}$ is the set of edges that are incident to $j$.

Finally, each capacitive node $j\in\mathcal{C}$ modeled by \eqref{eq:model_capacitive_nodes} can be written as an ISO-PHS by defining 
\begin{equation}\label{eq:capacitor_dynamics_PH}
\begin{aligned}
		% States and co-states mixing
		{x}_j &= C_jp_j,~
		% Control port mixing
		{u}_j=0,~ \zeta_j=0,~
		% Interaction port mixing
		{d}_j= \sum_{i\in \mathcal{I}_j}q_i,~{z}_j=p_j,\\	
		% J, R, R, G, K mixing
		J_j &= 0, \,
		R_j = 0,~
		{\mathcal{R}}_j(x_j)  =0,~G_j(x_j) =0, \\
		K_j & =1,~ Q_i=\frac{1}{C_i},
\end{aligned}
\end{equation}
where  $\mathcal{I}_j$ is defined as before.
% }

\section{Proof of Proposition~\ref{prop:pressure_control_pump_general}} \label{app:proof:pressure_control_pump_general}

Following \cite{Donaire09}, we begin by introducing a change of coordinates from $q_{\mathrm{P},i}$ to $\chi_i$ as (see \eqref{eq:controller_pump_specific})
\begin{equation}\label{eq:change_coord_donaire}
    \chi_i=q_{\mathrm{P},i}+r_i.
\end{equation}
Then, by  computing $\dot{\chi}_{i}$, we get 
\begin{align}
	J_{\mathrm{P},i}\dot{{\chi}}_{i} & = J_{\mathrm{P},i}\dot{q}_{\mathrm{P},i}+J_{\mathrm{P},i}\dot{r}_i = \nu_i + u_{\mathrm{P},i},
\end{align}
where $\nu_i$ is given in \eqref{eq:controller_pump_specific}. Following the IDA-PBC  design methodology \cite{Ortega04}, we assign $u_{\mathrm{P},i}$ as in \eqref{eq:controller_pump_specific} to obtain
\begin{equation}
J_{\mathrm{P},i}\dot{{\chi}}_i=-R^\mathrm{p}_i \chi_i-(p_{\mathrm{P},i}-p_{\mathrm{P},i}^*).
\end{equation}
By propagating the coordinate transformation \eqref{eq:change_coord_donaire} to the $p_{\mathrm{P},i}$-dynamics, we can write the closed-loop dynamics as in \eqref{eq:closed_loop_dyn_press_control_general}.
%
% \begin{subequations}\label{eq:pump_close_loop_explicit_pressure}
% \begin{align}
%     J_{\mathrm{P},i}\dot{{\chi}}_i & =-R^\mathrm{p}_i \chi_i-(p_{\mathrm{P},i}-p_{\mathrm{P},i}^*),\\
%     C_{\mathrm{P},i}\dot{p}_{\mathrm{P},i} & = \chi_i-r_i+d_i,\\
%     Q_{\mathrm{I},i}\dot{r}_i & = (p_{\mathrm{P},i}-p_{\mathrm{P},i}^*),
%     \end{align}
% \end{subequations}
%
% which are as in \eqref{eq:closed_loop_dyn_press_control_general}.
%
In order to show that \eqref{eq:closed_loop_dyn_press_control_general} is EIP with supply rate $({d}_i-\bar{{d}}_i) ({z}_i-\bar{{z}}_i)$ and positive definite storage function $H^\mathrm{p}_i$ in \eqref{eq:H_p_P}, let ${d}_i$ be fixed to an arbitrary equilibrium value $\bar{{d}}_i$ with associated equilibria $\bar{{z}}_i$ for the output and $\vec{\bar{x}}_i^\mathrm{p}$ (see \eqref{eq:equilibrium_pump_close_loop_pressure}) for the state vector. Since $\vec{\bar{x}}^\mathrm{p}_i$ satisfies  $\vec{f}^\mathrm{p}_i(\vec{\bar{x}}_i^\mathrm{p})+\vec{K}_i^\mathrm{p}{\bar{d}}_i=0$, we can write 
\eqref{eq:closed_loop_dyn_press_control_general} equivalently as
\begin{equation}\label{eq:proof_pump_press_controller_shifted_form}
    \dot{\x}_i^\mathrm{p} = \vec{f}_i^\mathrm{p}(\x_i^\mathrm{p})-\vec{f}_i^\mathrm{p}(\vec{\bar{x}}_i^\mathrm{p})+\vec{K}_i^\mathrm{p}({d}_i-\bar{{d}}_i).
\end{equation}
For the time derivative of $H^\mathrm{p}_i$ in \eqref{eq:H_p_P}, it holds that
\begin{subequations}\label{eq:V_dot_press_control_generic}
\begin{align}
&	\dot{{H}}^\mathrm{p}_i  = \nabla^\top {H}^\mathrm{p}_i\dot{\x}^\mathrm{p}_i = -\psi_i^\mathrm{p}(\x_i^\mathrm{p}) +  ({z}_i-\bar{{z}}_i) ({d}_i-\bar{{d}}_i),\\
&	\psi_i^\mathrm{p}(\x_i^\mathrm{p})  = R^\mathrm{p}_i (\chi_i-\bar{\chi}_i)^2,
	\end{align}
\end{subequations}
where we have used the identity 
$$(\K^\mathrm{p}_i)^\top \nabla H^\mathrm{p}_i=	{z}_i-\bar{{z}}_i=p_{\mathrm{P},i}-\bar{p}_{\mathrm{P},i}.$$ 
Since $R^\mathrm{p}_i>0$, it follows that $\psi_i\mathrm{p}(\x_i^\mathrm{p})\geq  0$ implying $\dot{H}^\mathrm{p}_i\leq ({z}_i-{\bar{z}}_i) ({d}_i-{\bar{d}}_i)$. Hence, EIP is established. 
%
% Due to the integral action xxx, it can directly be verified from \eqref{eq:closed_loop_dyn_press_control_general} that for any feasible equilibrium value $\bar{d}_i$, $\bar{\x}_i^\mathrm{p}$ is as in \eqref{eq:equilibrium_pump_close_loop_pressure}.

\section{Proof of Proposition~\ref{prop:flow_control_pump_general}} \label{app:proof:flow_control_pump_general}

By combining \eqref{eq:modeling:pipe_equations} with the controller \eqref{eq:PI_control_flow_pump_generic}, the closed-loop system \eqref{eq:closed_loop_sys_pump_flow_control_a} follows directly.
In order to show that system \eqref{eq:closed_loop_sys_pump_flow_control_a} is EIP with supply rate $({d}_i-\bar{{d}}_i) ({z}_i-\bar{{z}}_i)$ and positive definite storage function $H^\mathrm{f}_i$ in \eqref{eq:H_f}, let ${d}_i$ be fixed to an arbitrary equilibrium value $\bar{{d}}_i$ with associated equilibria $\bar{{z}}_i$ for the output and $\vec{\bar{x}}_i^\mathrm{f}$ (see \eqref{eq:equilibrium_pump_close_loop_flow}) for the state vector. Since  $\vec{\bar{x}}^\mathrm{f}_i$ satisfies  $\vec{f}_i(\vec{\bar{x}}_i^\mathrm{f})+\vec{K}_i^\mathrm{f}{\bar{d}}_i=0$, we can write \eqref{eq:closed_loop_sys_pump_flow_control_a} equivalently as
\begin{equation}\label{eq:proof_pump_controller_shifted_form}
    \dot{\x}_i^\mathrm{f} = \vec{f}_i^\mathrm{f}(\x_i^\mathrm{f})-\vec{f}_i^\mathrm{f}(\vec{\bar{x}}_i^\mathrm{f})+\vec{K}_i^\mathrm{f}({d}_i-\bar{{d}}_i).
\end{equation}
For the time-derivative of $H_i^\mathrm{f}$ in \eqref{eq:H_f}, it holds that
\begin{subequations}
\begin{align}
		&\dot{H}^\mathrm{f}_i  =\nabla^\top H^\mathrm{f}_i\dot{\x}^\mathrm{f}  =-\psi_i^\mathrm{f}(\x_i^\mathrm{f}) +({z}_i-\bar{{z}}_i)({d}_i-\bar{{d}}_i)  \\
       & \psi_i^\mathrm{f}(\x_i^\mathrm{f}) = (q_i-\bar{q}_i)(\lambda_i(q_i)-\lambda_i(\bar{q}_i)) \nonumber\\
       &\phantom{\psi_i^\mathrm{f}(\x_i^\mathrm{f}) =} +2\frac{R_{\mathrm{P},i}Q_{\mathrm{I},i}}{\kappa_i^\mathrm{f}}(q_{\mathrm{P},i}-\bar{q}_{\mathrm{P},i})^2,
	\end{align}
\end{subequations}
% 	\begin{align}
% 		\dot{H}^\mathrm{f}_i  =\nabla^\top H^\mathrm{f}_i\dot{\x}^\mathrm{f} & =-(q_i-\bar{q}_i)(\lambda_i(q_i)-\lambda_i(\bar{q}_i)) \nonumber \\
% & \phantom{=} -2\tfrac{R_{\mathrm{P},i}Q_{\mathrm{I},i}}{\kappa_i^\mathrm{f}}(q_{\mathrm{P},i}-\bar{q}_{\mathrm{P},i})^2\\
% & \phantom{=}+({z}_i-\bar{{z}}_i)({d}_i-\bar{{d}}_i),
% 	\end{align}
%
where we have used the identity 
$$(\K^\mathrm{f}_i)^\top \nabla H^\mathrm{f}_i=	{z}_i-\bar{{z}}_i=q_{i}-\bar{q}_{i}.$$ 
Since $\lambda_i(q_i)$ is strictly increasing and \eqref{eq:conds_stab_pump_flow} holds, it follows that $\psi_i\mathrm{f}(\x_i^\mathrm{f})\geq  0$ implying $\dot{H}^\mathrm{f}_i\leq ({z}_i-{\bar{z}}_i) ({d}_i-{\bar{d}}_i)$. Hence, EIP is established. 
% Due to the integral action of the controller, it is straightforward to see from \eqref{eq:closed_loop_sys_pump_flow_control_a} that if   ${d}_i$ is fixed to a feasible equilibrium value ${\bar{d}}_i$, then $\vec{\bar{x}}_i^\mathrm{f}$ is such that $\bar{q}_i=\bar{q}_{\mathrm{P},i}=q_i^*$.

\section{Proof of Proposition~\ref{prop:control_law_uv}} \label{app:proof:flow_control_valve_general}

Following the same reasoning as in the proof of Proposition~\ref{prop:flow_control_pump_general} and using $\bar{u}_i=\bar{r}_i$, $\pm \hat{\mu}_i(q_i)\bar{r}_i$, we can write the closed-loop system \eqref{eq:cl_generic_valve_control_a} equivalently as
\begin{subequations}\label{eq:proof_valve_controller_cl_sys_shifted}
\begin{align}
    J_i\dot{q}_i  =& -(\lambda_i(q_i)-\lambda_i(\bar{q}_i))+\hat{\mu}_i(q_i)K_{\mathrm{P},i}\hat{y}_i -\hat{\mu}_i(q_i)(r_i-\bar{r}_i)\nonumber \\
    \phantom{J_i\dot{q}_i  =} & -\bar{r}_i(\hat{\mu}_i(q_i)-\hat{\mu}_i(\bar{q}_i))+(d_i-\bar{d}_i)\\
Q_{\mathrm{I},i}\dot{r}_i  =&\hat{\mu}_i(q_i)(q_i-q_i^*).
\end{align}
\end{subequations}
Note that \eqref{eq:proof_valve_controller_cl_sys_shifted} is equivalent to
\begin{equation}\label{eq:proof_pipe_valve_shifted}
    \dot{\x}_i^\mathrm{v} = \vec{f}_i^\mathrm{v}(\x_i^\mathrm{v})-\vec{f}_i^\mathrm{v}(\bar{x}_i^\mathrm{v})+ \vec{K}_i^\mathrm{v}(d_i-\bar{d}_i),
\end{equation}
for any arbitrary equilibrium pair $(\bar{d}_i,\bar{z}_i)$ and associated equilibrium state vector $\bar{\x}_i^\mathrm{v}$ (see \eqref{eq:equilibrium_valve_close_loop_flow}).
For the time-derivative of $H_i^\mathrm{v}$ in \eqref{eq:H_v}, it holds that
\begin{align}\label{eq:H_v_dot_generic}
    \dot{H}_i^\mathrm{v}   & =\nabla^\top  H_i^\mathrm{v}\dot{\x}_i^\mathrm{v}= -\psi^\mathrm{v}(\x_i^\mathrm{v}) +(z_i-\bar{z}_i)(d_i-\bar{d}_i),\\
    \psi^\mathrm{v}(\x_i^\mathrm{v})  & =(q_i-\bar{q}_i) (\lambda_i(q_i)-\lambda_i(\bar{q}_i)) +K_{\mathrm{P},i}\hat{y}_i^2 \nonumber \\
    & ~~~ +\bar{r}_i (q_i-\bar{q}_i)(\hat{\mu}_i(q_i)-\hat{\mu}_i(\bar{q}_i)), \label{eq:dissipation_x_v}
\end{align}
where we have used the identity 
$$(\K^\mathrm{v}_i)^\top \nabla H^\mathrm{v}_i=	z_i-\bar{{z}}_i=q_i-\bar{q}_i.$$ 
Since both $\lambda_i(q_i)$ and $\hat{\mu}_i(q_i)$ are strictly increasing (see Assumption~\ref{assumption:incompressibility_pos_values} and \eqref{eq:modeling:valve_substitution}), $\bar{u}_i=\bar{r}_i>0$ per definition (see Section~\ref{sec:modeling:valve}), and $K_{\mathrm{P},i}>0$, it follows that $\psi_i(\x_i^\mathrm{v})\geq  0$ implying $\dot{H}_i^\mathrm{v} \leq (z_i-\bar{z}_i)(d_i-\bar{d}_i)$. Hence, EIP is established.

\section{Proof of Lemma~\ref{lemma:actuated_property_fufilled}} \label{app:proof_actuated_property_fulfilled}

In the following, we successively consider the closed-loop systems according to their order in Problem~\ref{problem} (a)--(i). Thus, we begin with $i\in \mathcal{D}_\mathrm{form}$. By combining the open-loop DGU model \eqref{eq:modeling:circulation_equations} with $u_{\mathrm{P},i}$ as in \eqref{eq:controller_pump_specific} and fixing $u_{\mathrm{v},i}=\bar{u}_{\mathrm{v},i}>0$ (due to $s_{\mathrm{v},i}=1$), we can write the closed-loop as in \eqref{eq:models_cl}  
\begin{subequations}\label{eq:closed_loop_D_form}
    \begin{align}
        \frac{\mathrm{d}}{\mathrm{d}t}\underbrace{\begin{bmatrix}
            J_iq_i\\
            J_{\mathrm{P},i}\chi_i\\
            C_{\mathrm{P},i}p_{\mathrm{P},i}\\
            Q_{\mathrm{I},i}r_i
        \end{bmatrix}}_{\vec{\hat{x}_i}} & =   \underbrace{\begin{bmatrix}p_{\mathrm{P},i}-\lambda_i(q_i)-\hat{\mu}_i(q_i)\bar{u}_{\mathrm{v},i}\\
        -R_i^\mathrm{p}\chi_i-\left(p_{\mathrm{P},i}-p_{\mathrm{P},i}^* \right)\\
        \chi_i-r_i\\
        (p_{\mathrm{P},i}-p_{\mathrm{P},i}^*)
        \end{bmatrix}}_{\vec{\hat{f}}_i(\vec{\hat{x}}_i)}+\underbrace{\begin{bmatrix}
        1\\
        0\\
        0\\
        0 \end{bmatrix}}_{\vec{\hat{K}}_i}d_i,\\
        z_i & = \underbrace{ \begin{bmatrix}
    \tfrac{1}{J_i} & 0 & 0 & 0 
    \end{bmatrix}}_{\vec{\hat{T}_i}}\vec{\hat{x}_i},
    \end{align}
\end{subequations}
with $d_i$ as in  \eqref{eq:modeling:circulation_equations}. 
To show that \eqref{eq:closed_loop_D_form} is EIP, we follow the same reasoning as in the proofs of Proposition~\ref{prop:pressure_control_pump_general} and \ref{prop:flow_control_pump_general}. The storage function 
$\hat{H}_i$ is as in \eqref{eq:storage_function_close_loop_subsystems} with
\begin{equation}
    \vec{\hat{Q}}_i=\text{diag}^{-1}(J_i,J_{\mathrm{P},i},C_{\mathrm{P},i},Q_{\mathrm{I},i}).
\end{equation}
The time derivative of $\hat{H}_i$ along solutions of \eqref{eq:closed_loop_D_form}  satisfies
\begin{align}\label{eq:time_der_stora_D_form}
    &   \dot{\hat{H}}_i=\nabla^\top \hat{H}_i\dot{\vec{\hat{x}}}_i=-\hat{\psi}_i(\vec{\hat{x}}_i) +(z_i-\bar{z}_i)(d_i-\bar{d}_i),\\
    &\hat{\psi}_i(\vec{\hat{x}}_i) =(q_i-\bar{q}_i)\left(\lambda_i(q_i)-\lambda_i(\bar{q}_i)\right) +R_i^\mathrm{P}(\chi_i-\bar{\chi}_i)^2 \nonumber \\
     & \phantom{\hat{\psi}_i(\vec{\hat{x}}_i)_i =} +(q_i-\bar{q}_i)\bar{u}_{\mathrm{v},i}(\hat{\mu}_i(q_i)-\hat{\mu}(\bar{q}_i)) , \label{eq:dissipation_D_form}
\end{align}
where we have used the identity 
$$(\hat{\K}_i)^\top \nabla \hat{H}_i=	z_i-\bar{{z}}_i=q_i-\bar{q}_i=\vec{\hat{T}}_i(\vec{\hat{x}}_i-\bar{\vec{\hat{x}}}_i).$$ 
As $\lambda_i$ and $\hat{\mu}_i$ are strictly increasing and $\bar{u}_{\mathrm{v},i},R_i^\mathrm{P}>0$, it follows that $\hat{\psi}_i(\vec{\hat{x}}_i)\geq 0$ and  $\dot{\hat{H}}_i\leq (z_i-\bar{z}_i)(d_i-\bar{d}_i)$. Hence, \eqref{eq:closed_loop_D_form} is EIP.  Finally, observe that due to the integral action \eqref{eq:controller_pump_specific_integral}, it holds for any feasible equilibrium value of $\vec{\hat{x}}_i$ that $\bar{p}_{\mathrm{P},i}=p_{\mathrm{P},i}^*$.

\medskip

Next, we consider $i\in \mathcal{D}_\mathrm{valve}$. By combining the open-loop DGU model \eqref{eq:modeling:circulation_equations} with $u_{\mathrm{P},i}$ and $u_{\mathrm{v},i}$ as in \eqref{eq:controller_pump_specific} and \eqref{eq:PI_controller_valve_generic}, respectively, we can write the closed-loop as in \eqref{eq:models_cl}
\begin{subequations}\label{eq:closed_loop_D_valve}
    \begin{equation}
        \resizebox{0.98\hsize}{!}{$\frac{\mathrm{d}}{\mathrm{d}t}\underbrace{\begin{bmatrix}
            J_iq_i\\
            J_{\mathrm{P},i}\chi_i\\
            C_{\mathrm{P},i}p_{\mathrm{P},i}\\
            Q_{\mathrm{I},i}^{\alpha}r_i^{\alpha}\\
            Q_{\mathrm{I},i}^{\beta}r_i^{\beta}
        \end{bmatrix}}_{\vec{\hat{x}_i}}  =   \underbrace{\begin{bmatrix}p_{\mathrm{P},i}-\lambda_i(q_i)-\hat{\mu}_i(q_i)\left( -K_{\mathrm{P},i}\hat{y}_i+r_i^\beta\right)\\
        -R_i^\mathrm{p}\chi_i-\left(p_{\mathrm{P},i}-p_{\mathrm{P},i}^* \right)\\
        \chi_i-r_i^\alpha\\
        (p_{\mathrm{P},i}-p_{\mathrm{P},i}^*)\\
        \hat{\mu}_i(q_i)(q_i-q_i^*)
        \end{bmatrix}}_{\vec{\hat{f}}_i(\vec{\hat{x}}_i)}+\underbrace{\begin{bmatrix}1\\
        0\\
        0\\
        0\\
        0\end{bmatrix}}_{\vec{\hat{K}}_i}d_i,$}
    \end{equation}
    \begin{equation}
        z_i=\underbrace{\begin{bmatrix}
    \tfrac{1}{J_i} & 0 & 0 & 0 & 0 
    \end{bmatrix}}_{\vec{\hat{T}_i}}\vec{\hat{x}}_i,
    \end{equation}
\end{subequations}
with  $d_i$ as in  \eqref{eq:modeling:circulation_equations}. Note that we have used the indices $\alpha$ and $\beta$ to distinguish between the integral actions of \eqref{eq:controller_pump_specific} and \eqref{eq:PI_controller_valve_generic}, respectively. 
To show that \eqref{eq:closed_loop_D_valve} is EIP, we proceed as before. The storage function $\hat{H}_i$ is as in \eqref{eq:storage_function_close_loop_subsystems} with
\begin{equation}
    \vec{\hat{Q}}_i=\text{diag}^{-1}(J_i,J_{\mathrm{P},i},C_{\mathrm{P},i},Q_{\mathrm{I},i}^{\alpha},Q_{\mathrm{I},i}^{\beta}).
\end{equation}
 The time derivative of $\hat{H}_i$ along solutions of \eqref{eq:closed_loop_D_valve} satisfies
\begin{align}
    &   \dot{\hat{H}}_i=\nabla^\top \hat{H}_i\dot{\vec{\hat{x}}}_i = -\hat{\psi}_i(\hat{x}_i) +(z_i-\bar{z}_i)(d_i-\bar{d}_i),\\
   &\hat{\psi}_i(\hat{\x}_i) =(q_i-\bar{q}_i) (\lambda_i(q_i)-\lambda_i(\bar{q}_i))  +R_i^\mathrm{P}(\chi_i-\bar{\chi}_i)^2 \nonumber \\
    & \phantom{\hat{\psi}_i(\hat{\x}_i) =} +\bar{r}_i^\beta(q_i-\bar{q}_i)(\hat{\mu}_i(q_i)-\hat{\mu}_i(\bar{q}_i))  + K_{\mathrm{P},i}\hat{y}_i^2 \label{eq:dissipation_D_valve}
\end{align}
where we have used the identity 
$$(\hat{\K}_i)^\top \nabla \hat{H}_i=	z_i-\bar{{z}}_i=q_i-\bar{q}_i=\vec{\hat{T}}_i(\vec{\hat{x}}_i-\bar{\vec{\hat{x}}}_i).$$
With the same reasoning as for \eqref{eq:dissipation_x_v} and $R_i^\mathrm{P}>0$, it follows that $\hat{\psi}_i(\vec{\hat{x}}_i)\geq 0$ and $\dot{\hat{H}}_i\leq (z_i-\bar{z}_i)(d_i-\bar{d}_i)$. Hence, EIP is established.  Finally, observe that due to the integral actions \eqref{eq:controller_pump_specific_integral} and \eqref{eq:controller_valve_integral} and Assumption~\ref{assumption:valve_equilibrium}, it holds for any feasible equilibrium value of $\vec{\hat{x}}_i$ that $\bar{q}_i=q_i^*$ and  $\bar{p}_{\mathrm{P},i}=p_{\mathrm{P},i}^*$. 

\medskip

For any $i\in \mathcal{D}_\mathrm{VSP}$, we combine the open-loop DGU model \eqref{eq:modeling:circulation_equations} with $u_{\mathrm{P},i}$ as in  \eqref{prop:flow_control_pump_general} and fix $u_{\mathrm{v},i}=\bar{u}_{\mathrm{v},i}>0$ (due to $s_{\mathrm{v},i}=1$) to write the closed-loop as in \eqref{eq:models_cl}
\begin{subequations}\label{eq:closed_loop_D_VSP}
        \begin{equation}
        \resizebox{0.98\hsize}{!}{$\frac{\mathrm{d}}{\mathrm{d}t}\underbrace{\begin{bmatrix}
            J_iq_i\\
            J_{\mathrm{P},i}q_{\mathrm{P},i}\\
            C_{\mathrm{P},i}p_{\mathrm{P},i}\\
            Q_{\mathrm{I},i}r_i
        \end{bmatrix}}_{\vec{\hat{x}_i}}  =   \underbrace{\begin{bmatrix}p_{\mathrm{P},i}-\lambda_i(q_i)-\hat{\mu}_i(q_i)\bar{u}_{\mathrm{v},i}\\
       -p_{\mathrm{P},i}-R_{\mathrm{P},i}q_{\mathrm{P},i}-K_{\mathrm{P},i}p_{\mathrm{P},i}-r_i\\
        q_{\mathrm{P},i}-q_i\\
        (q_i-q_i^*)\end{bmatrix}}_{\vec{\hat{f}}_i(\vec{\hat{x}}_i)}
        +\underbrace{
        \begin{bmatrix}
        1\\
        0\\
        0\\
        0 \end{bmatrix}}_{\vec{\hat{K}}_i}d_i$},
    \end{equation}
\begin{equation}
        z_i  = \underbrace{ \begin{bmatrix}
    \tfrac{1}{J_i} & 0 & 0 & 0 
    \end{bmatrix}}_{\vec{\hat{T}_i}}\vec{\hat{x}_i},
\end{equation}
\end{subequations}
with  $d_i$ as in \eqref{eq:modeling:circulation_equations}.
To show that  \eqref{eq:closed_loop_D_VSP} is EIP, we note that $\hat{H}_i$ is given by \eqref{eq:storage_function_close_loop_subsystems} with $\vec{\hat{Q}}_i=\vec{Q}_i^\mathrm{f}$ from \eqref{eq:H_f}. The time derivative of $\hat{H}_i$ along solutions of \eqref{eq:closed_loop_D_VSP}  satisfies
\begin{align}
    &   \dot{\hat{H}}_i=\nabla^\top \hat{H}_i\dot{\vec{\hat{x}}}_i =-\hat{\psi}_i(\vec{\hat{x}}_i) +(z_i-\bar{z}_i)(d_i-\bar{d}_i),\\
     &\hat{\psi}_i(\vec{\hat{x}}_i) = (q_i-\bar{q}_i)\left(\lambda_i(q_i)-\lambda_i(\bar{q}_i)\right) \nonumber \\
    &\phantom{\hat{\psi}_i(\vec{\hat{x}}_i) =} +(q_i-\bar{q}_i)\bar{u}_{\mathrm{v},i}(\hat{\mu}_i(q_i)-\hat{\mu}(\bar{q}_i))\nonumber\\
    &\phantom{\hat{\psi}_i(\vec{\hat{x}}_i) =} +2\tfrac{R_{\mathrm{P},i}Q_{\mathrm{I},i}}{\kappa_i^\mathrm{f}}(q_{\mathrm{P},i}-\bar{q}_{\mathrm{P},i})^2, \label{eq:dissipation_D_vsp}
\end{align}
where we have used the identity
$$(\vec{\hat{K}}_i)^\top \nabla \hat{H}_i=	z_i-\bar{{z}}_i=q_i-\bar{q}_i=\vec{\hat{T}}_i(\vec{\hat{x}}_i-\bar{\vec{\hat{x}}}_i),$$
As $\lambda_i$ and $\hat{\mu}_i$ are strictly increasing, $\bar{u}_{\mathrm{v},i}>0$, and \eqref{eq:conds_stab_pump_flow} holds, it follows that $\hat{\psi}_i(\vec{\hat{x}}_i)\geq 0$ and $\dot{\hat{H}}_i\leq (z_i-\bar{z}_i)(d_i-\bar{d}_i)$. Hence,  \eqref{eq:closed_loop_D_VSP} is EIP.  Finally, observe that due to the integral action \eqref{eq:PI_control_flow_pump_integral}, it holds for any feasible equilibrium value of $\vec{\hat{x}}_i$ that $\bar{q}_i=\bar{q}_{\mathrm{P},i}=q_{i}^*$. 

\medskip

For describing the consumers $\mathcal{L}$ in closed-loop, we recall that  the open-loop model of any $i\in \mathcal{L}$ is identical to that of any DGU $i\in \mathcal{D}$, i.e., to \eqref{eq:modeling:circulation_equations}. For any $i\in \mathcal{L}_\mathrm{boost}$, we assign $u_{\mathrm{P},i}$ and $u_{\mathrm{v},i}$  as in \eqref{eq:controller_pump_specific} and \eqref{eq:PI_controller_valve_generic}, respectively. Then, the resulting closed-loop is equivalent to \eqref{eq:closed_loop_D_valve} with the same EIP and equilibrium properties. 
Next, we consider $i\in \mathcal{L}_\mathrm{valve}$. In this case, the associated pump can either be turned off or is not present. Thus, the open-loop of any $i\in \mathcal{L}_\mathrm{valve}$ is equivalent to that of a control valve in series with a pipe element, i.e., to \eqref{eq:modeling:mixing_equations}. By closing the loop with $u_{\mathrm{v},i}$ as in \eqref{eq:PI_controller_valve_generic}, we obtain a closed-loop system equivalent to \eqref{eq:cl_generic_valve_control_a}, which is clearly of the form \eqref{eq:models_cl}. Moreover, EIP and $\bar{q}_i=q_{i}^*$ directly follow from  Proposition~\ref{prop:control_law_uv} and Assumption~\ref{assumption:valve_equilibrium}. 
Lastly, for any $i\in \mathcal{L}_\mathrm{VSP}$, we fix $u_{\mathrm{v},i}=\bar{u}_{\mathrm{v},i}>0$ and assign $u_{\mathrm{P},i}$ as in  \eqref{prop:flow_control_pump_general}. Then, the resulting closed-loop is equivalent to \eqref{eq:closed_loop_D_VSP} with the same EIP and equilibrium properties. 

\medskip

Next, we consider $i\in \mathcal{P}_\mathrm{boost}$, i.e., an arbitrary pipe in series with a booster pump. By combining the open-loop pipe model \eqref{eq:modeling:pipe_equations} with $u_{\mathrm{P},i}$ as in \eqref{eq:controller_pump_specific}, we can write the closed-loop dynamics as in \eqref{eq:models_cl}
\begin{subequations}\label{eq:closed_loop_P_boost}
    \begin{equation}
        \resizebox{0.9\hsize}{!}{$\frac{\mathrm{d}}{\mathrm{d}t}\underbrace{\begin{bmatrix}
            J_iq_i\\
            J_{\mathrm{P},i}\chi_i\\
            C_{\mathrm{P},i}p_{\mathrm{P},i}\\
            Q_{\mathrm{I},i}r_i
        \end{bmatrix}}_{\vec{\hat{x}_i}}  =   \underbrace{\begin{bmatrix}p_{\mathrm{P},i}-\lambda_i(q_i)\\
        -R_i^\mathrm{p}\chi_i-\left(p_{\mathrm{P},i}-p_{\mathrm{P},i}^* \right)\\
        \chi_i-r_i\\
        (p_{\mathrm{P},i}-p_{\mathrm{P},i}^*)
        \end{bmatrix}}_{\vec{\hat{f}}_i(\vec{\hat{x}}_i)}+\underbrace{\begin{bmatrix}1\\
        0\\
        0\\
        0\end{bmatrix}}_{\vec{\hat{K}}_i}d_i,$}
    \end{equation}
    \begin{equation}
        z_i=\underbrace{\begin{bmatrix}
    \tfrac{1}{J_i} & 0 & 0 & 0 
    \end{bmatrix}}_{\vec{\hat{T}_i}}\vec{\hat{x}}_i,
    \end{equation}
\end{subequations}
with $d_i$ as in \eqref{eq:modeling:pipe_equations}. 
To show that \eqref{eq:closed_loop_P_boost} is EIP, we note that $\hat{H}_i$ is given by \eqref{eq:storage_function_close_loop_subsystems} with
\begin{equation}
    \vec{\hat{Q}}_i=\text{diag}^{-1}(J_i,J_{\mathrm{P},i},C_{\mathrm{P},i},Q_{\mathrm{I},i}).
\end{equation}
The time derivative of $\hat{H}_i$ along solutions of \eqref{eq:closed_loop_P_boost} satisfies
\begin{align}
     &   \dot{\hat{H}}_i=\nabla^\top \hat{H}_i\dot{\vec{\hat{x}}}_i = -\hat{\psi}_i(\hat{x}_i) +(z_i-\bar{z}_i)(d_i-\bar{d}_i),\\
    &\hat{\psi}_i(\hat{\x}_i) =(q_i-\bar{q}_i) (\lambda_i(q_i)-\lambda_i(\bar{q}_i)) +R_i^\mathrm{P}(\chi_i-\bar{\chi}_i)^2 \label{eq:dissipation_P_boost}
\end{align}
 where we have used the identity 
$$(\hat{\K}_i)^\top \nabla \hat{H}_i=	z_i-\bar{{z}}_i=q_i-\bar{q}_i=\vec{\hat{T}}_i(\vec{\hat{x}}_i-\bar{\vec{\hat{x}}}_i).$$ 
As $\lambda_i$ is strictly increasing and $R_i^\mathrm{P}>0$, it follows that $\hat{\psi}_i(\vec{\hat{x}}_i)\geq 0$ and $\dot{\hat{H}}_i\leq (z_i-\bar{z}_i)(d_i-\bar{d}_i)$. Hence, \eqref{eq:closed_loop_P_boost} is EIP. Finally, observe that due to the integral action \eqref{eq:controller_pump_specific_integral}, it holds for any feasible equilibrium value  $\vec{\hat{x}}_i$ that $\bar{p}_{\mathrm{P},i}=p_{\mathrm{P},i}^*$.

\medskip

The final type of actuated edges corresponds to mixing valves $i\in \mathcal{M}$. By combining the open-loop model \eqref{eq:modeling:mixing_equations} with  $u_{\mathrm{v},i}$ as in \eqref{eq:PI_controller_valve_generic}, we obtain a closed-loop system equivalent to \eqref{eq:cl_generic_valve_control_a}, which is clearly of the form \eqref{eq:models_cl}. Moreover, EIP and $\bar{q}_i=q_{i}^*$ directly follow from  Proposition~\ref{prop:control_law_uv} and Assumption~\ref{assumption:valve_equilibrium}. 

\medskip

Finally, we consider pressure holding units $i\in \mathcal{H}$, which are the only actuated nodes. By combining the open-loop model \eqref{eq:model_pressure_holding} with $u_{\mathrm{P},i}$ as in \eqref{eq:controller_pump_specific}, we obtain a closed-loop system equivalent to \eqref{eq:closed_loop_dyn_press_control_general}. Clearly, this system can be written as in \eqref{eq:models_cl}. Moreover, EIP and $\bar{p}_{\mathrm{P},i}=p_{\mathrm{P},i}^*$ follow directly from Proposition~\ref{prop:pressure_control_pump_general}.

\section{Proof of Lemma~\ref{lemma:EIP_unactuated}}\label{app:proof_eip_unactuated}

For any pipe $i\in \mathcal{P}\setminus \mathcal{P}_\mathrm{boost}$ without booster pump, we use the open-loop model \eqref{eq:modeling:pipe_equations} and set $u_{\mathrm{P},i}=0$ to obtain a model of form \eqref{eq:models_cl}.
\begin{subequations}\label{eq:pipe_no_booster}
\begin{align}
   \frac{\mathrm{d}}{\mathrm{d}t}\underbrace{J_iq_i}_{{\hat{x}}_i} & = \underbrace{-\lambda_i(q_i)}_{{\hat{f}}_i({\hat{x}}_i)}+\underbrace{[1]}_{{\hat{K}}_i}d_i,\\
   z_i & = q_i= \underbrace{[\frac{1}{J_i}]}_{{\hat{T}}_i}{\hat{x}}_i,
\end{align}
\end{subequations}
with $d_i$ as in \eqref{eq:modeling:pipe_equations}. 
EIP of \eqref{eq:pipe_no_booster} follows directly by using $\hat{H}_i$ in \eqref{eq:storage_function_close_loop_subsystems} as storage function with ${\hat{Q}}_i=\tfrac{1}{J_i}$ and noting that
\begin{subequations}\label{eq:H_dot_pipes_noboost}
\begin{align}
    \dot{\hat{H}}_i & =-\hat{\psi}_i(\hat{x}_i)+(z_i-\bar{z}_i)(d_i-\bar{d}_i),\\
    \hat{\psi}_i(\hat{x}_i) & = (q_i-\bar{q}_i)(\lambda_i(q_i)-\lambda_i(\bar{q}_i))\geq0.
\end{align}
\end{subequations}
%
% As $\lambda_i$ is strictly increasing, $\hat{\psi}_i(\vec{\hat{x}}_i)\leq 0$ and $\dot{\hat{H}}_i\leq (z_i-\bar{z}_i)(d_i-\bar{d}_i)$. Hence, \eqref{eq:pipe_no_booster} is EIP.
%
Writing the model  \eqref{eq:model_capacitive_nodes} of any capacitive node $i\in \mathcal{C}$ as in \eqref{eq:models_cl} is trivial.
% \begin{subequations}\label{eq:EIP_form_capacitive}
%     \begin{align}
%         \frac{\mathrm{d}}{\mathrm{d}t}\underbrace{C_ip_i}_{\vec{\hat{x}}_i} & = \underbrace{[1]}_{\vec{\hat{K}}_i}d_i,\\
%         z_i & = p_i=\underbrace{[\frac{1}{C_i}]}_{\vec{\hat{T}}_i}\vec{\hat{x}}_i,
%     \end{align}
% \end{subequations}
% where $d_i=\sum_{j\in \mathcal{I}_i}$, as in \eqref{eq:capacitor_dynamics_PH}. 
Furthermore, EIP of system \eqref{eq:model_capacitive_nodes}  follows directly by using $\hat{H}_i$ in \eqref{eq:storage_function_close_loop_subsystems} as a storage function  with ${\hat{Q}}_i=\tfrac{1}{C_i}$ and noting that
\begin{equation}
    \dot{\hat{H}}_i(\hat{x}_i)=(z_i-\bar{z}_i)(d_i-\bar{d}_i).
\end{equation}

\section{Proof of Theorem~\ref{thm:stability_analysis}} \label{app:proof_stability}

Let $(\bar{\vec{\hat{x}}}_\mathcal{E},\bar{\vec{\hat{x}}}_\Delta)$ denote a feasible equilibrium of the closed-loop system \eqref{eq:equivalent_ODE}.  Due to Lemma~\ref{lemma:actuated_property_fufilled},  this equilibrium is such that each bullet point in Problem~\ref{problem} is fulfilled. Consider the following Lyapunov function candidate:
\begin{equation}
    V(\vec{\hat{x}}_\mathcal{E},\vec{\hat{x}}_\Delta)=\sum_{i\in \mathcal{E}\cup \Delta }\hat{H}_i(\vec{\hat{x}}_i),
\end{equation}
where each $\hat{H}_i$ is defined in Lemmas~\ref{lemma:actuated_property_fufilled} or \ref{lemma:EIP_unactuated}. Note that $V(\vec{\hat{x}}_\mathcal{E},\vec{\hat{x}}_\Delta)>0$ for all $(\vec{\hat{x}}_\mathcal{E},\vec{\hat{x}}_\Delta)\neq (\bar{\vec{\hat{x}}}_\mathcal{E},\bar{\vec{\hat{x}}}_\Delta)$ and $V(\bar{\vec{\hat{x}}}_\mathcal{E},\bar{\vec{\hat{x}}}_\Delta)=0$. The time derivative of $V$ along solutions of \eqref{eq:equivalent_ODE} on the invariant set $\mathbb{M}$ in \eqref{eq:invariant_set} is given by
\begin{align}
    & \dot{V}(\vec{\hat{x}}_\mathcal{E},\vec{\hat{x}}_\Delta)  =\sum_{ i\in \mathcal{E}\cup \Delta}\nabla \hat{H}_i^\top(\vec{\hat{x}}_i) \dot{\vec{\hat{x}}}_i\nonumber \\
    &  =\sum_{i\in \mathcal{E}\cup \Delta}\nabla \hat{H}_i^\top(\vec{\hat{x}}_i) \left(\vec{\hat{f}}_i(\vec{\hat{x}}_i)-\vec{\hat{f}}_i(\bar{\vec{\hat{x}}}_i) \right). \label{eq:dot_V_overall}
\end{align}
%
% {\color{red}
To get this equality, we have used three facts about any $i\in \mathcal{E}\cup \Delta$. {\bf (I)} That for  any any feasible equilibrium input ${d}_i$,   $\bar{\vec{\hat{x}}}_i$ satisfies $\vec{\hat{f}}_i(\bar{\vec{\hat{x}}}_i)+\vec{\hat{K}}_i\bar{{d}}_i=0$, implying that $\dot{\vec{\hat{x}}}_i=\vec{\hat{f}}_i(\vec{\hat{x}}_i)-\vec{\hat{f}}_i(\bar{\vec{\hat{x}}}_i)+\vec{\hat{K}}_i(\vec{\hat{d}}_i-\bar{\vec{\hat{d}}}_i)$.  {\bf (II)}  That due to   Lemmas~\ref{lemma:actuated_property_fufilled}  and \ref{lemma:EIP_unactuated},  $\nabla \hat{H}_i^\top(\vec{\hat{x}}_i) \hat{K}_i=({z}_i-\bar{{z}}_i)^\top$. {\bf (III)} That   Lemma~\ref{lemma:interconnection} implies 
$$\sum_{i\in \{\mathcal{D},\mathcal{L},\mathcal{P},\mathcal{M},\HH,\mathcal{C},\mathcal{K} \}}({z}_i-\bar{{z}}_i)^\top ({d}_i-\bar{{d}}_i) =0.$$
% }
%
Stability of $(\bar{\vec{\hat{x}}}_\mathcal{E},\bar{\vec{\hat{x}}}_\Delta)$ follows directly from \eqref{eq:dot_V_overall}, as its right-hand side is upper bounded by zero according to the EIP properties established in Lemmas~\ref{lemma:actuated_property_fufilled} and \ref{lemma:EIP_unactuated}. 

We move on to show that $(\bar{\vec{\hat{x}}}_\mathcal{E},\bar{\vec{\hat{x}}}_\Delta)$ is in fact asymptotically stable. Considering LaSalle's invariance principle, at this point it is sufficient to show that $(\bar{\vec{\hat{x}}}_\mathcal{E},\bar{\vec{\hat{x}}}_\Delta)$ is the largest invariant set of \eqref{eq:equivalent_ODE} where  $\dot{V}$ is zero. 
Next we characterize the conditions under which $ \dot{V}(\vec{\hat{x}})=0$. For that we split the summands in the right-hand side of \eqref{eq:dot_V_overall} as we explain next.

Firstly, from the proof of Lemma~\ref{lemma:actuated_property_fufilled} we have that
\begin{align}\label{eq:sum_v_dot_sources}
    \sum_{i\in \mathcal{D}}\nabla \hat{H}_i^\top(\vec{\hat{x}}_i) \left(\vec{\hat{f}}_i(\vec{\hat{x}}_i)-\vec{\hat{f}}_i(\bar{\vec{\hat{x}}}_i)\right) & =-\sum_{i\in \mathcal{D}}\hat{\psi}_i(\vec{\hat{x}}_i),
\end{align}
where $\hat{\psi}_i$ is given in \eqref{eq:dissipation_D_form}, $\eqref{eq:dissipation_D_valve}$ and \eqref{eq:dissipation_D_vsp} for $i$ in $\mathcal{D}_\mathrm{form}$, $\mathcal{D}_\mathrm{valve}$ and $\mathcal{D}_\mathrm{VSP}$, respectively. For $i\in \mathcal{D}_\mathrm{form}$, the mononoticity of $\lambda_i$ and $\hat{\mu}_i$, and the condition $\bar{u}_{\mathrm{v},i}>0$ (see Assumption~\ref{assumption:stem_position}), imply that $ \hat{\psi}_i(\vec{\hat{x}}_i)=0$ if and only if
\begin{equation}
   \vec{\hat{x}}_i\in \Xi_i=\{\vec{\hat{x}}_i:~q_i=\bar{q}_i,~\chi_i=\bar{\chi}_i=0 \}.
\end{equation}
For $i\in \mathcal{D}_\mathrm{valve}$, the monotonicity of $\lambda_i$ in combination with assumptions~\ref{assumption:stem_position} and \ref{assumption:valve_equilibrium}, imply that $\hat{\psi}_i(\vec{\hat{x}}_i)=0$ if and only if
\begin{equation}
    \vec{\hat{x}}_i\in \Xi_i=\{\vec{\hat{x}}_i:~q_i=\bar{q}_i=q_i^*,~\chi_i=\bar{\chi}_i=0 \}.
\end{equation}
For $i\in \mathcal{D}_{\mathrm{VSP}}$, the monotonicity of $\lambda_i$ and $\hat{\mu}_i$, and the condition $\bar{u}_{\mathrm{v},i}>0$ (see again assumptions~\ref{assumption:stem_position} and \ref{assumption:valve_equilibrium}), imply that $ \hat{\psi}_i(\vec{\hat{x}}_i)=0$ if and only if
\begin{equation}
\vec{\hat{x}}_i\in \Xi_i=\{\vec{\hat{x}}_i:~q_i=\bar{q}_i=q_i^*,~q_{\mathrm{P},i}=\bar{q}_{\mathrm{P},i}=q_i^* \}.
\end{equation}

Following an analogous reasoning for the set of consumers $\mathcal{L}$, we have from Lemma~\ref{lemma:actuated_property_fufilled} that
\begin{align}\label{eq:sum_v_dot_consumers}
    \sum_{i\in \mathcal{L}}\nabla \hat{H}_i^\top(\vec{\hat{x}}_i) \left(\vec{\hat{f}}_i(\vec{\hat{x}}_i)-\vec{\hat{f}}_i(\bar{\vec{\hat{x}}}_i)\right) & =-\sum_{i\in \mathcal{L}}\hat{\psi}_i(\vec{\hat{x}}_i),
\end{align}
where $\hat{\psi}_i(\vec{\hat{x}}_i)=\hat{\psi}_j(\vec{\hat{x}}_j)$  for any  $i\in \mathcal{L}_\mathrm{boost}$ and $j\in \mathcal{D}_\mathrm{valve}$; $\hat{\psi}_i=\psi_i^\mathrm{v}$  for any $i\in \mathcal{L}_\mathrm{valve}$, with $\psi_i^\mathrm{v}$   in \eqref{eq:dissipation_x_v}; and   $\hat{\psi}_i=\hat{\psi}_j$ for any $i\in \mathcal{L}_\mathrm{VSP}$ and $j\in \mathcal{D}_\mathrm{VSP}$. Then, $\hat{\psi}_i(\vec{\hat{x}}_i)=0$ for any $i\in \mathcal{L}_\mathrm{boost}$   if and only if
\begin{equation}
 \vec{\hat{x}}_i\in \Xi_i=\{\vec{\hat{x}}_i:~q_i=\bar{q}_i=q_i^*,~\chi_i=\bar{\chi}_i=0 \},
\end{equation}
$\hat{\psi}_i(\vec{\hat{x}}_i)=0$  for any $i\in \mathcal{L}_\mathrm{valve}$ if and only if
\begin{equation}
\vec{\hat{x}}_i\in \Xi_i=\{\vec{\hat{x}}_i:~q_i=\bar{q}_i=q_i^*\},
\end{equation}
and $\hat{\psi}_i(\vec{\hat{x}}_i)=0$ for $i\in \mathcal{L}_\mathrm{VSP}$  if and only if
\begin{equation}
\vec{\hat{x}}_i\in \Xi_i=\{\vec{\hat{x}}_i:~q_i=\bar{q}_i=q_i^*,~q_{\mathrm{P},i}=\bar{q}_{\mathrm{P},i}=q_i^* \}.
\end{equation}

Moving on to the pipes with booster pumps $\mathcal{P}_\mathrm{boost}$, it holds  due to Lemma~\ref{lemma:actuated_property_fufilled} that
\begin{align}\label{eq:sum_v_dot_boost_pipe}
    \sum_{i\in \mathcal{P}_\mathrm{boost}}\!\!\!\!\nabla \hat{H}_i^\top(\vec{\hat{x}}_i) \left(\vec{\hat{f}}_i(\vec{\hat{x}}_i)-\vec{\hat{f}}_i(\bar{\vec{\hat{x}}}_i)\right) & =-\!\!\!\!\!\!\sum_{i\in \mathcal{P}_\mathrm{boost}}\hat{\psi}_i(\vec{\hat{x}}_i),
\end{align}
where $\hat{\psi}_i$ is in \eqref{eq:dissipation_P_boost}. Then,  for each $i\in \mathcal{P}_\mathrm{boost}$ it holds that $ \hat{\psi}_i(\vec{\hat{x}}_i)=0$ if and only if
\begin{equation}
   \vec{\hat{x}}_i\in \Xi_i=\{\vec{\hat{x}}_i:~q_i=\bar{q}_i,~\chi_i=\bar{\chi}_i=0 \}.
\end{equation}

Next we consider the set of mixing valves $\mathcal{M}$. Due to Lemma~\ref{lemma:actuated_property_fufilled}  it holds that 
\begin{align}\label{eq:sum_v_dot_mixingv}
    \sum_{i\in \mathcal{M}}\nabla \hat{H}_i^\top(\vec{\hat{x}}_i) \left(\vec{\hat{f}}_i(\vec{\hat{x}}_i)-\vec{\hat{f}}_i(\bar{\vec{\hat{x}}}_i)\right) & =-\sum_{i\in \mathcal{M}}\hat{\psi}_i(\vec{\hat{x}}_i).
\end{align}
Here $\hat{\psi}_i=\hat{\psi}_j$ for any $i\in \mathcal{M}$ and $j\in \mathcal{L}_\mathrm{valve}$. Then, considering Assumption~\ref{assumption:valve_equilibrium}, it holds that $\hat{\psi}_i(\vec{\hat{x}}_i)=0$ for any $i\in \mathcal{M}$ if and only if
\begin{equation}
\vec{\hat{x}}_i\in \Xi_i=\{\vec{\hat{x}}_i:~q_i=\bar{q}_i=q_i^*\}.
\end{equation}

Moving on to the set of pipes without booster pumps $\mathcal{P}\setminus \mathcal{P}_\mathrm{boost}$, we get from Lemma~\ref{lemma:EIP_unactuated} and its proof that
\begin{equation}\label{eq:sum_v_dot_pipes_no_booster}
  % \resizebox{.95\hsize}{!}{
  \sum_{i\in \mathcal{P}\setminus \mathcal{P}_\mathrm{boost}}\!\!\!\!\!\!\!\!\nabla \hat{H}_i^\top(\vec{\hat{x}}_i) \left(\vec{\hat{f}}_i(\vec{\hat{x}}_i)-\vec{\hat{f}}_i(\bar{\vec{\hat{x}}}_i)\right) =-\!\!\!\!\!\!\!\!\sum_{i\in \mathcal{P}\setminus\mathcal{P}_\mathrm{boost}}\!\!\!\!\!\!\!\!\hat{\psi}_i(\vec{\hat{x}}_i),
\end{equation}
where $\hat{\psi}_i$ is in \eqref{eq:H_dot_pipes_noboost}. Then, $\hat{\psi}_i(\vec{\hat{x}}_i)=0$ for any $i\in \mathcal{P}\setminus \mathcal{P}_\mathrm{boost}$ if and only if
\begin{equation}
   \vec{\hat{x}}_i\in \Xi_i=\{\vec{\hat{x}}_i:~q_i=\bar{q}_i\}.
\end{equation}

For the set of pressure holding units $\mathcal{H}$ (in closed-loop with \eqref{eq:controller_pump_specific}), we have from Lemma~\ref{lemma:actuated_property_fufilled} (and Proposition~\ref{prop:pressure_control_pump_general}) that 
\begin{align}\label{eq:sum_v_dot_press_hold}
    \sum_{i\in \mathcal{H}}\nabla \hat{H}_i^\top(\vec{\hat{x}}_i) \left(\vec{\hat{f}}_i(\vec{\hat{x}}_i)-\vec{\hat{f}}_i(\bar{\vec{\hat{x}}}_i)\right) & =-\sum_{i\in \mathcal{H}}\hat{\psi}_i(\vec{\hat{x}}_i),
\end{align}
where $\hat{\psi}_i=\psi_i^\mathrm{p}$, with $\psi_i^\mathrm{p}$ in \eqref{eq:V_dot_press_control_generic}. Then, $ \hat{\psi}_i(\vec{\hat{x}}_i)=0$ for any $i\in \mathcal{H}$ if and only if
\begin{equation}
\vec{\hat{x}}_i\in \Xi_i=\{\vec{\hat{x}}_i:~\chi_i=\bar{\chi}_i=0\}.
\end{equation}

For the set of capacitive nodes $\mathcal{C}$ it holds from the proof of Lemma~\ref{lemma:EIP_unactuated} that
\begin{align}\label{eq:sum_v_dot_capacitive_nodes}
    \sum_{i\in \mathcal{C}}\nabla \hat{H}_i^\top(\vec{\hat{x}}_i) \left(\vec{\hat{f}}_i(\vec{\hat{x}}_i)-\vec{\hat{f}}_i(\bar{\vec{\hat{x}}}_i)\right) & =0.
\end{align}

Considering these developments, its possible  to characterize the condition  $\dot{V}(\vec{\hat{x}}_\mathcal{E},\vec{\hat{x}}_\Delta)=0$ as follows:
\begin{align*}
  \dot{V}(\vec{\hat{x}})=0 & \Leftrightarrow  \hat{\x}\in \Xi,\\
  \Xi & = \{\hat{\x}:~\hat{x}_i\in\Xi_i,~\forall i\in \mathcal{E}\cup\Delta\}.
\end{align*}
To show asymptotic stability of $(\bar{\vec{\hat{x}}}_\mathcal{E},\bar{\vec{\hat{x}}}_\Delta)$, let $(\vec{\hat{x}}_\mathcal{E},\vec{\hat{x}}_\Delta)\in \Xi$ be any solution of \eqref{eq:equivalent_ODE} in the manifold $\mathbb{M}$ that remains in $\Xi$ for all time. Then, we directly have that $q_i=\bar{q}_i$ for all $i\in \mathcal{E}$. Next we show that $q_i=\bar{q}_i$ for all $i\in \mathcal{E}$ implies $p_j=\bar{p}_j$  for all $j\in \Delta$ and that eventually these two conditions imply that $r_k=\bar{r}_k$ for all $k\in \mathcal{D}\cup \mathcal{L}\cup\mathcal{P}_\mathrm{boost}\cup \mathcal{M}\cup \mathcal{H}$ and hence that $(\vec{\hat{x}}_\mathcal{E},\vec{\hat{x}}_\Delta)=(\bar{\vec{\hat{x}}}_\mathcal{E},\bar{\vec{\hat{x}}}_\Delta)$.

From the dynamics \eqref{eq:model_capacitive_nodes} of any capacitive node $j\in \mathcal{C}$ we have that if $q_i=\bar{q}_i$ for all $i\in \mathcal{E}$, then $\dot{p}_j=0$ for all time. From the closed-loop dynamics \eqref{eq:closed_loop_dyn_press_control_general} of any pressure holding unit $k\in \mathcal{H}$, the condition $q_i=\bar{q}_i$ for all $i\in \mathcal{E}$ together with $\chi_k=0$  lead to $p_{\mathrm{P},k}=\bar{p}_{\mathrm{P},k}$. This means that if $(\vec{\hat{x}}_\mathcal{E},\vec{\hat{x}}_\Delta)\in \mathbb{M}\cap \Xi$ for all time, both $q_i=\bar{q}_i$ for all $i\in \mathcal{E}$  and $p_j=\bar{p}_j$ and for all $k\in \Delta$. A consequence of this is that $r_i=\bar{r}_i$ for any $i\in \mathcal{D}\cup \mathcal{L}\cup\mathcal{P}_\mathrm{boost}\cup \mathcal{M}\cup \mathcal{H}$. Indeed, see on the one hand in  the definition of the pumps and valves controllers (Proposition~\ref{prop:pressure_control_pump_general}, \ref{prop:flow_control_pump_general} and \ref{prop:control_law_uv}) that $\dot{r}_i$ would be zero.  On the other hand, since $r_i$ enters linearly in the dynamics of the---now at equilibrium variables $\chi_i$, $q_{\mathrm{P},i}$ or $q_i$, of the respective pumps and valves dynamics---then $r_i=\bar{r}_i$.  Therefore, $(\bar{\vec{\hat{x}}}_\mathcal{E},\bar{\vec{\hat{x}}}_\Delta)$ is the largest invariant set of the overall closed-loop DHN dynamics contained in $\Xi$. By LaSalle's invariance principle,  $\bar{\hat{x}}$ is an asymptotically stable equilibrium point.

%%%%%%%%%%%%%%%%%%%%%%%%%%%%%%%%%%%%%%%%%%%%%%%%%%%%%%%%%%%%%%%%%%%%%%%%%%%%%%%%

\bibliographystyle{IEEEtran}
\bibliography{PhD_references.bib}

\end{document}